\documentclass[final,12pt,3p,times]{elsarticle}

\def\etajet{\eta^{\rm jet}}
\def\phijet{\phi^{\rm jet}}
\def\pb1{pb$^{-1}$}
\def\cost{\vert\cos\theta^*\vert}
\def\kt{k_t}
\def\colab#1{#1 Collaboration}
\def\as{\alpha_s}
\def\mz{m_Z}
\def\asz{\as(\mz)}
\def\etal{et al.}
\def\etg{E_{\rm T}^{\gamma}}
\def\ptjet{p_{\rm T}^{\rm jet}}
\def\rapjet{y^{\rm jet}}
\def\etag{\eta^{\gamma}}
\def\phig{\phi^{\gamma}}
\def\delphj{\Delta\phi^{\gamma{\rm j}}}
\def\mgjn{m^{\gamma{\rm j}}}
\def\ctgjn{\cos\theta^{\gamma{\rm j}}}

\def\etad{\eta}
\def\phid{\phi}
\def\etagdet{\eta^{\gamma}}
\def\etisop{E_{\rm T,part}^{\rm iso}}
\def\etisod{E_{\rm T,det}^{\rm iso}}

\def\qq{q\bar q}

\def\figdir{./}

\usepackage{mcite}
\usepackage{epsfig}

\usepackage{preprintcover}

\PreprintCoverPaperTitle{\bf Dynamics of isolated-photon plus jet production\\ 
  in {\boldmath $pp$} collisions at {\boldmath $\sqrt s=7$}~TeV with the
  ATLAS detector}

\PreprintIdNumber{CERN-PH-EP-2013-092}

\PreprintCoverAbstract{The dynamics of isolated-photon plus jet
  production in $pp$ collisions at a centre-of-mass energy of $7$ TeV
  has been studied with the ATLAS detector at the LHC using an
  integrated luminosity of $37$~\pb1. Measurements of isolated-photon
  plus jet bin-averaged cross sections are presented as functions of
  photon transverse energy, jet transverse momentum and jet
  rapidity. In addition, the bin-averaged cross sections as functions
  of the difference between the azimuthal angles of the photon and the
  jet, the photon--jet invariant mass and the scattering angle in the
  photon--jet centre-of-mass frame have been
  measured. Next-to-leading-order QCD calculations are compared to the
  measurements and provide a good description of the data, except for
  the case of the azimuthal opening angle.}

\PreprintJournalName{Nuclear Physics B}

\begin{document}

\begin{frontmatter}

\title{\bf Dynamics of isolated-photon plus jet production\\ in
  {\boldmath $pp$} collisions at {\boldmath $\sqrt s=7$}~TeV with the
  ATLAS detector}

\author{The ATLAS Collaboration}

\address{}

\begin{abstract}

The dynamics of isolated-photon plus jet production in $pp$ collisions
at a centre-of-mass energy of $7$ TeV has been studied with the ATLAS
detector at the LHC using an integrated luminosity of
$37$~\pb1. Measurements of isolated-photon plus jet bin-averaged cross
sections are presented as functions of photon transverse energy, jet
transverse momentum and jet rapidity. In addition, the bin-averaged
cross sections as functions of the difference between the azimuthal
angles of the photon and the jet, the photon--jet invariant mass and
the scattering angle in the photon--jet centre-of-mass frame have been
measured. Next-to-leading-order QCD calculations are compared to the
measurements and provide a good description of the data, except for
the case of the azimuthal opening angle.

\end{abstract}


\end{frontmatter}

\section{Introduction}
\label{intro}
The production of prompt photons in association with a jet in
proton--proton collisions, $pp\rightarrow\gamma+{\rm jet}+{\rm X}$,
provides a testing ground for perturbative QCD (pQCD) in a cleaner
environment than in jet production, since the photon originates
directly from the hard interaction. The measurements of angular
correlations between the photon and the jet can be used to probe the
dynamics of the hard-scattering process. Since the dominant production
mechanism in $pp$ collisions at the LHC is through the 
$qg\rightarrow q\gamma$ process, measurements of prompt-photon plus
jet production have been used to constrain the gluon density in the
proton~\cite{np:b860:311,epl:101:61002}. Furthermore, precise
measurements of photon plus jet production are also useful for the
tuning of the Monte Carlo (MC) models. In addition, these events
constitute the main reducible background in the identification of
Higgs bosons decaying to a photon pair.

\begin{figure}[h]
\vfill
\setlength{\unitlength}{1.0cm}
\begin{picture} (18.0,4.9)
\put (1.0,0.0){\epsfig{figure=\figdir 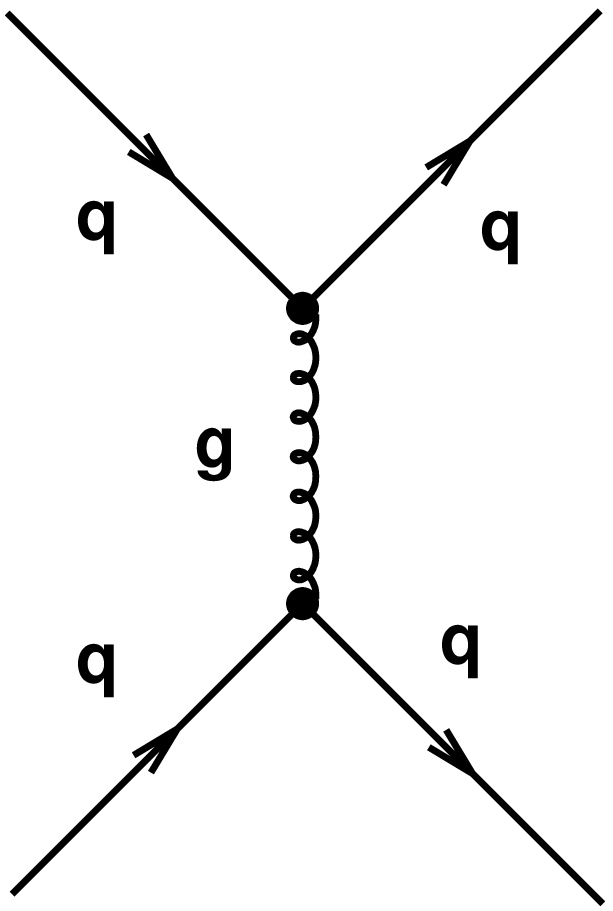,width=3cm}}
\put (5.0,0.0){\epsfig{figure=\figdir 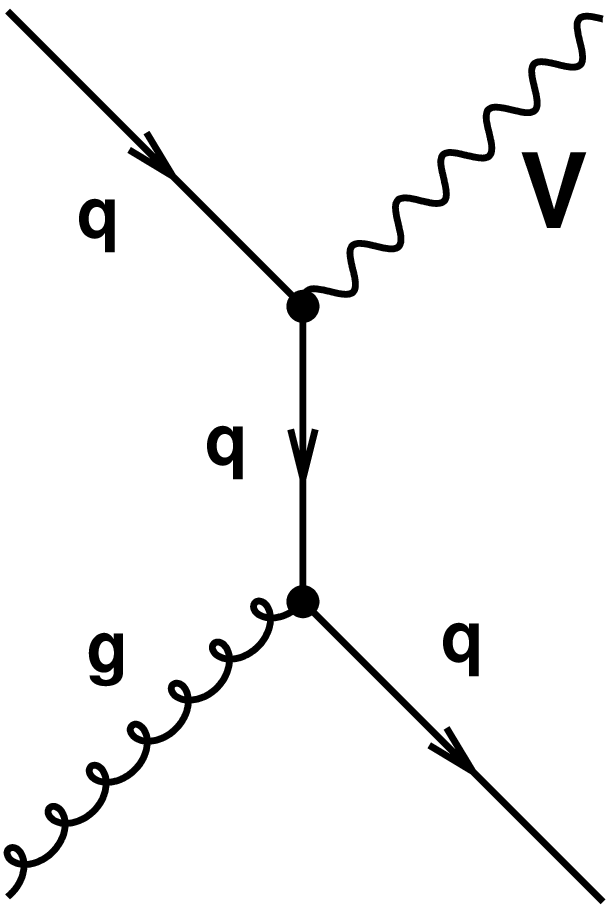,width=3cm}}
\put (9.0,0.0){\epsfig{figure=\figdir 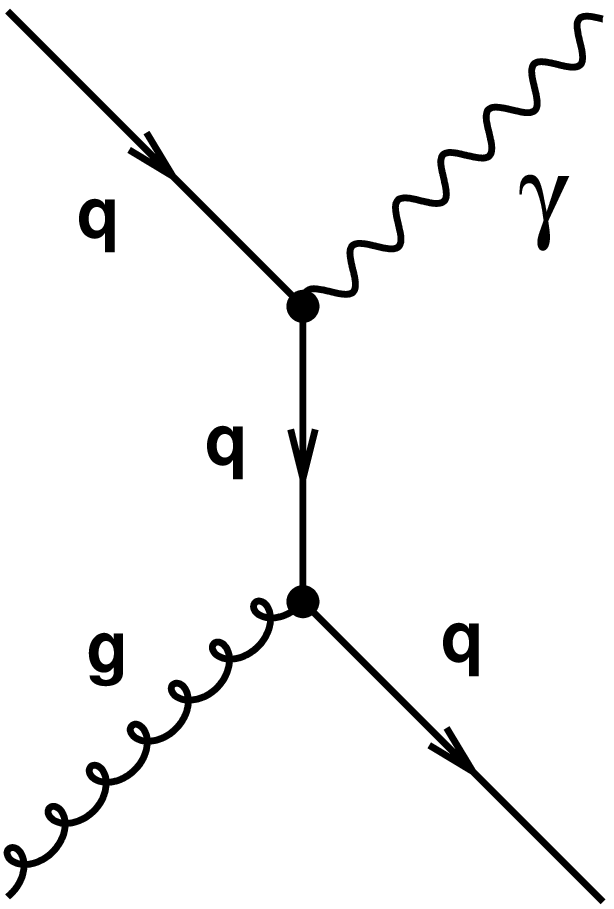,width=3cm}}
\put (13.0,0.0){\epsfig{figure=\figdir 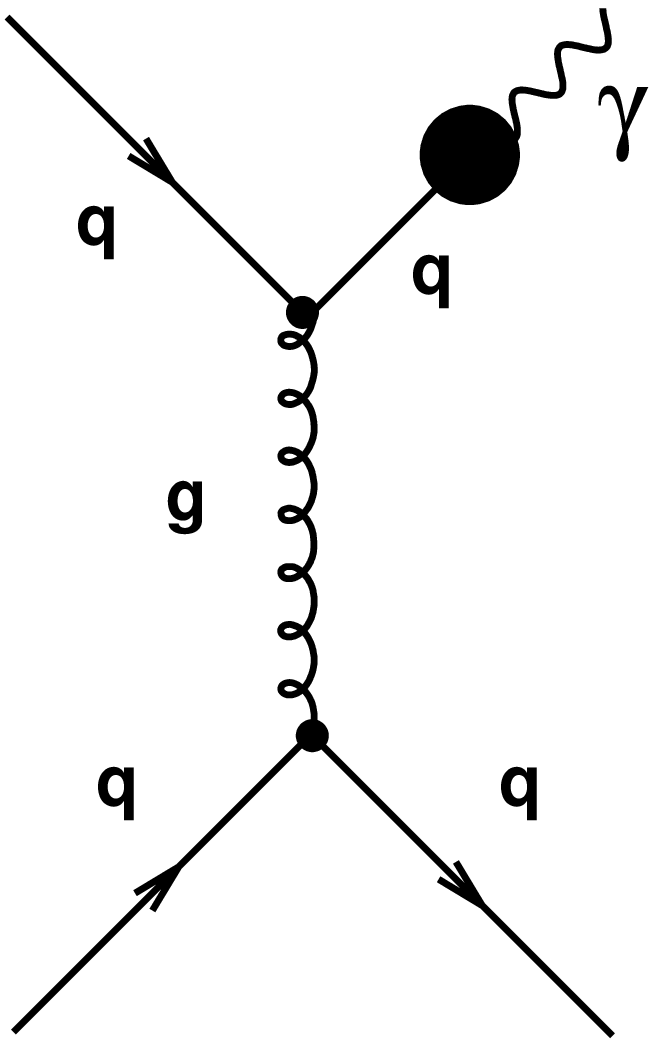,width=3.7cm}}
\put (2.0,-0.5){{\bf\small (a)}}
\put (6.0,-0.5){{\bf\small (b)}}
\put (10.0,-0.5){{\bf\small (c)}}
\put (14.0,-0.5){{\bf\small (d)}}
\end{picture}
\vspace{0.1cm}
\caption
{Examples of Feynman diagrams for 
  (a) dijet production, 
  (b) $V+{\rm jet}$ production with $V=W$ or $Z$, 
  (c) $\gamma+{\rm jet}$ production through direct-photon processes
  and 
  (d) $\gamma+{\rm jet}$ production through fragmentation processes.}
\label{fig01}
\vfill
\end{figure}

The dynamics of the underlying processes in $2\rightarrow 2$ hard
collinear scattering can be investigated using the variable
$\theta^*$, where $\cos\theta^*\equiv\tanh(\Delta y/2)$ and $\Delta y$
is the difference between the rapidities\footnote{ The ATLAS reference
  system is a Cartesian right-handed coordinate system, with the
  nominal collision point at the origin. The anticlockwise beam
  direction defines the positive $z$-axis, while the positive $x$-axis
  is defined as pointing from the collision point to the centre of the
  LHC ring and the positive $y$-axis points upwards. The azimuthal
  angle $\phi$ is measured around the beam axis, and the polar angle
  $\theta$ is measured with respect to the $z$-axis. Pseudorapidity is
  defined as $\eta=-\ln\tan(\theta/2)$, rapidity is defined as
  $y=0.5\ln[(E+p_z)/(E-p_z)]$, where $E$ is the energy and $p_z$ is
  the $z$-component of the momentum, and transverse energy is defined
  as $E_{\rm T}=E\sin\theta$.} of the two final-state particles. The
variable $\theta^*$ coincides with the scattering angle in the
centre-of-mass frame, and its distribution is sensitive to the spin of
the exchanged particle. For processes dominated by $t$-channel gluon
exchange, such as dijet production in $pp$ collisions shown in
Fig.~\ref{fig01}(a), the differential cross section behaves as
$(1-|\cos\theta^*|)^{-2}$ when $|\cos\theta^*|\rightarrow 1$. In
contrast, processes dominated by $t$-channel quark exchange, such as
$W/Z+{\rm jet}$ production shown in Fig.~\ref{fig01}(b), are expected
to have an asymptotic $(1-|\cos\theta^*|)^{-1}$ behaviour. This
fundamental prediction of QCD can be tested in photon plus jet
production at the centre-of-mass energy of the LHC.

At leading order (LO) in pQCD, the process 
$pp\rightarrow\gamma+{\rm jet}+{\rm X}$ proceeds via two production
mechanisms: direct photons (DP), which originate from the hard
process, and fragmentation photons (F), which arise from the
fragmentation of a coloured high transverse momentum ($p_{\rm T}$)
parton~\cite{pr:d76:034003,pr:d79:114024}. The direct-photon
contribution, as shown in Fig.~\ref{fig01}(c), is expected to exhibit
a $(1-|\cos\theta^*|)^{-1}$ dependence when 
$|\cos\theta^*|\rightarrow 1$, whereas that of fragmentation
processes, as shown in Fig.~\ref{fig01}(d), is predicted to be the
same as in dijet production, namely $(1-|\cos\theta^*|)^{-2}$. For
both processes, there are also $s$-channel contributions which are,
however, non-singular when $|\cos\theta^*|\rightarrow 1$. As a result,
a measurement of the cross section for prompt-photon plus jet
production as a function of $|\cos\theta^*|$ provides a handle on the
relative contributions of the direct-photon and fragmentation
components as well as the possibility to test the dominance of
$t$-channel quark exchange, such as that shown in
Fig.~\ref{fig01}(c).

Measurements of prompt-photon production in a final state with
accompanying hadrons necessitates of an isolation requirement on the
photon to avoid the large contribution from neutral-hadron decays into
photons. The production of inclusive isolated photons in $pp$
collisions has been studied previously by
ATLAS~\cite{pr:d83:052005,pl:b706:150} and
CMS~\cite{prl:106:082001,pr:d84:052011}. Recently, the differential
cross sections for isolated photons in association with jets as
functions of the photon transverse energy in different regions of
rapidity of the highest transverse-momentum (leading) jet were
measured by ATLAS~\cite{pr:d85:092014}. The analysis presented in this
paper is based on the same data sample and similar selection criteria
as in the previous publication, but extends the study by measuring
also cross sections in terms of the leading-jet and photon-plus-jet
properties. The goal of the analysis presented here is to study the
kinematics and dynamics of the isolated-photon plus jet system by
measuring the bin-averaged cross sections as functions of the
leading-photon transverse energy ($\etg$), the leading-jet transverse
momentum ($\ptjet$) and rapidity ($\rapjet$), the difference between
the azimuthal angles of the photon and the jet ($\delphj$), the
photon--jet invariant mass ($\mgjn$) and $\ctgjn$, where the variable
$\theta^*$ is referred to as $\theta^{\gamma{\rm j}}$ here and
henceforth. The photon was required to be isolated by using the same
isolation criterion as in previous
measurements~\cite{pr:d83:052005,pl:b706:150,pr:d85:092014} based on
the amount of transverse energy inside the cone given by
$\sqrt{(\eta-\etag)^2+(\phi-\phig)^2}\leq \Delta R=0.4$, centred
around the photon direction (defined by $\etag$ and $\phig$). The jets
were defined using the anti-$\kt$ jet algorithm~\cite{jhep:04:063}
with distance parameter $R=0.6$. The measurements were performed in
the phase-space region of $\etg>45$~GeV, $|\etag|<2.37$ (excluding the
region $1.37<|\etag|<1.52$), $\ptjet>40$~GeV, $|\rapjet|<2.37$ and
$\Delta R_{\gamma{\rm j}}^2=(\etag-\etajet)^2+(\phig-\phijet)^2>1$. 
The measurements of $d\sigma/d\mgjn$ and $d\sigma/d|\ctgjn|$ were
performed for $|\etag+\rapjet|<2.37$, $|\ctgjn|<0.83$ and
$\mgjn>161$~GeV; these additional requirements select a region where
the $\mgjn$ and $|\ctgjn|$ distributions are not distorted by the
restrictions on the transverse momenta and rapidities of the photon
and the jet. Next-to-leading-order (NLO) QCD calculations were
compared to the measurements. Photon plus jet events constitute an
important background in the identification of the Higgs decaying into
diphotons; the $\cost$ distribution for the diphoton events has been
used~\cite{1307.1432} to study the spin of the new ``Higgs-like''
particle observed by ATLAS~\cite{pl:b716:1} and
CMS~\cite{pl:b716:30}. To understand the photon plus jet background in
terms of pQCD and to aid in better constraining the contributions of
direct-photon and fragmentation processes in the MC models, a
measurement of the bin-averaged cross section as a function of
$|\ctgjn|$ was also performed without the restrictions on $\mgjn$ or
on $|\etag+\rapjet|$. Predictions from both leading-logarithm
parton-shower MC models and NLO QCD calculations were compared to this
measurement.

\section{The ATLAS detector}
\label{detector}
The ATLAS experiment~\cite{jinst:3:s08003} uses a multi-purpose
particle detector with a forward-backward symmetric cylindrical
geometry and nearly $4\pi$ coverage in solid angle. 

The inner detector covers the pseudorapidity range $|\etad|<2.5$ and
consists of a silicon pixel detector, a silicon microstrip detector
and, for $|\etad|<2$, a transition radiation tracker. The inner
detector is surrounded by a thin superconducting solenoid providing a
$2$~T magnetic field and is used to measure the momentum of
charged-particle tracks.

The electromagnetic calorimeter is a lead liquid-argon (LAr) sampling
calorimeter. It is divided into a barrel section, covering the
pseudorapidity region $|\etad|<1.475$, and two end-cap sections,
covering the pseudorapidity regions $1.375<|\etad|<3.2$. It consists
of three shower-depth layers in most of the pseudorapidity range. The
first layer is segmented into narrow strips in the $\etad$ direction
(width between $0.003$ and $0.006$ depending on $\etad$, with the
exception of the regions $1.4<|\etad|<1.5$ and $|\etad|>2.4$). This
high granularity provides discrimination between single-photon showers
and two overlapping showers coming from, for example, a $\pi^0$
decay. The second layer of the electromagnetic calorimeter, which
collects most of the energy deposited in the calorimeter by the photon
shower, has a cell granularity of $0.025\times0.025$ in
$\etad\times\phid$. A third layer collects the tails of the
electromagnetic showers. An additional thin LAr presampler covers
$|\etad|<1.8$ to correct for energy loss in material in front of the
calorimeter. The electromagnetic energy scale is calibrated using
$Z\rightarrow ee$ events with an uncertainty less than
$1\%$~\cite{epj:c72:1909}.

A hadronic sampling calorimeter is located outside the electromagnetic
calorimeter. It is made of scintillator tiles and steel in the barrel
section ($|\etad|<1.7$) and of two end-caps of copper and LAr
($1.5<|\etad|<3.2$). The forward region ($3.1<|\etad|<4.9$) is
instrumented with a copper/tungsten LAr calorimeter for both
electromagnetic and hadronic measurements. Outside the ATLAS
calorimeters lies the muon spectrometer, which identifies and measures
the deflection of muons up to $|\etad|=2.7$, in a magnetic field
generated by superconducting air-core toroidal magnet systems.

Events containing photon candidates were selected by a three-level
trigger system. The first-level trigger (level-1) is hardware-based
and uses a trigger cell granularity of $0.1\times 0.1$ in
$\etad\times\phid$. The algorithms of the second- and third-level
triggers are implemented in software and exploit the full granularity
and precision of the calorimeter to refine the level-1 trigger
selection, based on improved energy resolution and detailed
information on energy deposition in the calorimeter cells.

\section{Data selection}
\label{datsel}
The data used in this analysis were collected during the
proton--proton collision running period of 2010, when the LHC operated
at a centre-of-mass energy of $\sqrt s=7$~TeV. This data set was
chosen to study the dynamics of isolated-photon plus jet production
down to $\etg=45$~GeV.

Only events taken in stable beam conditions and passing detector and
data-quality requirements were considered. Events were recorded using
a single-photon trigger, with a nominal transverse energy threshold of
$40$~GeV; this trigger was used to collect events in which the photon
transverse energy, after reconstruction and calibration, was greater
than $45$~GeV. The total integrated luminosity of the collected sample
amounts to $37.1\pm 1.3$~\pb1~\cite{1302.4393}.

The selection criteria applied by the trigger to shower-shape
variables computed from the energy profiles of the showers in the
calorimeters are looser than the photon identification criteria
applied in the offline analysis; for isolated photons with
$\etg>43$~GeV and pseudorapidity $|\etagdet|< 2.37$, the trigger
efficiency is close to $100\%$.

The sample of isolated-photon plus jet events was selected using
offline criteria similar to those reported in the previous
publication~\cite{pr:d85:092014} and described below.

Events were required to have a reconstructed primary vertex, with at
least five associated charged-particle tracks with $p_T>150$~MeV,
consistent with the average beam-spot position. This requirement
reduced non-collision backgrounds. The effect of this requirement on
the signal was found to be negligible. The remaining fraction of
non-collision backgrounds was estimated to be less than
$0.1\%$~\cite{pr:d83:052005,pl:b706:150}.

During the 2010 data-taking period, there were on average $2$--$3$
proton--proton interactions per bunch crossing. The effects of the
additional $pp$ interactions (pile-up) on the photon isolation and jet
reconstruction are described below.

\subsection{Photon selection}
\label{photsel}
The selection of photon candidates is based on the reconstruction of
isolated electromagnetic clusters in the calorimeter with transverse
energies exceeding $2.5$~GeV. Clusters were matched to
charged-particle tracks based on the distance in ($\etad$,$\phid$)
between the cluster centre and the track impact point extrapolated to
the second layer of the LAr calorimeter. Clusters matched to tracks
were classified as electron candidates, whereas those without matching
tracks were classified as unconverted photon candidates. Clusters
matched to pairs of tracks originating from reconstructed conversion
vertices in the inner detector or to single tracks with no hit in the
innermost layer of the pixel detector were classified as converted
photon candidates~\cite{atlas-phys-pub-2011-007}. The overall
reconstruction efficiency for unconverted (converted) photons with
transverse energy above 20~GeV and pseudorapidity in the range
$|\etag|<2.37$, excluding the transition region $1.37<|\etag|<1.52$
between calorimeter sections, was estimated to be $99.8\
(94.3)\%$~\cite{atlas-phys-pub-2011-007}. The final energy
measurement, for both converted and unconverted photons, was made
using only the calorimeter, with a cluster size depending on the
photon classification. In the barrel, a cluster corresponding to
$3\times 5$ ($\etad\times\phid$) cells in the second layer was used
for unconverted photons, while a cluster of $3\times 7$ cells was used
for converted photon candidates to compensate for the opening angle
between the conversion products in the $\phi$ direction due to the
magnetic field. In the end-cap, a cluster size of $5\times 5$ was used
for all candidates. A dedicated energy calibration~\cite{0901.0512}
was then applied separately for converted and unconverted photon
candidates to account for upstream energy loss and both lateral and
longitudinal leakage. Photons reconstructed near regions of the
calorimeter affected by readout or high-voltage failures were
rejected, eliminating around $5\%$ of the selected candidates.

Events with at least one photon candidate with calibrated
$\etg>45$~GeV and $|\etagdet|<2.37$ were selected. The candidate was
excluded if $1.37<|\etagdet|<1.52$. The same shower-shape and
isolation requirements as described in previous
publications~\cite{pr:d83:052005,pl:b706:150,pr:d85:092014} were
applied to the candidates; these requirements are referred to as
``tight'' identification criteria. The selection criteria for the
shower-shape variables are independent of the photon-candidate
transverse energy, but vary as a function of the photon
pseudorapidity, to take into account significant changes in the total
thickness of the upstream material and variations in the calorimeter
geometry or granularity. They were optimised independently for
unconverted and converted photons to account for the different
developments of the showers in each case. The application of these
selection criteria suppresses background from jets misidentified as
photons.

The photon candidate was required to be isolated by restricting the
amount of transverse energy around its direction. The transverse
energy deposited in the calorimeters inside a cone of radius 
$\Delta R=0.4$ centred around the photon direction is denoted by
$\etisod$. The contributions from those cells (in any layer) in a
window corresponding to $5\times 7$ cells of the second layer of the
electromagnetic calorimeter around the photon-shower barycentre are
not included in the sum. The mean value of the small leakage of the
photon energy outside this region, evaluated as a function of the
photon transverse energy, was subtracted from the measured value of
$\etisod$. The typical size of this correction is a few percent of the
photon transverse energy. The measured value of $\etisod$ was further
corrected by subtracting the estimated contributions from the
underlying event and additional inelastic $pp$ interactions. This
correction was computed on an event-by-event basis and amounted on
average to $900$~MeV~\cite{pl:b706:150}. After all these corrections,
$\etisod$ was required to be below $3$~GeV for a photon to be
considered isolated.

The relative contribution to the total cross section from
fragmentation processes decreases after the application of this
requirement, though it remains non-negligible especially at low
transverse energies. The isolation requirement significantly reduces
the main background, which consists of multi-jet events where one jet
typically contains a $\pi^0$ or $\eta$ meson that carries most of the
jet energy and is misidentified as an isolated photon because it
decays into an almost collinear photon pair.

A small fraction of events contain more than one photon candidate
passing the selection criteria. In such events, the highest-$\etg$
(leading) photon was kept for further study.

\subsection{Jet selection}
\label{jetsel}
Jets were reconstructed from three-dimensional topological clusters
built from calorimeter cells, using the anti-$\kt$ algorithm with
distance parameter $R=0.6$. The jet four-momenta were computed from
the sum of the topological cluster four-momenta, treating each as a
four-vector with zero mass. The jet four-momenta were then
recalibrated using a jet energy scale (JES) correction described in
Ref.~\cite{epj:c73:2304}. This calibration procedure corrected the
jets for calorimeter instrumental effects, such as inactive material
and noncompensation, as well as for the additional energy due to
multiple $pp$ interactions within the same bunch crossing. These jets
are referred to as detector-level jets. The uncertainty on the JES
correction in the central (forward) region, $|\etad|<0.8$
($2.1<|\etad|<2.8$), is less than $4.6\%$ ($6.5\%$) for all jets with
transverse momentum  $p_{\rm T}>20$~GeV and less than $2.5\%$ ($3\%$)
for jets with $60<p_{\rm T}<800$~GeV.

Jets reconstructed from calorimeter signals not originating from a
$pp$ collision were rejected by applying jet-quality
criteria~\cite{epj:c73:2304}. These criteria suppressed fake jets from
electronic noise in the calorimeter, cosmic rays and beam-related
backgrounds. Remaining jets were required to have calibrated
transverse momenta greater than $40$~GeV. Jets overlapping with the
candidate photon or with an isolated electron were discarded; if the
jet axis lay within a cone of radius $\Delta R=1\ (0.3)$ around the
leading-photon (isolated-electron) candidate, the jet was
discarded. The removal of electrons misidentified as jets suppresses
contamination from $W/Z$ plus jet events. In events with multiple jets
satisfying the above requirements, the jet with highest $\ptjet$
(leading jet) was retained for further study. The leading-jet rapidity
was required to be in the region $|\rapjet|<2.37$.

\subsection{Final photon plus jet sample}
The above requirements select approximately $124\,000$ events. The
fraction of events with multiple photons fulfilling the above
conditions is $3\cdot 10^{-4}$. The average jet multiplicity in the
data is $1.19$. The signal MC (see Section~\ref{mc}) predictions for
the jet multiplicity are $1.21$ in {\sc Pythia}~\cite{jhep:0605:026}
and $1.19$ in {\sc Herwig}~\cite{jhep:0101:010}.

\begin{figure}[h]
\vfill
\setlength{\unitlength}{1.0cm}
\begin{picture} (18.0,8.0)
\put (0.0,0.0){\epsfig{figure=\figdir 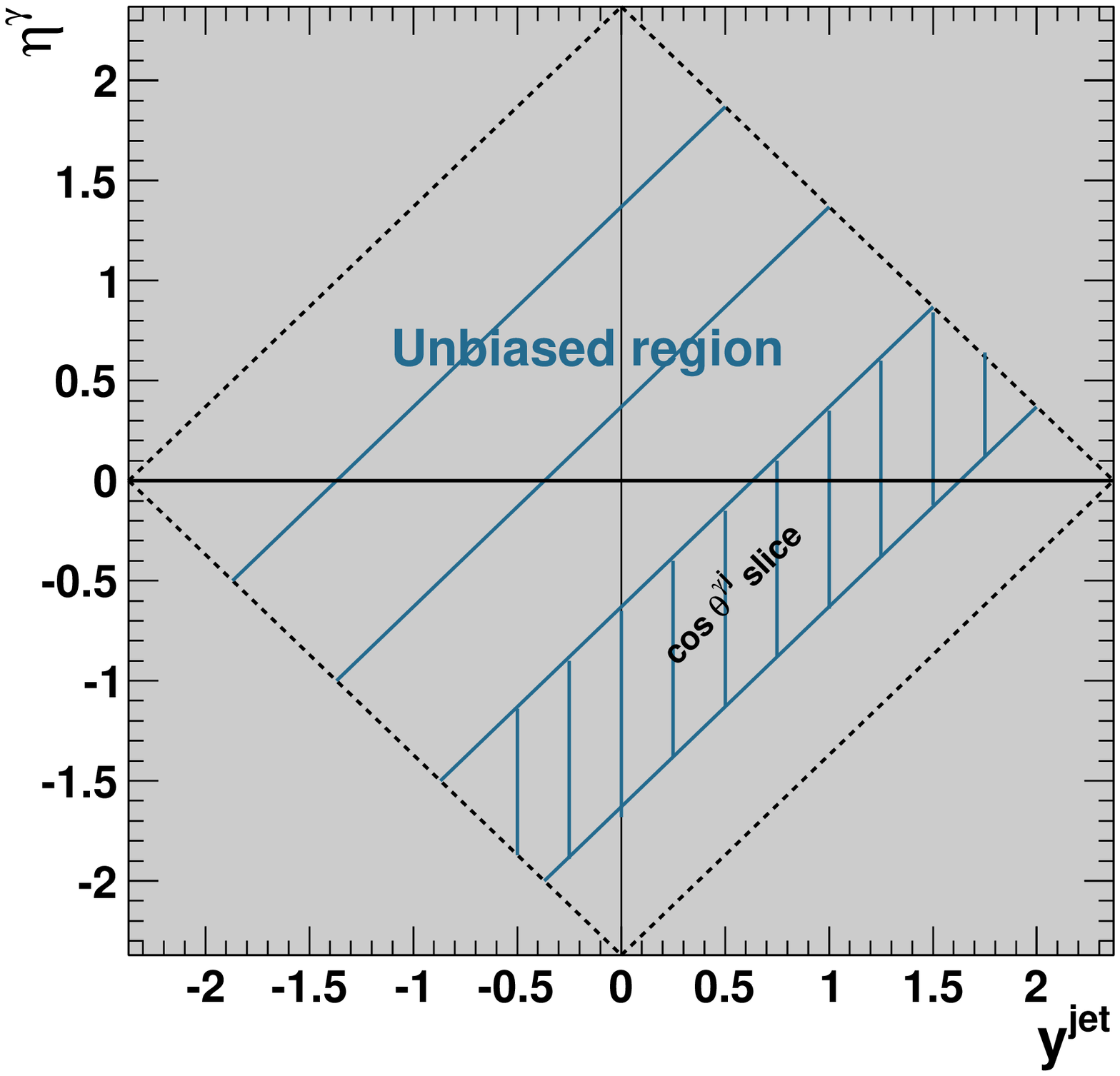,width=8cm}}
\put (8.0,0.0){\epsfig{figure=\figdir 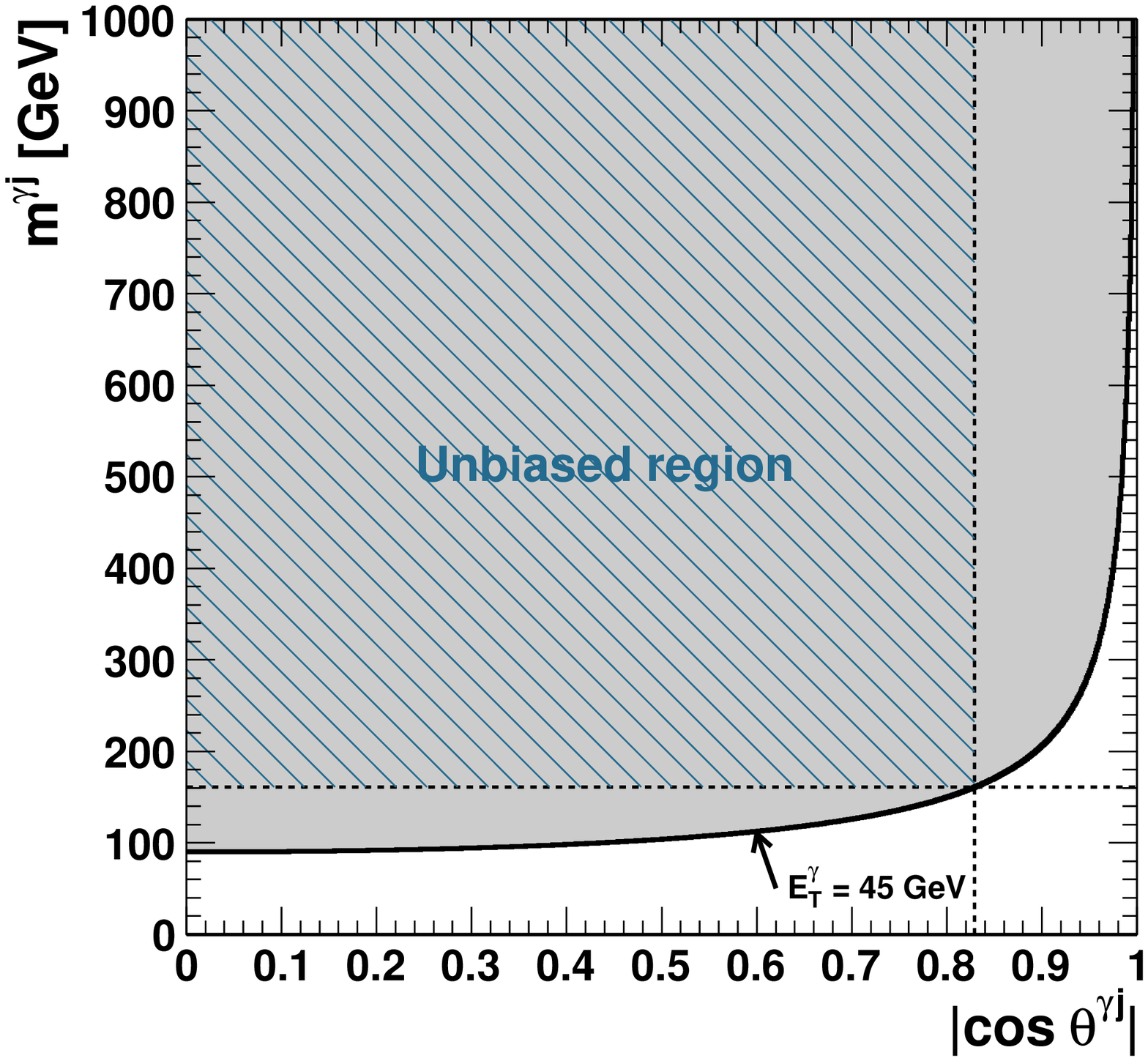,width=8cm}}
\put (4.0,0.0){{\bf\small (a)}}
\put (12.0,0.0){{\bf\small (b)}}
\end{picture}
\caption
{The selected regions in the (a) $\etagdet$-$y^{\rm jet}$ and (b)
  $\mgjn$-$|\ctgjn|$ planes.
In (a), the dashed lines correspond to: 
$\etagdet+y^{\rm jet}=2.37$ (first quadrant),
$\etagdet-y^{\rm jet}=2.37$ (second quadrant),
$\etagdet+y^{\rm jet}=-2.37$ (third quadrant) and
$\etagdet-y^{\rm jet}=-2.37$ (fourth quadrant).
In (b), the horizontal (vertical) dashed line corresponds to
$\mgjn=161$~GeV ($|\ctgjn|=0.83$) and the solid line corresponds to
$\etg=45$~GeV.}
\label{fig02}
\vfill
\end{figure}

For the measurements of the bin-averaged cross sections as functions
of $\mgjn$ and $|\ctgjn|$, additional requirements were imposed to
remove the bias due to the rapidity and transverse-momentum
requirements on the photon and the jet. Specifically, to have a
uniform coverage in both $\ctgjn$ and $\mgjn$, the
restrictions~$|\etagdet+\rapjet|<2.37$, $|\ctgjn|<0.83$ and
$\mgjn>161$~GeV were applied. The first two requirements restrict the
phase space to the inside of the square delineated by the dashed
lines, as shown in Fig.~\ref{fig02}(a); within this square, slices in
$\ctgjn$ have the same length along the $\etagdet+\rapjet$ axis. The
third requirement avoids the bias induced by the minimal requirement
on $\etg$, as shown in Fig.\ref{fig02}(b); the hatched area represents
the largest region in which unbiased measurements of both $|\ctgjn|$
and $\mgjn$ distributions can be performed. These requirements do not
remove the small bias due to the exclusion of the
$1.37<|\etagdet|<1.52$ region. The number of events selected in the
data after these additional requirements is approximately $26\,000$.

The contamination from jets produced in pile-up events in the selected
samples was estimated to be negligible.

\section{Monte Carlo simulations}
\label{mc}
Samples of simulated events were generated to study the
characteristics of signal and background. These MC samples were also
used to determine the response of the detector to jets of hadrons and
the correction factors necessary to obtain the particle-level cross
sections. In addition, they were used to estimate hadronisation
corrections to the NLO QCD calculations.

The MC programs {\sc Pythia}~6.423~\cite{jhep:0605:026} and {\sc
  Herwig}~6.510~\cite{jhep:0101:010} were used to generate the
simulated signal events. In both generators, the partonic processes
are simulated using leading-order matrix elements, with the inclusion
of initial- and final-state parton showers. Fragmentation into hadrons
was performed using the Lund string model~\cite{prep:97:31} in the
case of {\sc Pythia} and the cluster model~\cite{np:b238:492} in the
case of {\sc Herwig}. The modified leading-order
MRST2007~\cite{epj:c55:553,0807.2132} parton distribution functions
(PDFs) were used to parameterise the proton structure. Both samples
include a simulation of the underlying event, via the multiple-parton
interaction model in the case of {\sc Pythia} and via the {\sc Jimmy}
package~\cite{zp:c72:637} in the case of {\sc Herwig}. The
event-generator parameters, including those of the underlying-event
modelling, were set according to the AMBT1~\cite{ATLAS-CONF-2010-031}
and AUET1~\cite{ATL-PHYS-PUB-2010-014} tunes for {\sc Pythia} and {\sc
  Herwig}, respectively. All the samples of generated events were
passed through the  {\sc Geant}4-based~\cite{nim:a506:250} ATLAS
detector simulation program~\cite{epj:c70:823}. They were
reconstructed and analysed by the same program chain as the data.

The {\sc Pythia} simulation of the signal includes leading-order
photon plus jet events from both direct processes (the hard
subprocesses $qg\rightarrow q\gamma$ and $\qq\rightarrow g\gamma$) and
photon bremsstrahlung in QCD dijet events, which can be generated
simultaneously. On the other hand, the {\sc Herwig} signal sample was
obtained from the cross-section-weighted mixture of samples containing
only direct-photon plus jet or only bremsstrahlung-photon plus jet
events, since these processes cannot be generated simultaneously.

The multi-jet background was simulated by using all tree-level
$2\rightarrow 2$ QCD processes and removing photon plus jet events
from photon bremsstrahlung. The background from diphoton events was
estimated using {\sc Pythia} MC samples by computing the ratio of
diphoton to isolated-photon plus jet events and was found to be
negligible~\cite{pr:d85:092014}.

Particle-level jets in the MC simulation were reconstructed using the
anti-$\kt$ jet algorithm and were built from stable particles, which
are defined as those with a rest-frame lifetime longer than
$10$~ps. The particle-level isolation requirement on the photon was
applied to the transverse energy of all stable particles, except for
muons and neutrinos, in a cone of radius $\Delta R=0.4$ around the
photon direction after the contribution from the underlying event was
subtracted; in this case, the same underlying-event subtraction
procedure used on data was applied at the particle level. The
isolation transverse energy at particle level is denoted by
$\etisop$. The measured bin-averaged cross sections refer to
particle-level jets and photons that are isolated by requiring
$\etisop<4$~GeV~\cite{pr:d83:052005}.

For the comparison to the measurements (see Section~\ref{res}),
samples of events were generated at the particle level using the {\sc
  Sherpa} 1.3.1~\cite{jhep:0902:007} program interfaced with the
CTEQ6L1~\cite{jhep:0207:012} PDF set. The samples were generated with
LO matrix elements for photon plus jet final states with up to three
additional partons, supplemented with parton showers. Fragmentation
into hadrons was performed using a modified version of the cluster
model~\cite{epj:c36:381}.

\section{Signal extraction}

\subsection{Background subtraction and signal-yield estimation}
\label{bgks}
A non-negligible background contribution remains in the selected
sample, even after the application of the tight identification and
isolation requirements on the photon. This background comes
predominantly from multi-jet processes, in which a jet is
misidentified as a photon. This jet usually contains a light neutral
meson, mostly a $\pi^0$ decaying into two collimated photons, which
carries most of the jet energy. The very small contributions expected
from diphoton and $W/Z$ plus jet
events~\cite{pr:d83:052005,pr:d85:092014} are neglected.

The background subtraction does not rely on MC background samples but
uses instead a data-driven method based on signal-depleted control
regions. The background contamination in the selected sample was
estimated using the same two-dimensional sideband technique as in the
previous analyses~\cite{pr:d83:052005,pl:b706:150,pr:d85:092014} and
then subtracted bin-by-bin from the observed yield. In this method,
the photon was classified as:
\begin{itemize}
\item ``isolated'', if $\etisod<3$~GeV;
\item ``non-isolated'', if $\etisod>5$~GeV;
\item ``tight'', if it passed the tight photon identification criteria;
\item ``non-tight'', if it failed at least one of the tight requirements
  on the shower-shape variables computed from the energy deposits in
  the first layer of the electromagnetic calorimeter, but passed all
  the other tight identification criteria.
\end{itemize}
In the two-dimensional plane formed by $\etisod$ and the photon
identification variable, four regions were defined:
\begin{itemize}
\item{$A$}: the ``signal'' region, containing tight and isolated photon
  candidates;
\item{$B$}: the ``non-isolated'' background control region, containing
  tight and non-isolated photon candidates;
\item{$C$}: the ``non-identified'' background control region,
  containing isolated and non-tight photon candidates;
\item{$D$}: the background control region containing non-isolated and
  non-tight photon candidates.
\end{itemize}

The signal yield in region $A$, $N_A^{\rm sig}$, was estimated by
using the relation

\begin{equation}
N_A^{\rm sig}=N_A-R^{\rm bg}\cdot(N_B-\epsilon_B N_A^{\rm sig})\cdot
\frac{(N_C-\epsilon_C N_A^{\rm sig})}{(N_D-\epsilon_D N_A^{\rm sig})},
\label{eqone}
\end{equation}
where $N_K$, with $K=A,B,C,D$, is the number of events observed in
region $K$ and

$$R^{\rm bg}=\frac{N_A^{\rm bg}\cdot N_D^{\rm bg}}{N_B^{\rm bg}\cdot N_C^{\rm
    bg}}$$
is the so-called background correlation and was taken as 
$R^{\rm bg}=1$ for the nominal results; $N_K^{\rm bg}$ with
$K=A,B,C,D$ is the number of background events in each
region. Eq.~(\ref{eqone}) takes into account the expected number of
signal events in the three background control regions 
($N_K^{\rm sig}$) via the signal leakage fractions,
$\epsilon_K=N_K^{\rm sig}/N_A^{\rm sig}$ with $K=B,C,D$, which were
extracted from MC simulations of the signal. Since the simulation does
not accurately describe the electromagnetic shower profiles, a
correction factor for each simulated shape variable was applied to
better match the
data~\cite{pr:d83:052005,pl:b706:150}. Eq.~(\ref{eqone}) leads to a
second-order polynomial equation in $N_A^{\rm sig}$ that has only one
physical ($N_A^{\rm sig}>0$) solution.

This method was tested on a cross section-weighted combination of
simulated signal and background samples and found to accurately
determine the amount of signal in the mixture. The only hypothesis
underlying Eq.~(\ref{eqone}) is that the isolation and identification
variables are uncorrelated in background events, thus 
$R^{\rm bg}=1$. This assumption was verified both in simulated
background samples and in data in the background-dominated region
defined by $\etisod>10$~GeV. Deviations from unity were taken as
systematic uncertainties (see Section~\ref{syst}).

The signal purity, defined as $N_A^{\rm sig}/N_A$, is typically above
$0.9$ and is similar whether {\sc Pythia} or {\sc Herwig} is used to
extract the signal leakage fractions. The signal purity increases as
$\etg$, $\ptjet$ and $\mgjn$ increase, is approximately constant as a
function of $|\rapjet|$ and $\delphj$ and decreases as $|\ctgjn|$
increases.

\begin{figure}[p]
\vfill
\setlength{\unitlength}{1.0cm}
\begin{picture} (18.0,16.9)
\put (1.0,12.2){\epsfig{figure=\figdir 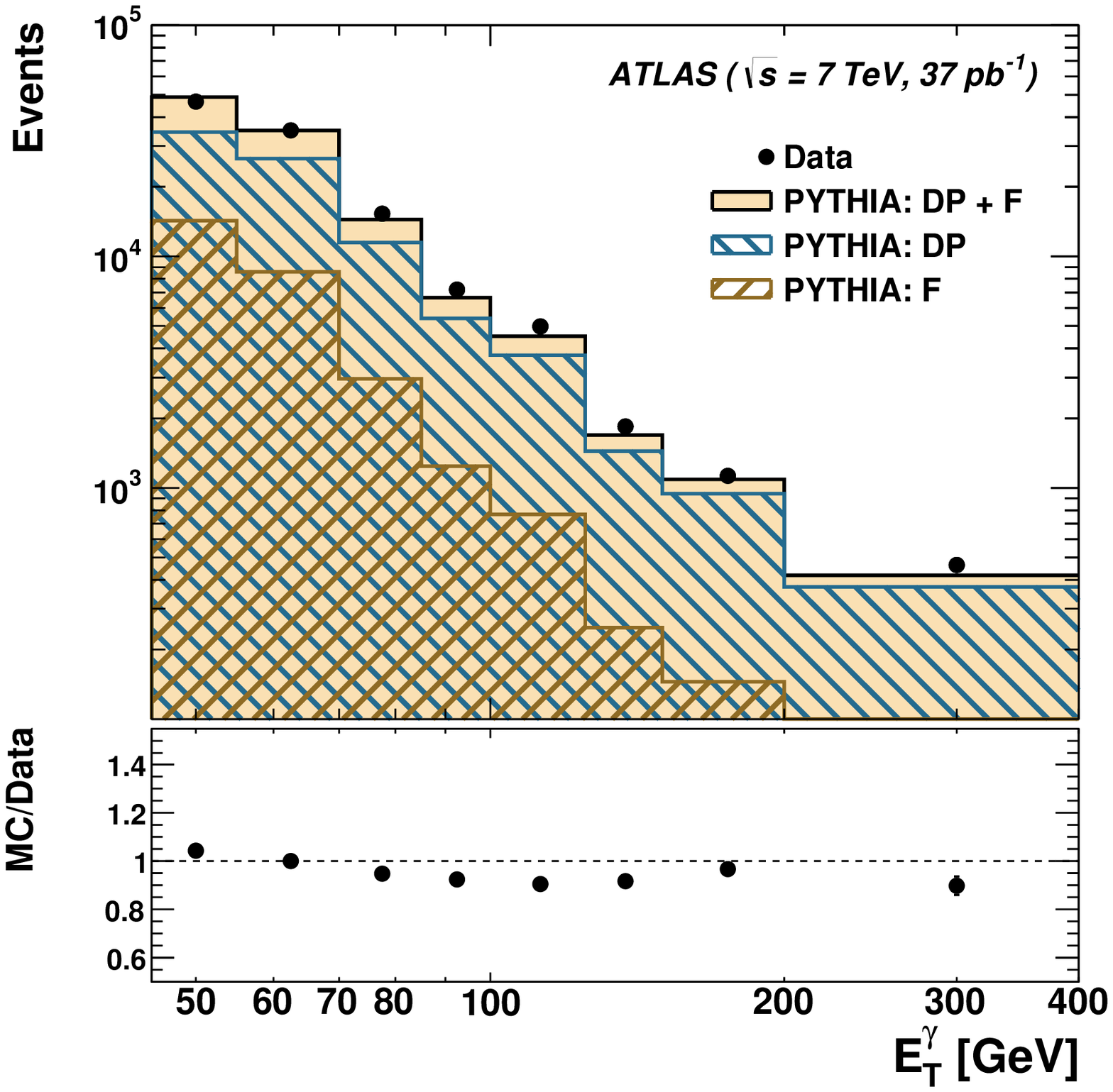,width=7cm}}
\put (9.0,12.2){\epsfig{figure=\figdir 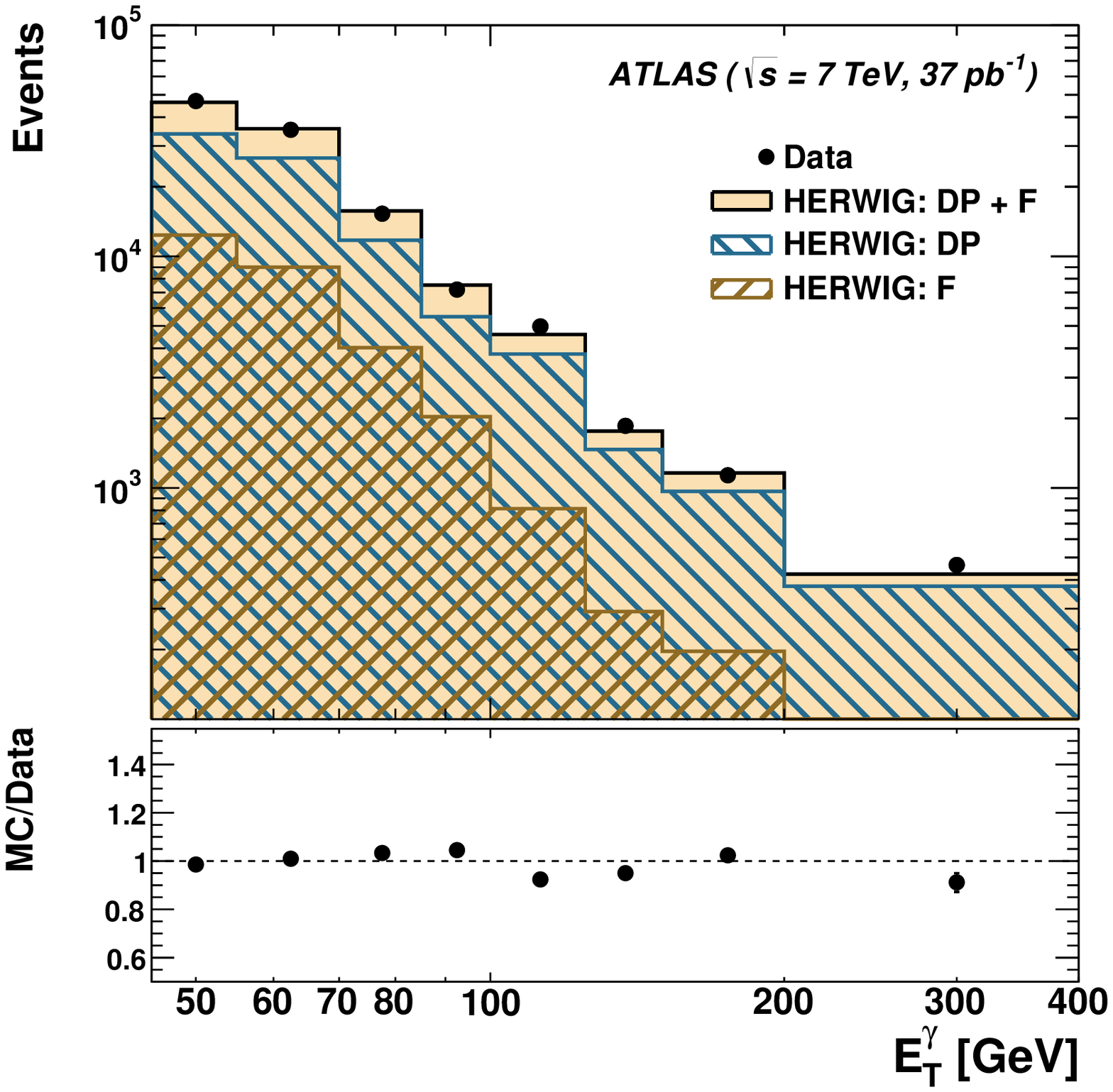,width=7cm}}
\put (1.0,5.6){\epsfig{figure=\figdir 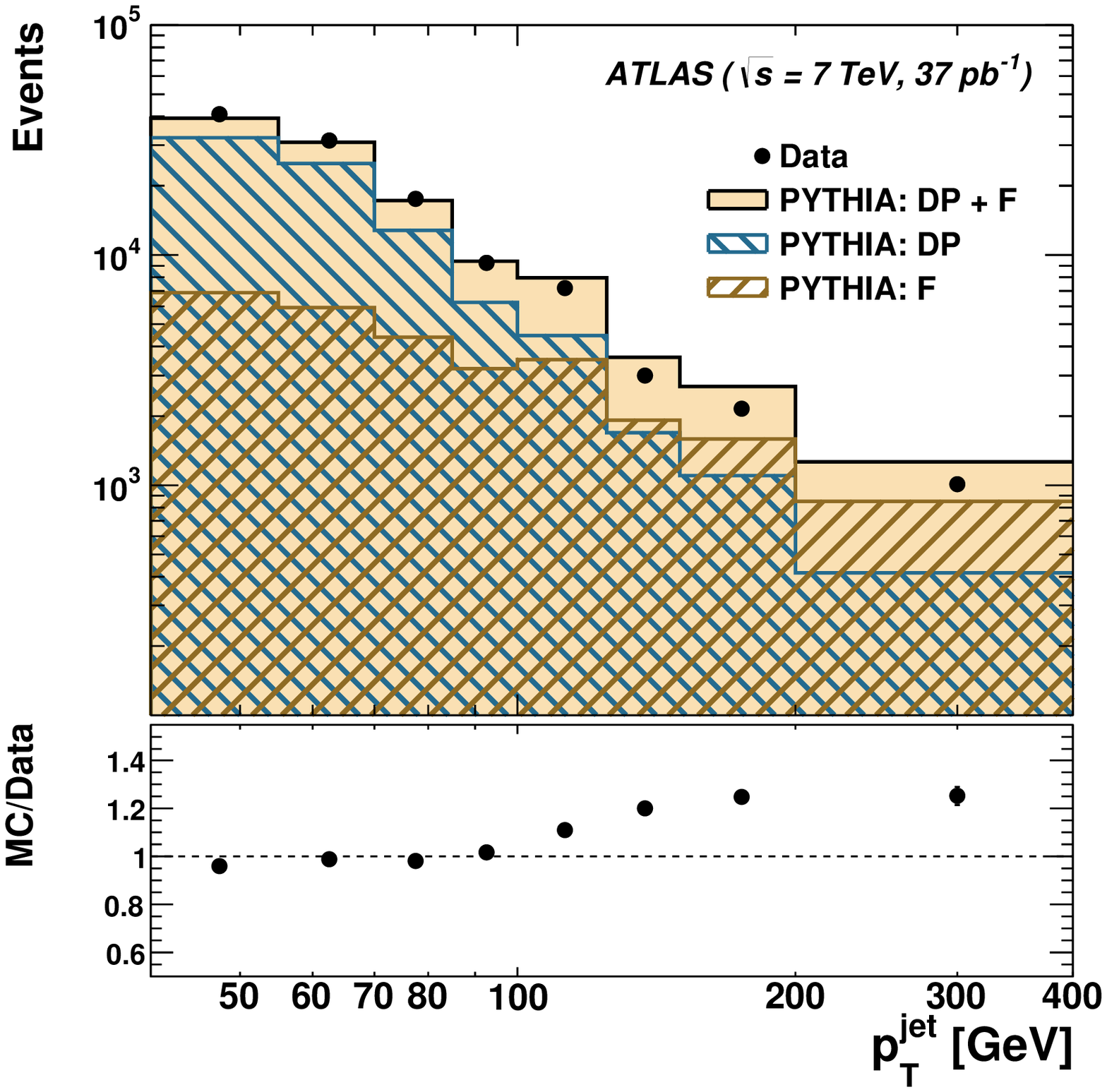,width=7cm}}
\put (9.0,5.6){\epsfig{figure=\figdir 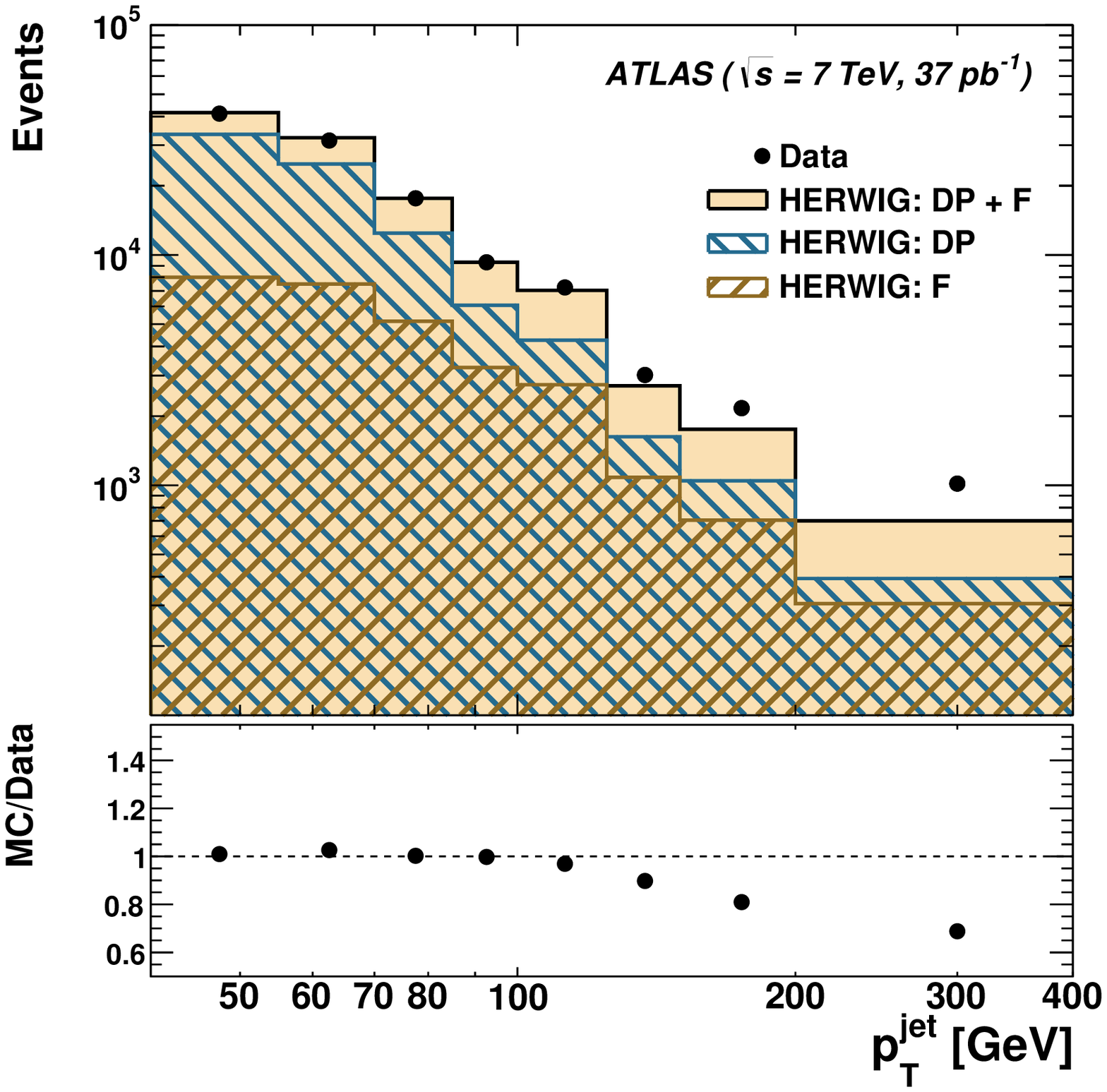,width=7cm}}
\put (1.0,-1.0){\epsfig{figure=\figdir 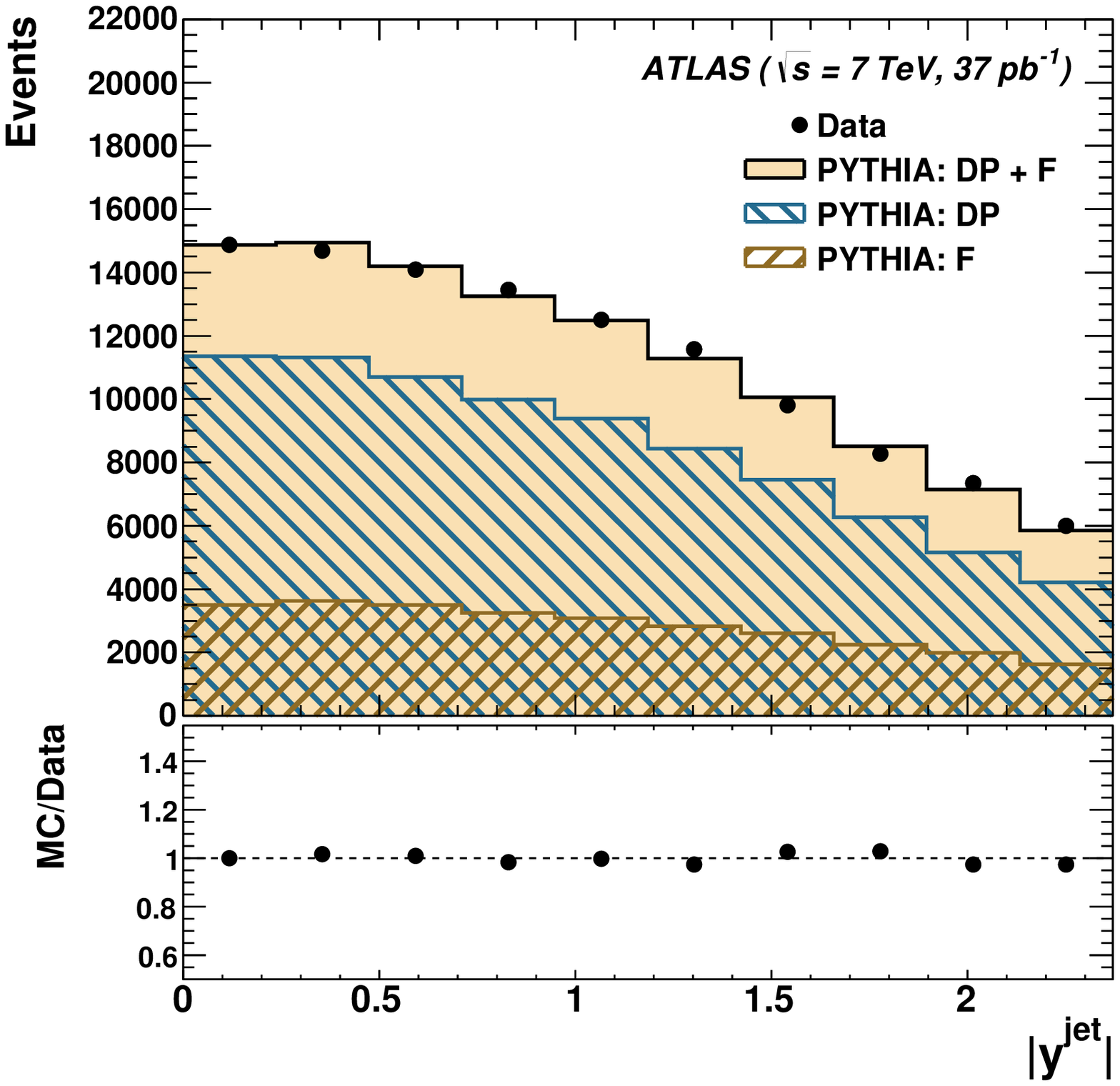,width=7cm}}
\put (9.0,-1.0){\epsfig{figure=\figdir 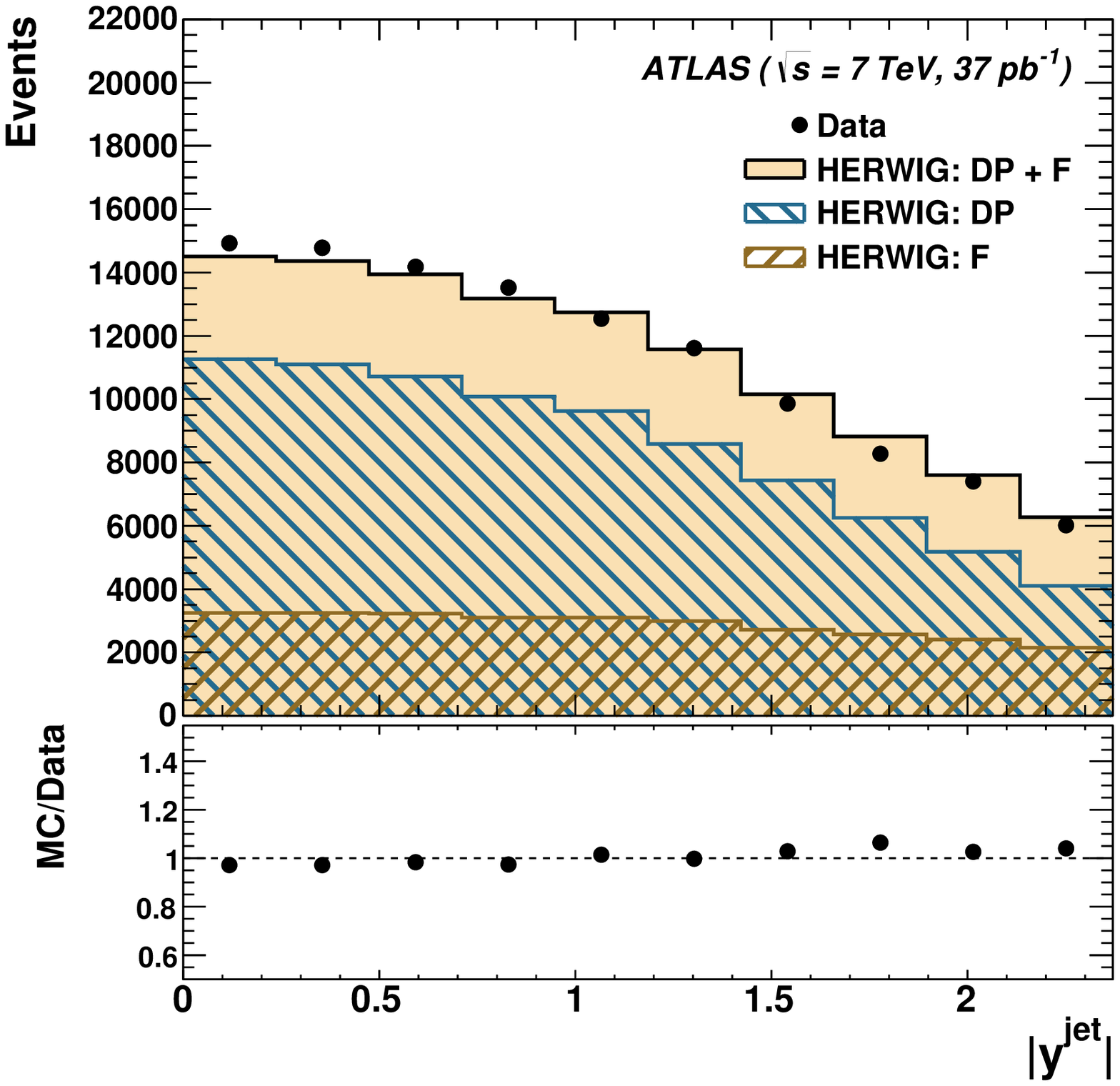,width=7cm}}
\put (4.0,12.75){{\bf\small (a)}}
\put (12.0,12.75){{\bf\small (b)}}
\put (4.0,6.15){{\bf\small (c)}}
\put (12.0,6.15){{\bf\small (d)}}
\put (4.0,-0.45){{\bf\small (e)}}
\put (12.0,-0.45){{\bf\small (f)}}
\end{picture}
\vspace{0.1cm}
\caption
{The estimated signal yield in data (dots) using the signal leakage
  fractions from (a,c,e) {\sc Pythia} or (b,d,f) {\sc Herwig} as
  functions of (a,b) $\etg$, (c,d) $\ptjet$ and (e,f) $|\rapjet|$. The
  error bars represent the statistical uncertainties that, for most of
  the points, are smaller than the marker size and, thus, not
  visible. For comparison, the MC simulations of the signal from {\sc
    Pythia} and {\sc Herwig} (shaded histograms) are also included in
  (a,c,e) and (b,d,f), respectively. The MC distributions are
  normalised to the total number of data events. The direct-photon
  (DP, right-hatched histograms) and fragmentation (F, left-hatched
  histograms) components of the MC simulations are also shown. The
  ratio of the MC predictions to the data are shown in the bottom part
  of the figures.}
\label{fig16}
\vfill
\end{figure}

\begin{figure}[p]
\vfill
\setlength{\unitlength}{1.0cm}
\begin{picture} (18.0,16.9)
\put (1.0,12.2){\epsfig{figure=\figdir 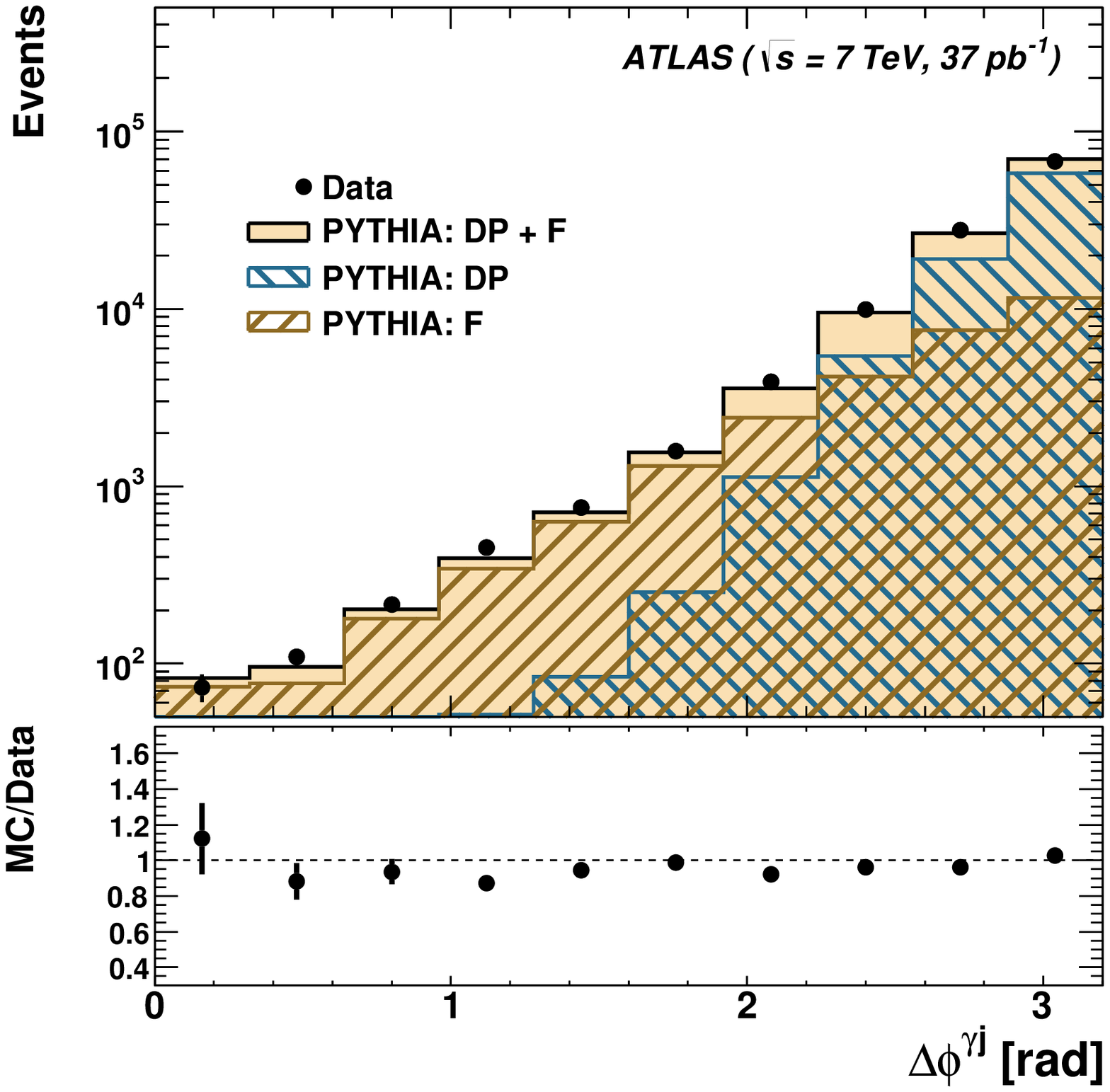,width=7cm}}
\put (9.0,12.2){\epsfig{figure=\figdir 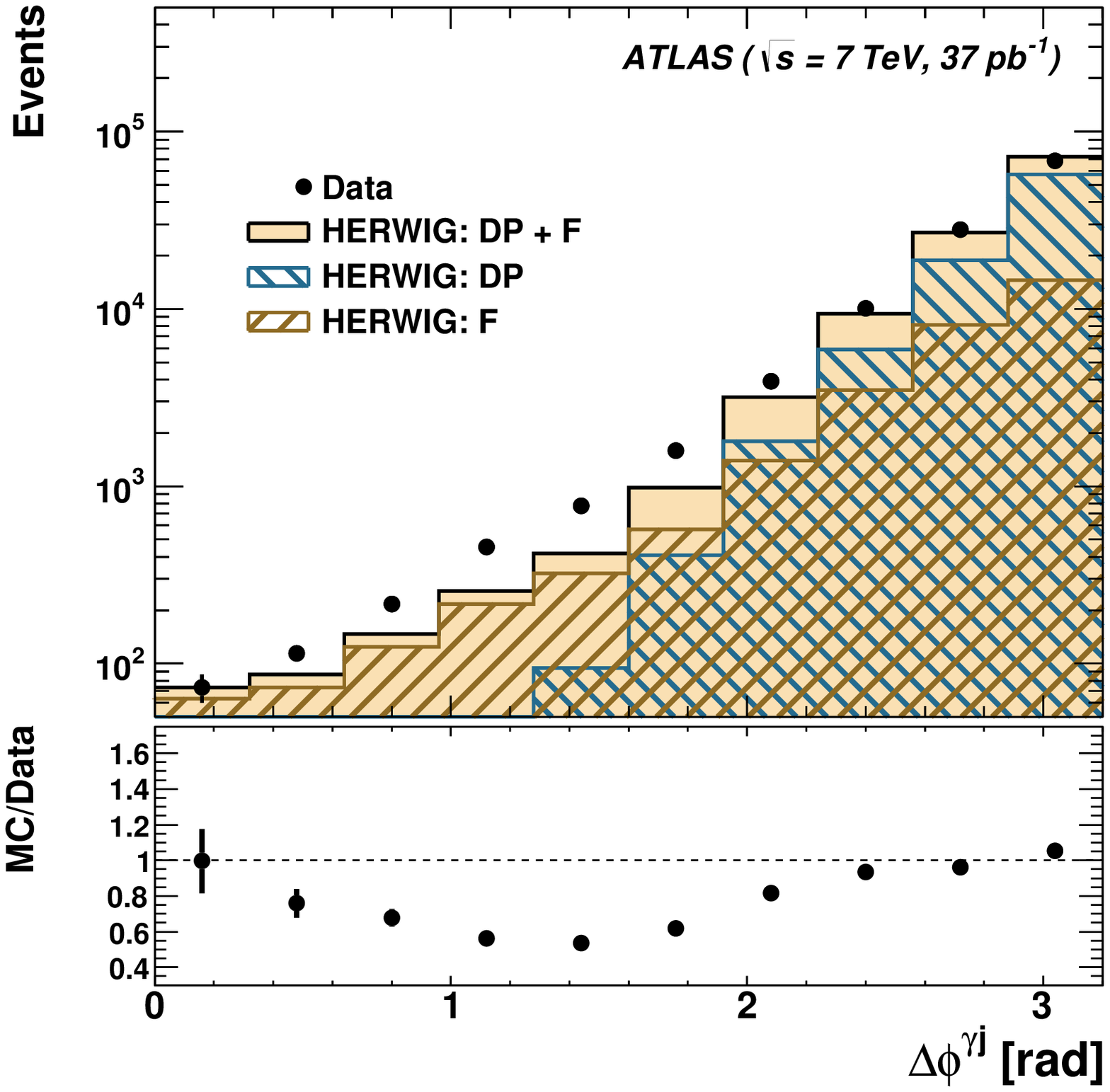,width=7cm}}
\put (1.0,5.6){\epsfig{figure=\figdir 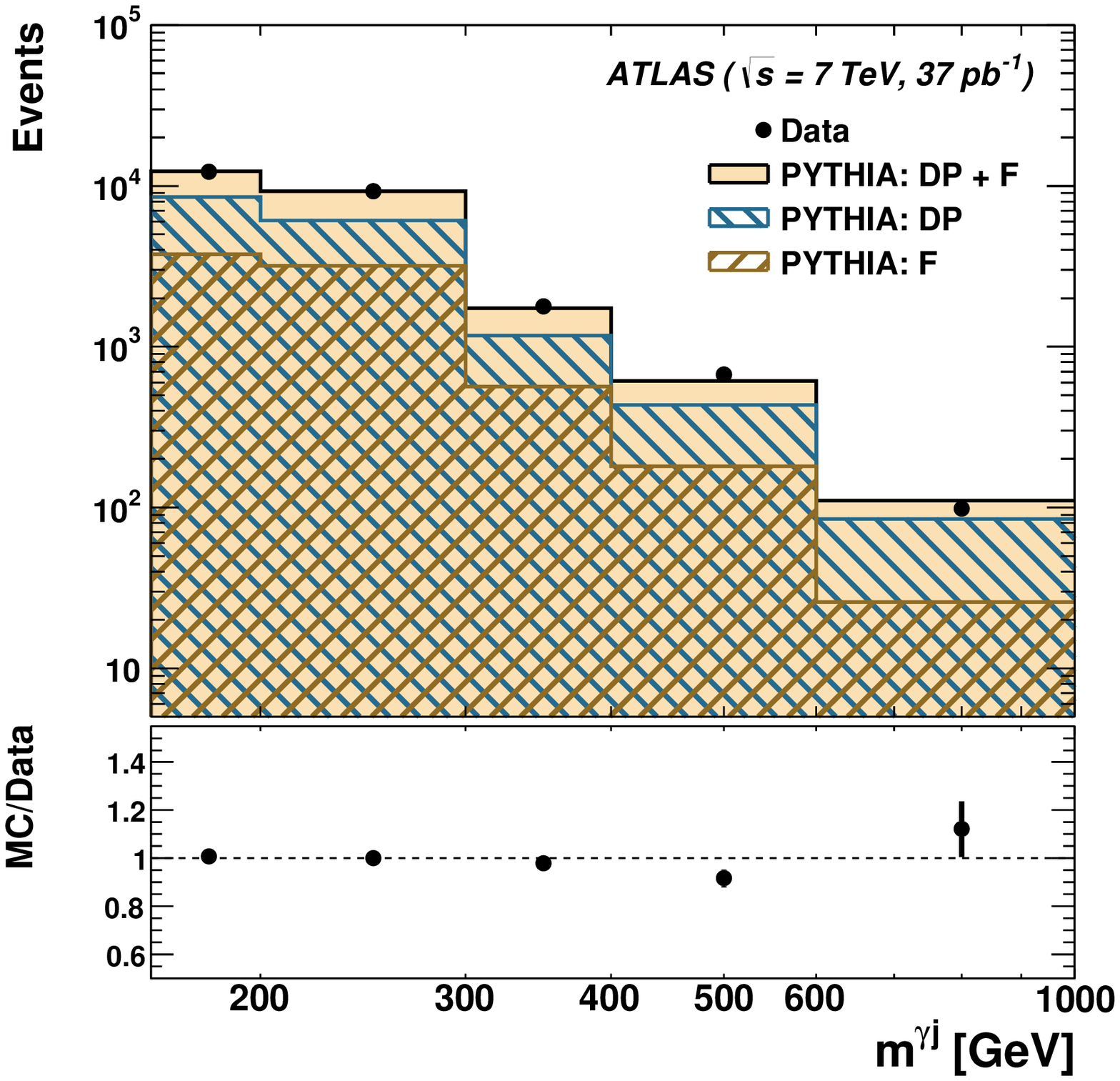,width=7cm}}
\put (9.0,5.6){\epsfig{figure=\figdir 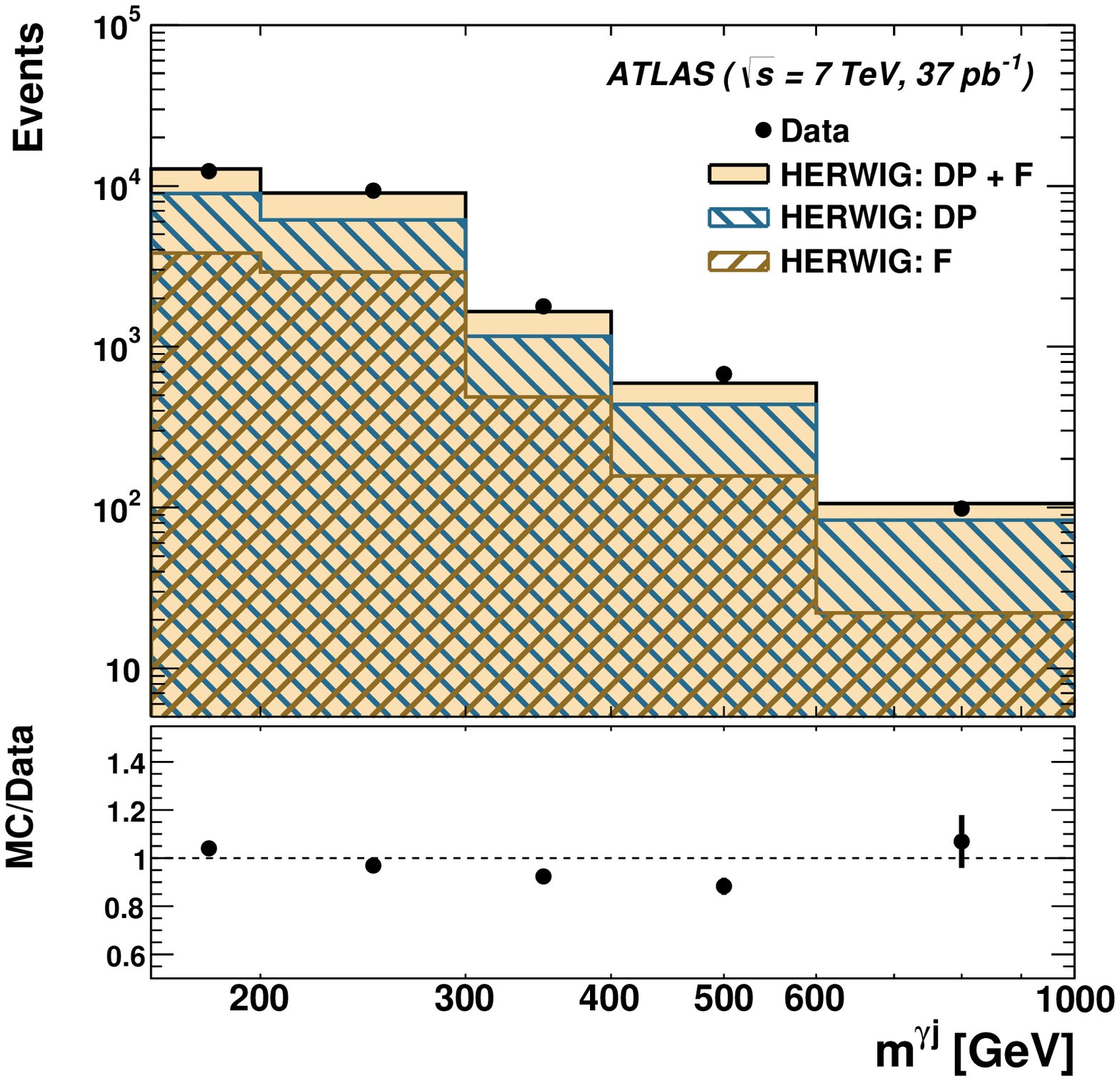,width=7cm}}
\put (1.0,-1.0){\epsfig{figure=\figdir 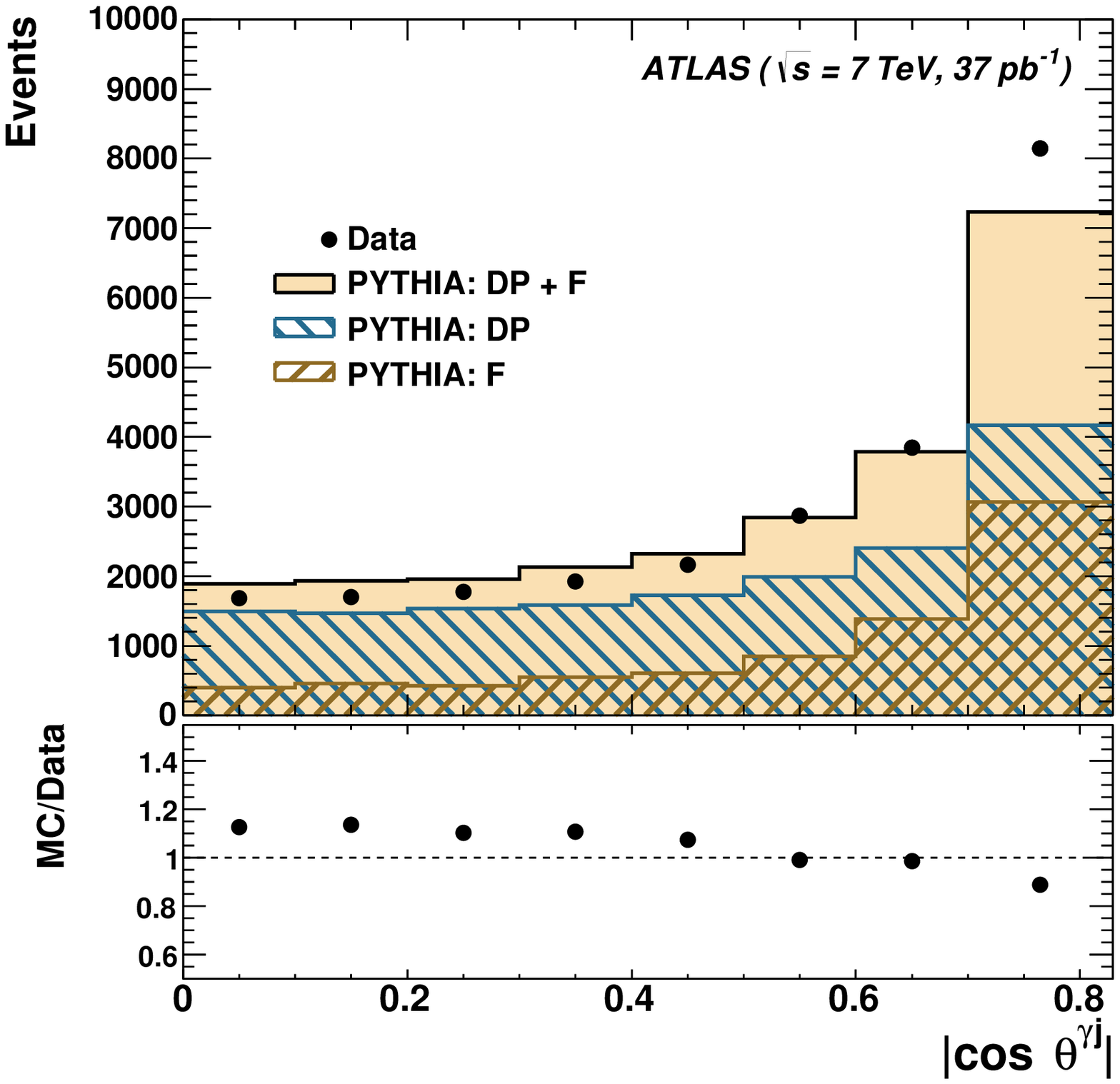,width=7cm}}
\put (9.0,-1.0){\epsfig{figure=\figdir 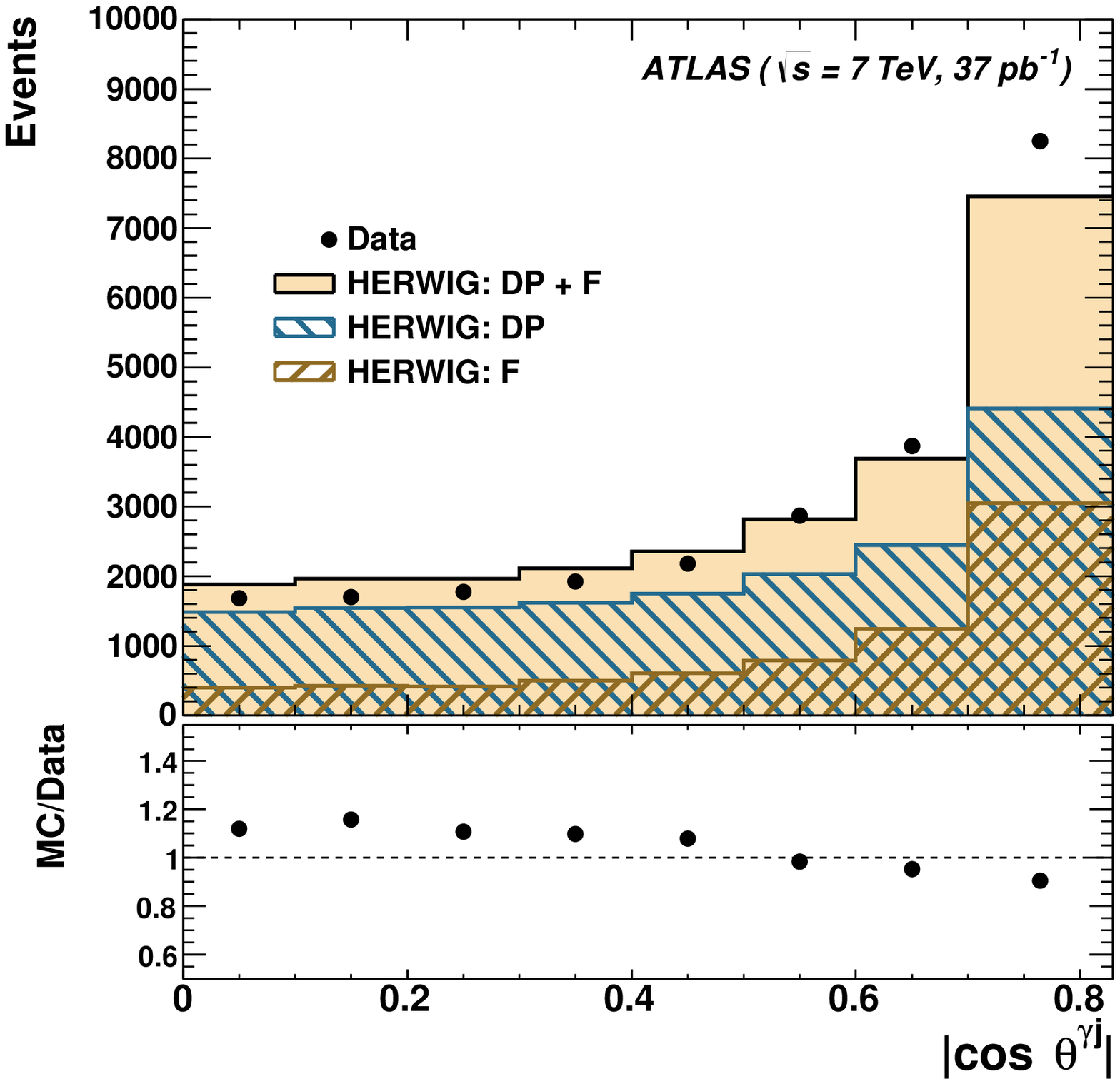,width=7cm}}
\put (4.0,12.75){{\bf\small (a)}}
\put (12.0,12.75){{\bf\small (b)}}
\put (4.0,6.15){{\bf\small (c)}}
\put (12.0,6.15){{\bf\small (d)}}
\put (4.0,-0.45){{\bf\small (e)}}
\put (12.0,-0.45){{\bf\small (f)}}
\end{picture}
\vspace{0.1cm}
\caption
{The estimated signal yield in data (dots) using the signal leakage
  fractions from (a,c,e) {\sc Pythia} or (b,d,f) {\sc Herwig} as
  functions of (a,b) $\delphj$, (c,d) $\mgjn$ and (e,f)
  $|\ctgjn|$. The distributions as functions of $\mgjn$ ($|\ctgjn|$)
  include requirements on $|\ctgjn|$ ($\mgjn$) and
  $|\etagdet+\rapjet|$ (see text). Other details as in the caption to
  Fig.~\ref{fig16}.}
\label{fig17}
\vfill
\end{figure}

\begin{figure}[t]
\vfill
\setlength{\unitlength}{1.0cm}
\begin{picture} (18.0,8.0)
\put (1.0,0.0){\epsfig{figure=\figdir 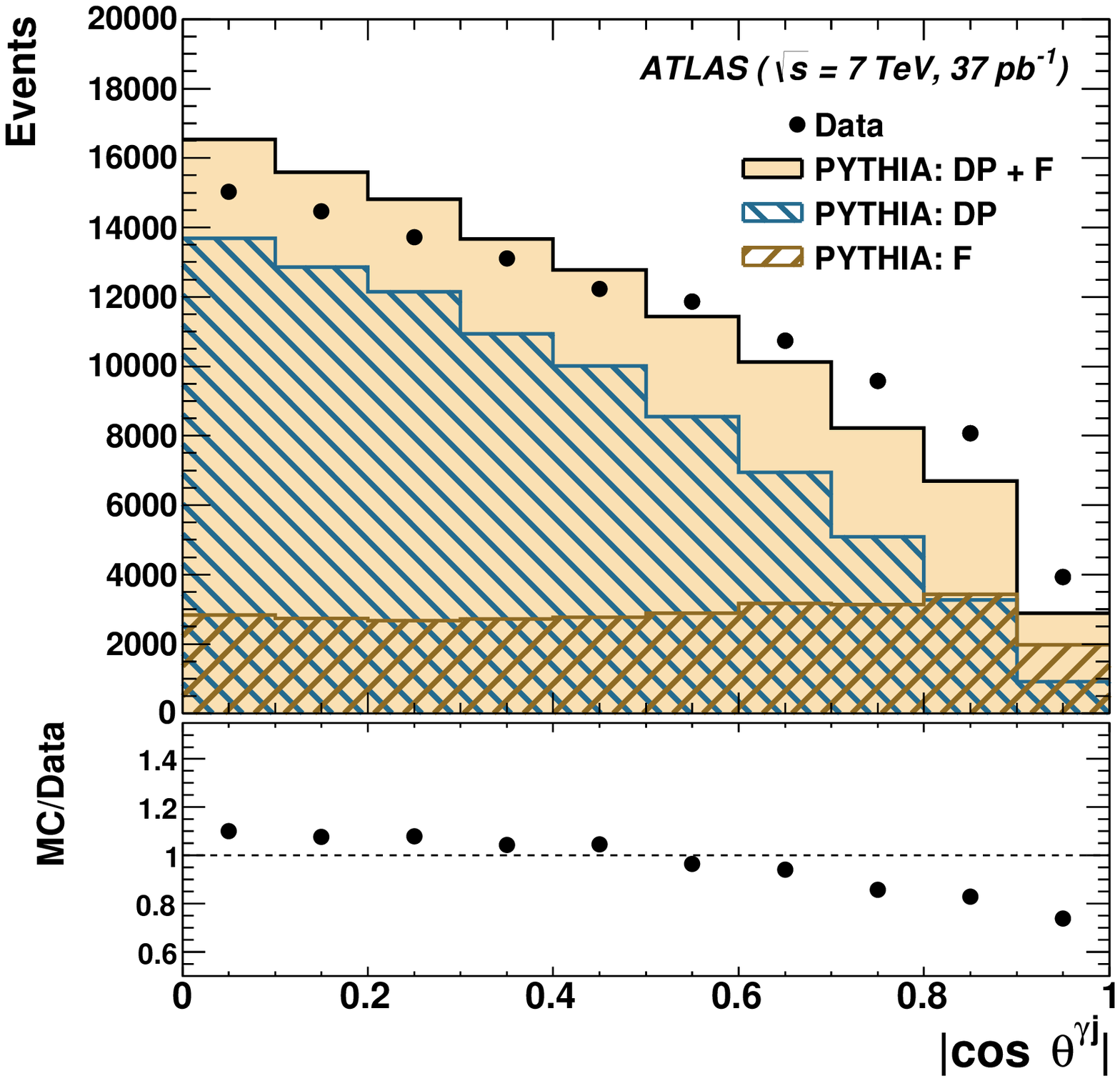,width=7cm}}
\put (9.0,0.0){\epsfig{figure=\figdir 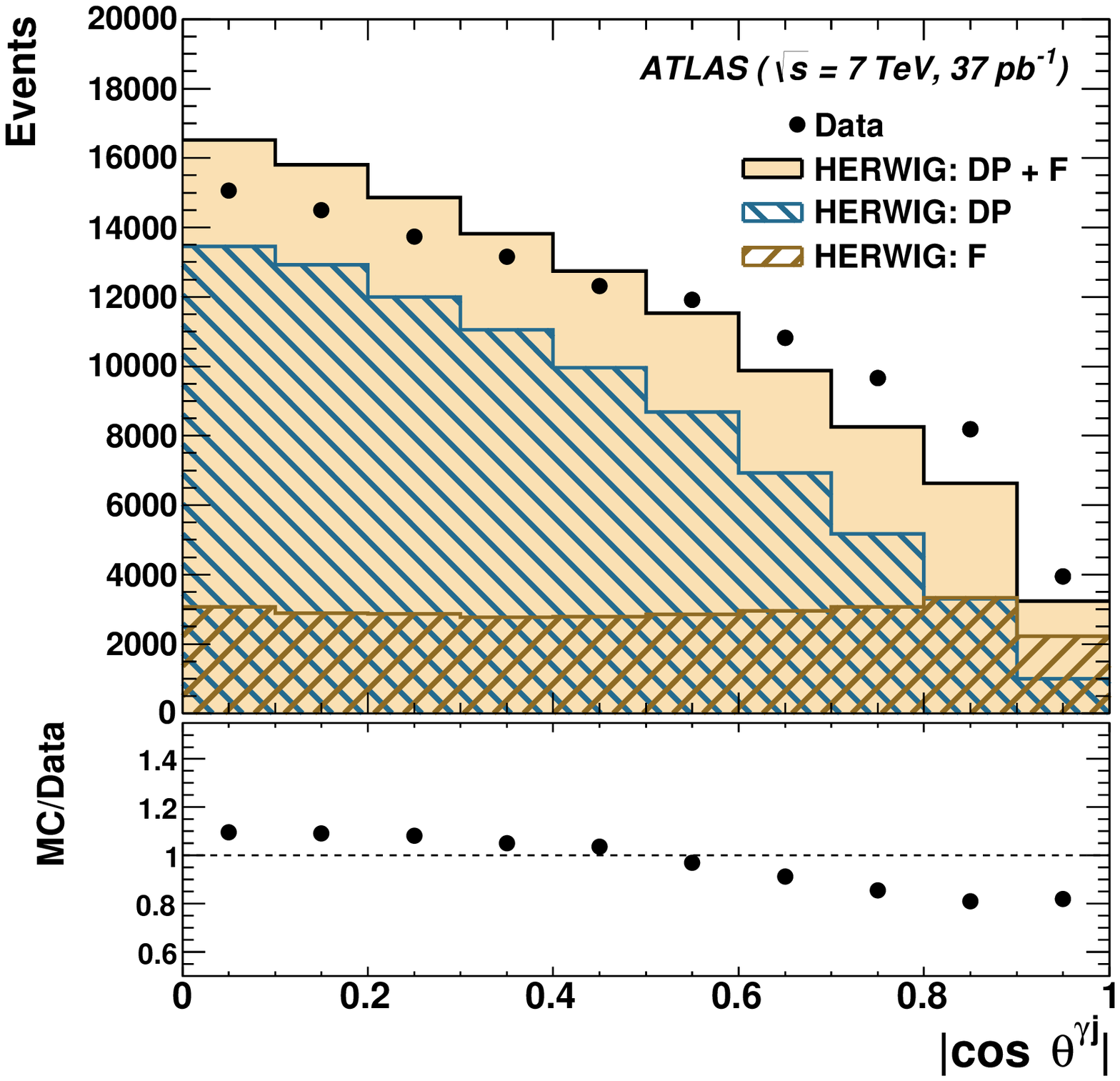,width=7cm}}
\put (4.0,0.55){{\bf\small (a)}}
\put (12.0,0.55){{\bf\small (b)}}
\end{picture}
\caption
{The estimated signal yield in data (dots) using the signal leakage
  fractions from (a) {\sc Pythia} or (b) {\sc Herwig} as functions of
  $|\ctgjn|$. These distributions do not include requirements on
  $\mgjn$ or $|\etagdet+\rapjet|$. Other details as in the caption to
  Fig.~\ref{fig16}.}
\label{fig1201}
\vfill
\end{figure}

The signal yield in data and the predictions of the signal MC
simulations are compared in Figs.~\ref{fig16}--\ref{fig1201}. Both
{\sc Pythia} and {\sc Herwig} give an adequate description of the
$\etg$, $|\rapjet|$ and $\mgjn$ data distributions. The measured
$\ptjet$ distribution is described well for $\ptjet\lesssim 100$~GeV;
for $\ptjet\gtrsim 100$~GeV, the simulation of {\sc Pythia} ({\sc
  Herwig}) has a tendency to be somewhat above (below) the data. The
simulation of {\sc Pythia} provides an adequate description of the
$\delphj$ data distribution, whereas that of {\sc Herwig} is somewhat
poorer. The $|\ctgjn|$ data distribution, with or without additional
requirements on $\mgjn$ or $|\etagdet+\rapjet|$, is not well described
by either {\sc Pythia} or {\sc Herwig}.

For most of these distributions, the shapes of the direct-photon and
fragmentation components in the signal MC simulations are somewhat
different. Therefore, in each case, the shape of the total MC
distribution depends on the relative fraction of the two
contributions. To obtain an improved description of the data by the
leading-order plus parton-shower MC samples, a fit to each data
distribution\footnote{ For the distribution of $\rapjet$, the result
  of the fit to that of $\ptjet$ was used.} was performed with the
weight of the direct-photon contribution, $\alpha$, as the free
parameter; the weight of the fragmentation contribution was given by
$1-\alpha$. In this context, the default admixture used in the MC
simulations would be represented by $\alpha=0.5$. The fitted values of
$\alpha$ were found to be different for each observable and in the
range $0.26$--$0.84$. It is emphasized that $\alpha$ does not
represent a physical observable and it was used solely for the purpose
of improving the description of the data by the LO
simulations. Nevertheless, an observable-dependent $\alpha$ may
approximate the effects of higher-order terms.\footnote{ In {\sc
    Pythia} and {\sc  Herwig}, the two components are simulated to
  LO. The NLO QCD radiative corrections are expected to affect
  differently the two components and their entanglement, making any
  distinction impossible. In fact, a variation was observed in the
  application of the same procedure at parton level: the optimal value
  of $\alpha$ resulting from a fit of the parton-level predictions of
  the two components in either {\sc Pythia} or {\sc Herwig} to the NLO
  QCD calculations (see Section~\ref{nlo}) depended on the
  observable.}

\begin{figure}[p]
\vfill
\setlength{\unitlength}{1.0cm}
\begin{picture} (18.0,16.9)
\put (1.0,12.2){\epsfig{figure=\figdir 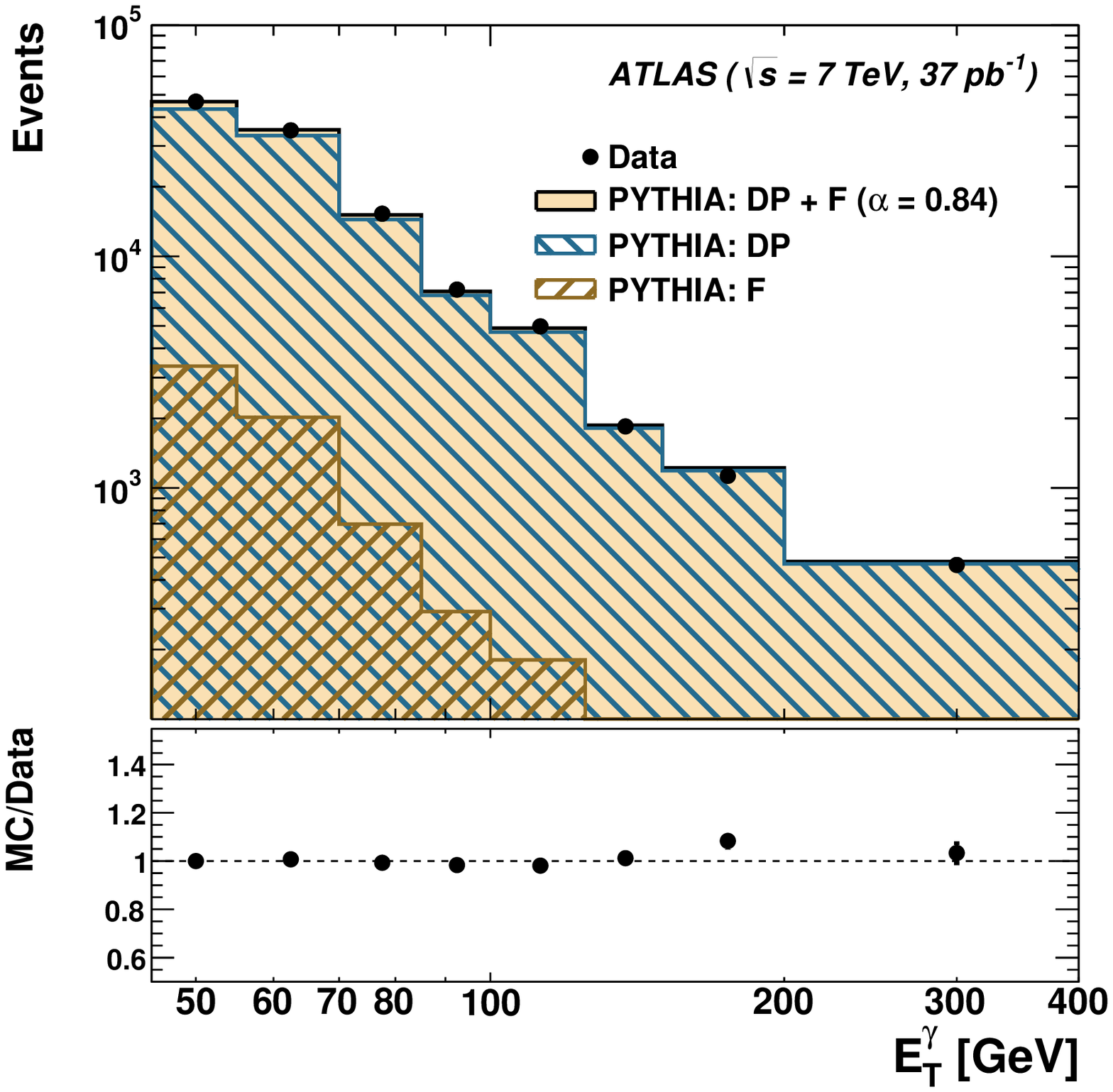,width=7cm}}
\put (9.0,12.2){\epsfig{figure=\figdir 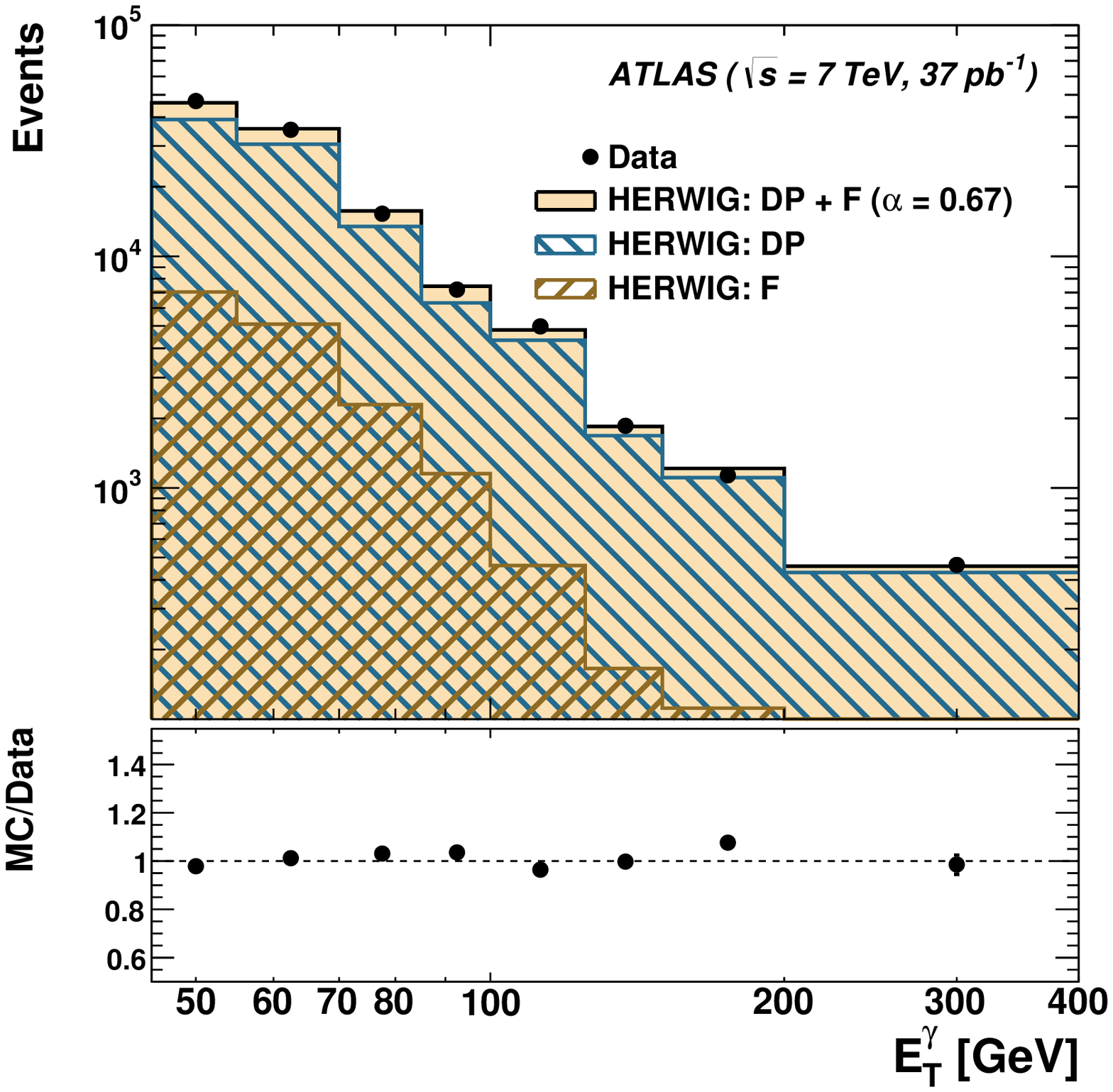,width=7cm}}
\put (1.0,5.6){\epsfig{figure=\figdir 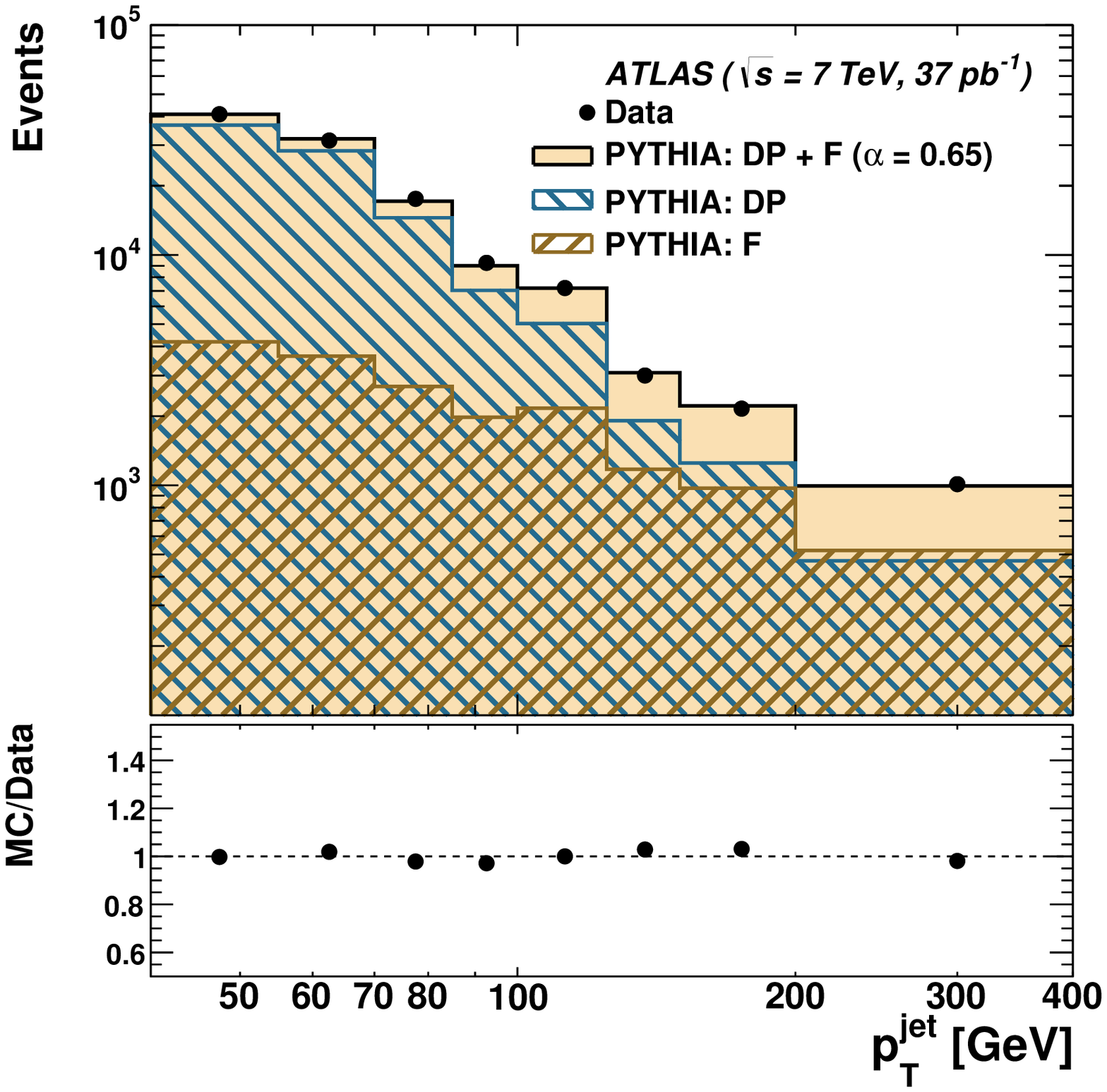,width=7cm}}
\put (9.0,5.6){\epsfig{figure=\figdir 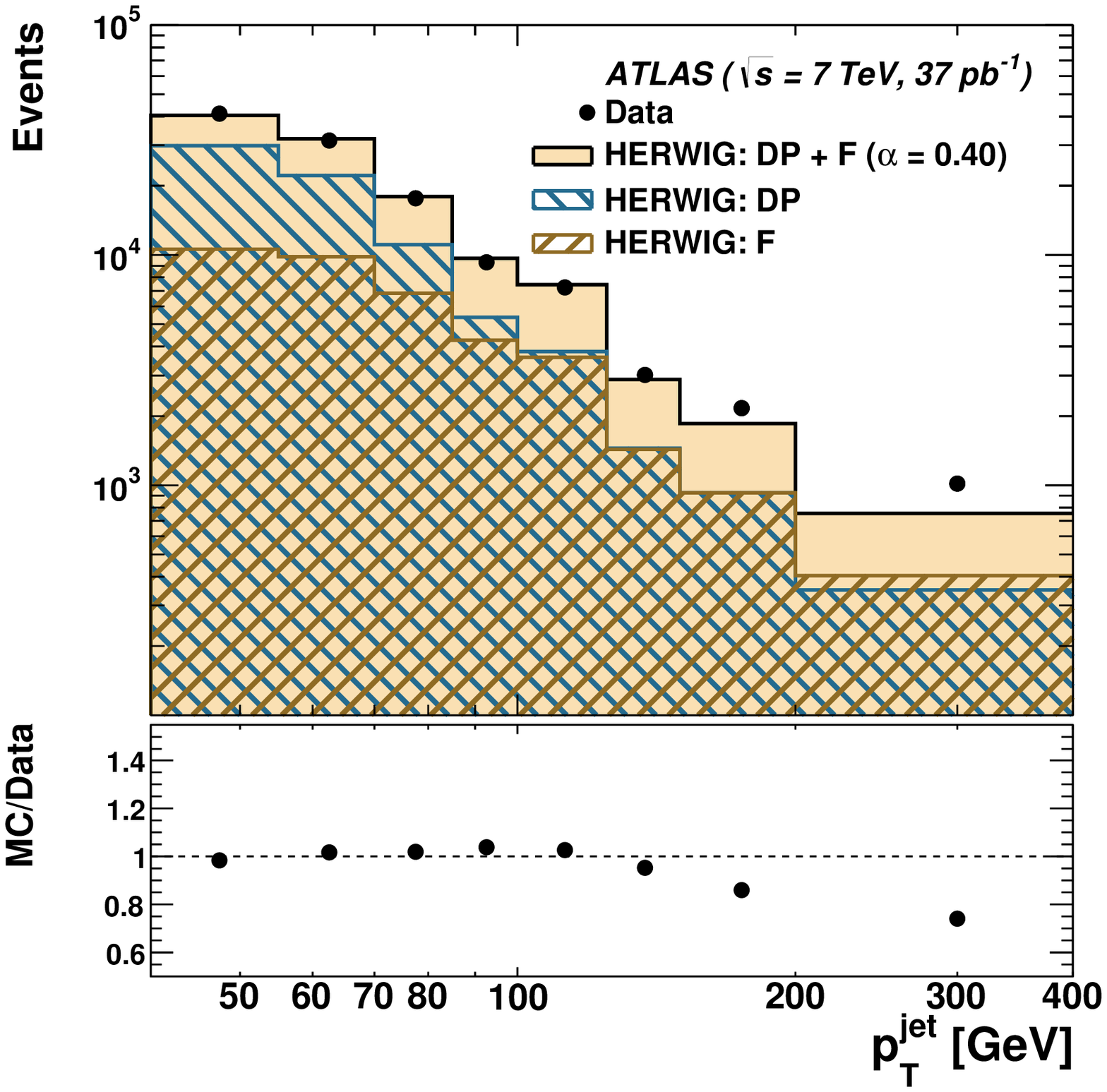,width=7cm}}
\put (1.0,-1.0){\epsfig{figure=\figdir 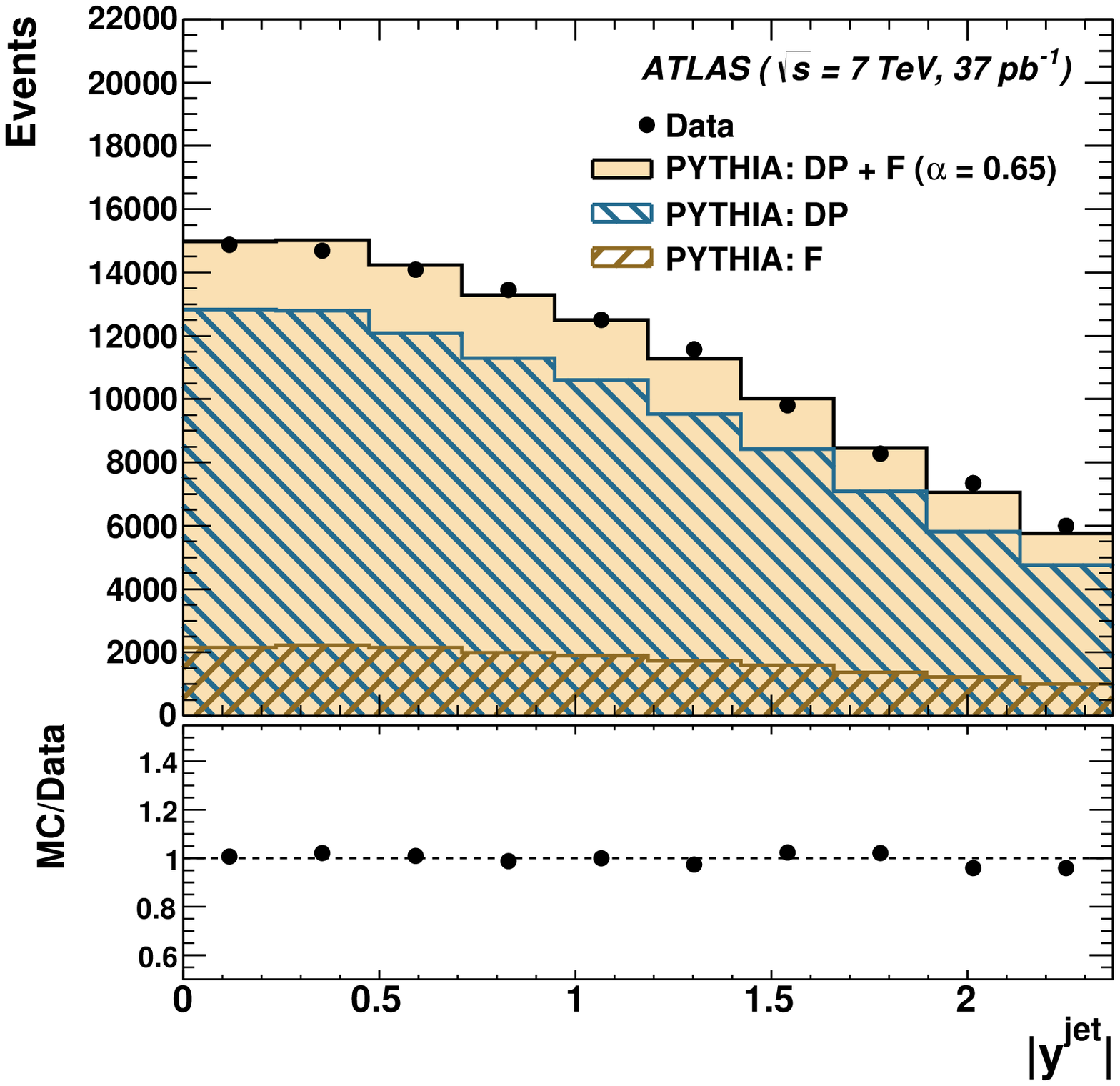,width=7cm}}
\put (9.0,-1.0){\epsfig{figure=\figdir 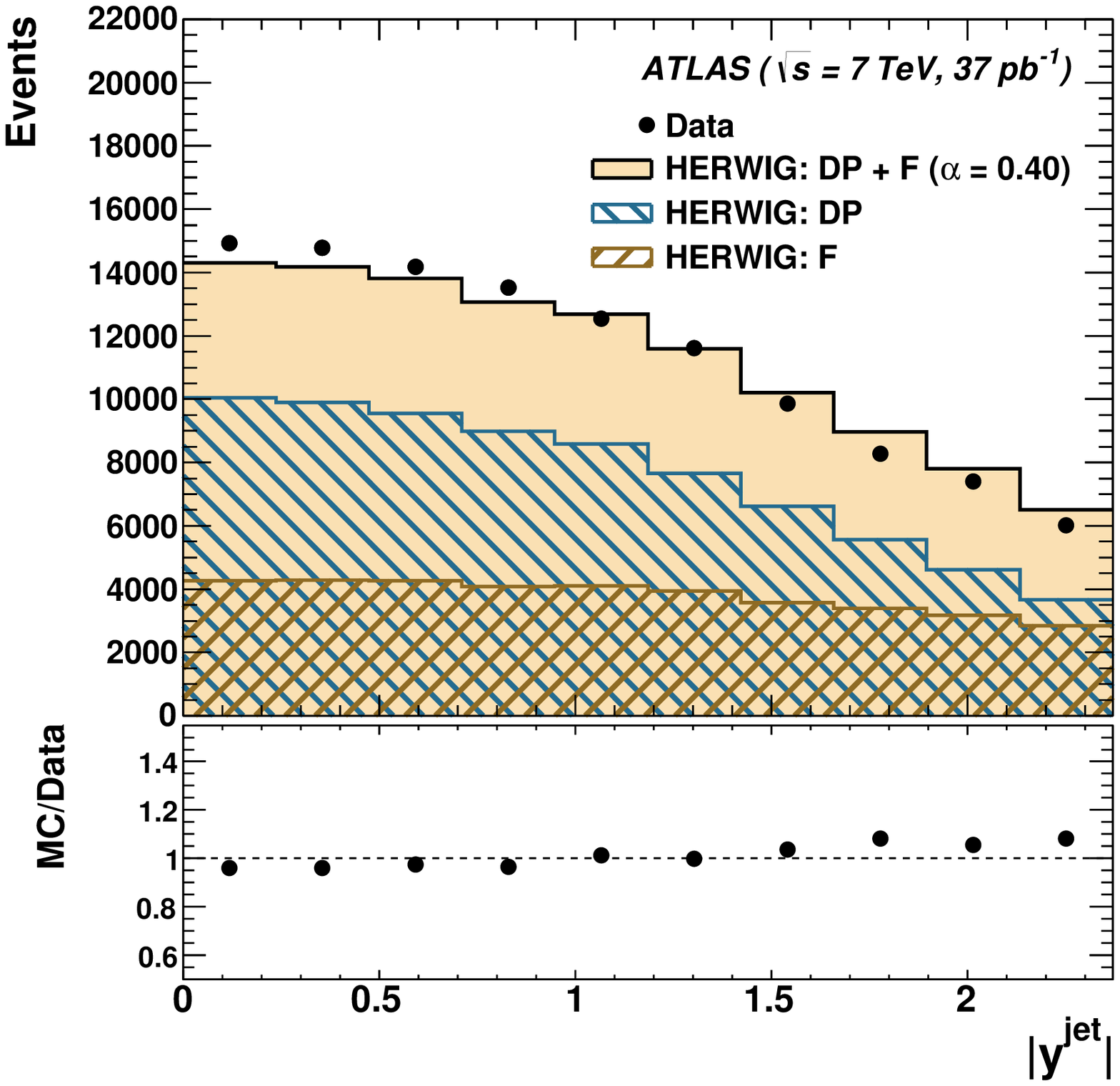,width=7cm}}
\put (4.0,12.75){{\bf\small (a)}}
\put (12.0,12.75){{\bf\small (b)}}
\put (4.0,6.15){{\bf\small (c)}}
\put (12.0,6.15){{\bf\small (d)}}
\put (4.0,-0.45){{\bf\small (e)}}
\put (12.0,-0.45){{\bf\small (f)}}
\end{picture}
\vspace{0.1cm}
\caption
{The estimated signal yield in data (dots) using the signal leakage
  fractions from (a,c,e) {\sc Pythia} or (b,d,f) {\sc Herwig} as
  functions of (a,b) $\etg$, (c,d) $\ptjet$ and (e,f) $|\rapjet|$. The
  direct-photon and fragmentation components of the MC simulations
  have been mixed using the value of $\alpha$ shown in each figure
  (see text). Other details as in the caption to Fig.~\ref{fig16}.}
\label{fig18}
\vfill
\end{figure}

\begin{figure}[p]
\vfill
\setlength{\unitlength}{1.0cm}
\begin{picture} (18.0,16.9)
\put (1.0,12.2){\epsfig{figure=\figdir 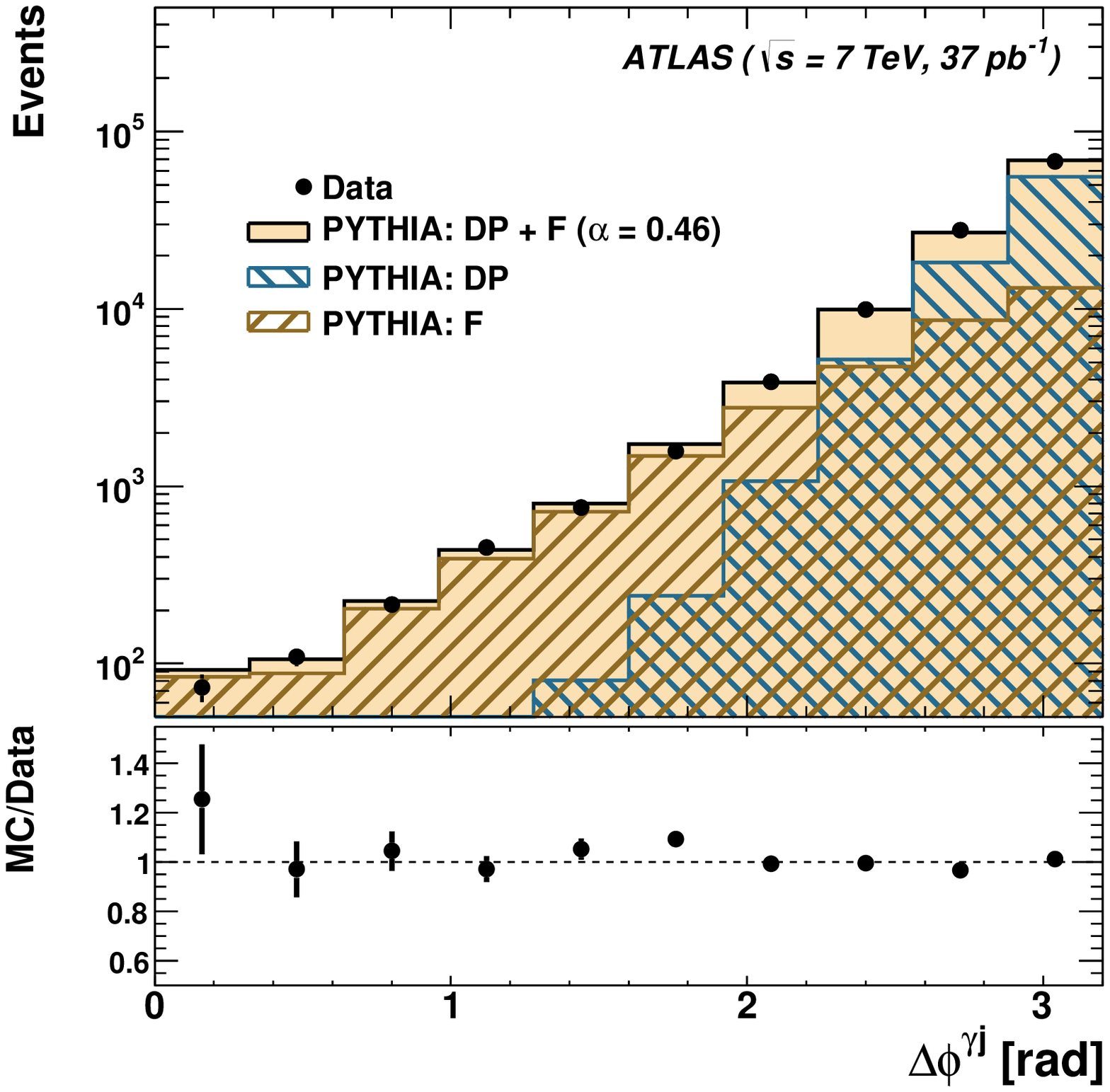,width=7cm}}
\put (9.0,12.2){\epsfig{figure=\figdir 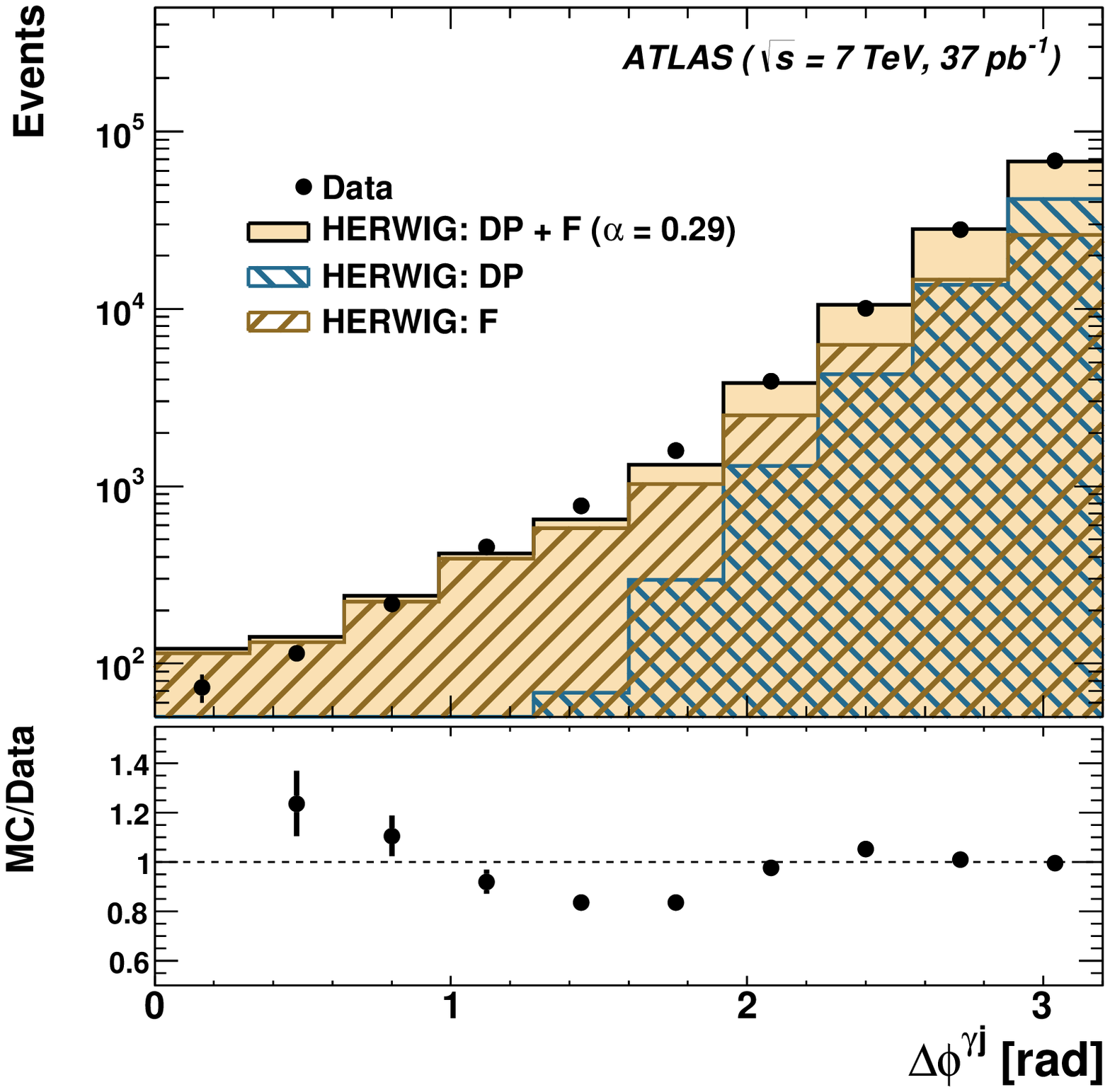,width=7cm}}
\put (1.0,5.6){\epsfig{figure=\figdir 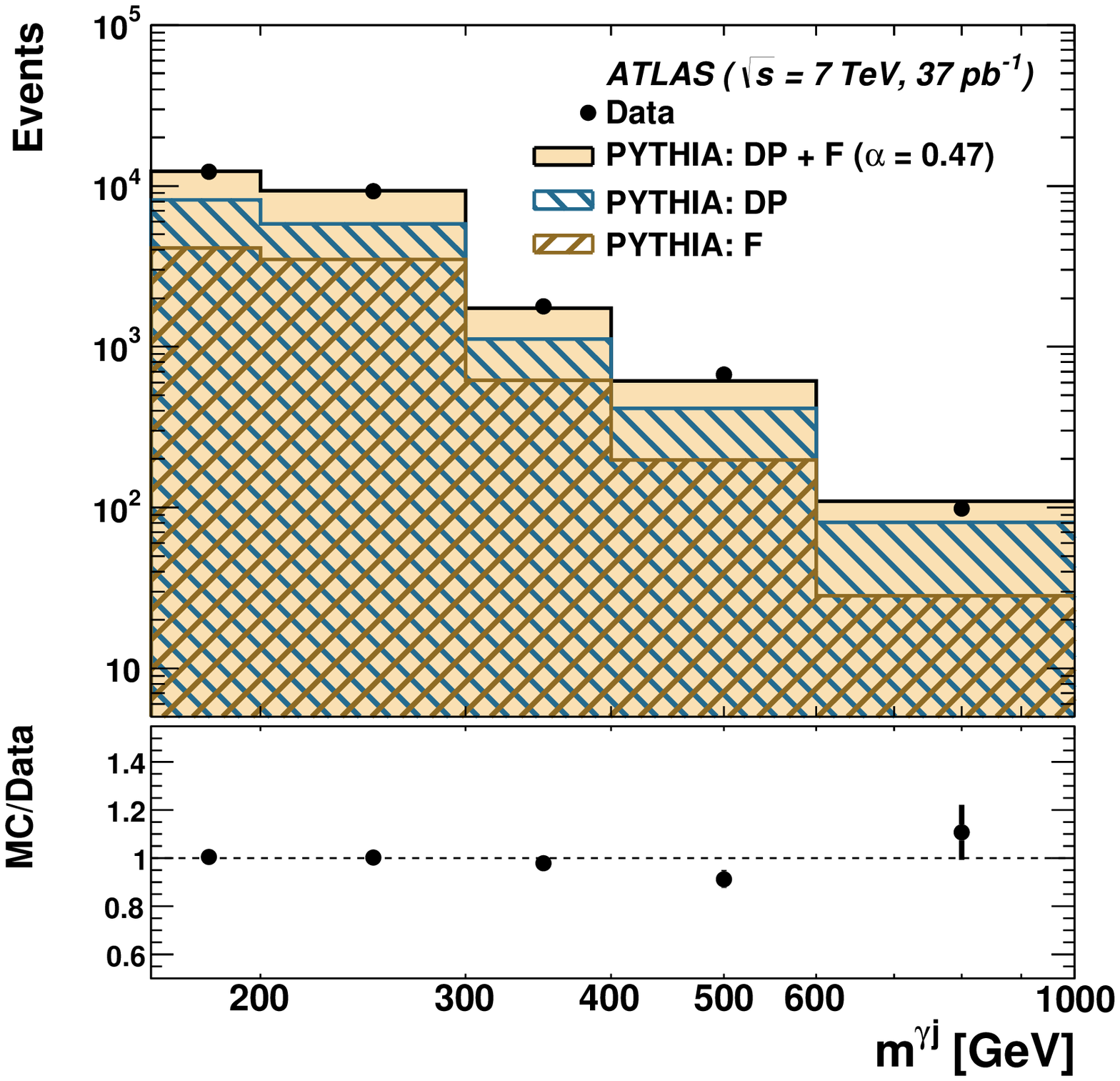,width=7cm}}
\put (9.0,5.6){\epsfig{figure=\figdir 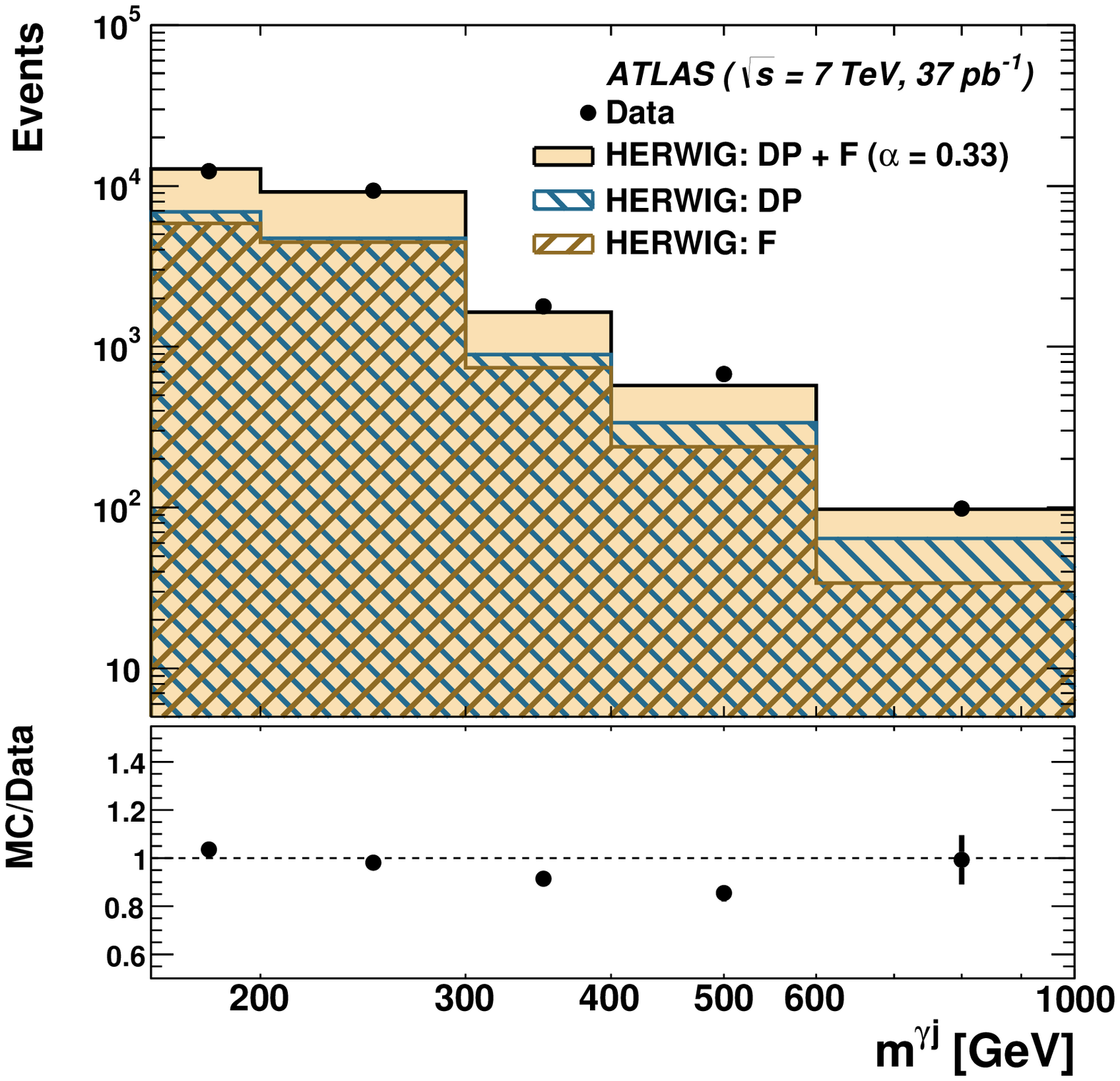,width=7cm}}
\put (1.0,-1.0){\epsfig{figure=\figdir 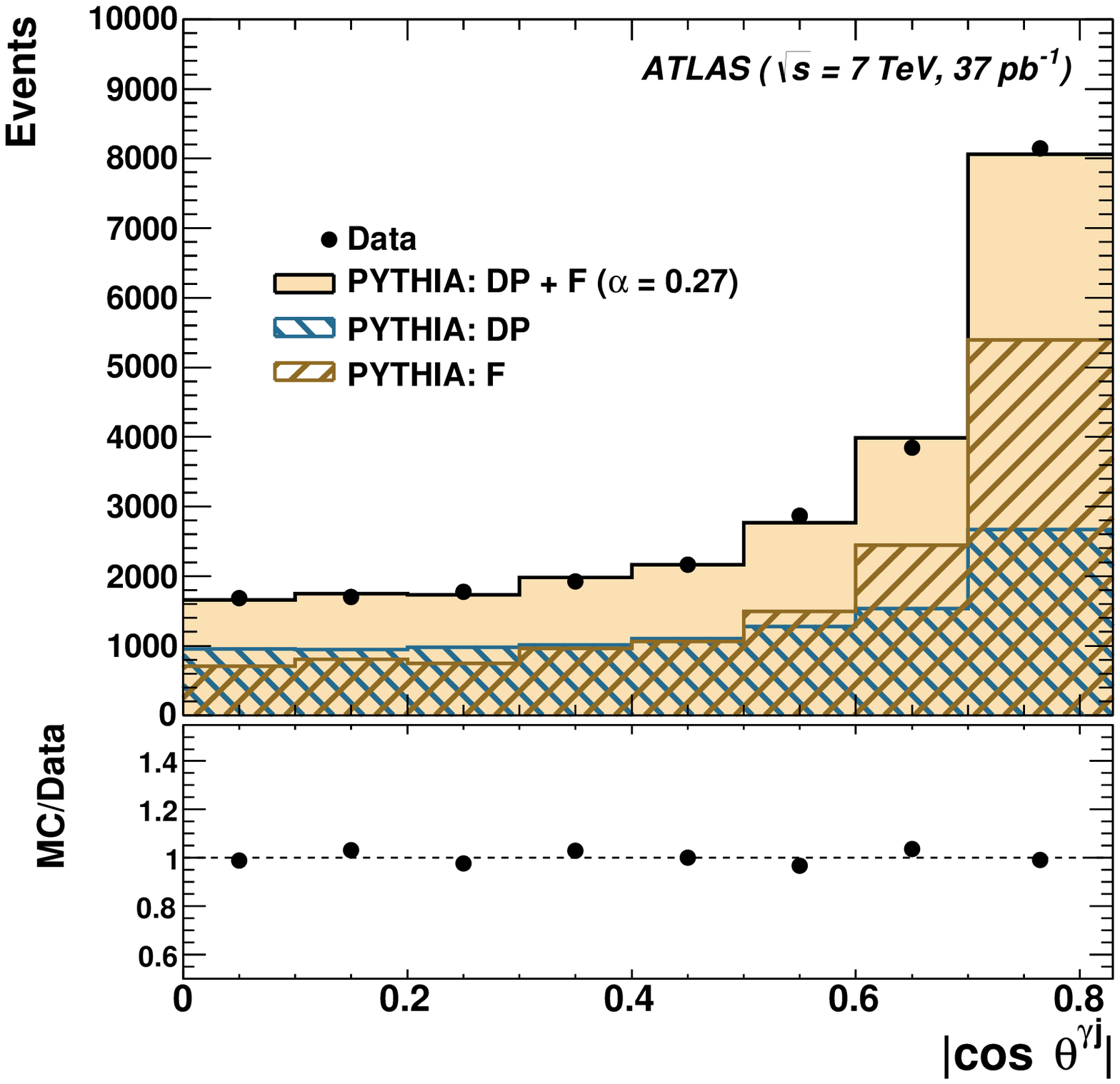,width=7cm}}
\put (9.0,-1.0){\epsfig{figure=\figdir 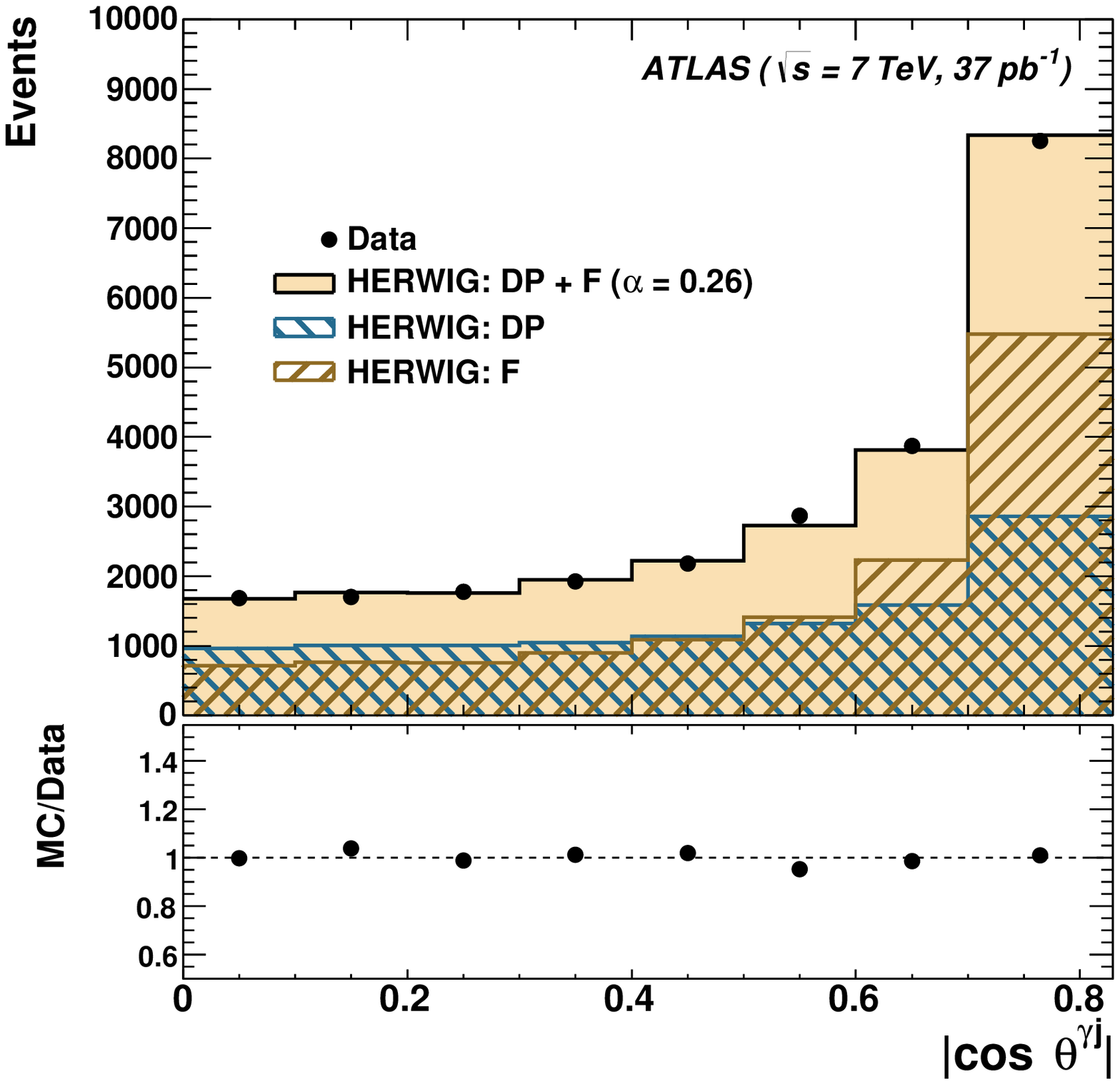,width=7cm}}
\put (4.0,12.75){{\bf\small (a)}}
\put (12.0,12.75){{\bf\small (b)}}
\put (4.0,6.15){{\bf\small (c)}}
\put (12.0,6.15){{\bf\small (d)}}
\put (4.0,-0.45){{\bf\small (e)}}
\put (12.0,-0.45){{\bf\small (f)}}
\end{picture}
\vspace{0.1cm}
\caption
{The estimated signal yield in data (dots) using the signal leakage
  fractions from (a,c,e) {\sc Pythia} or (b,d,f) {\sc Herwig} as
  functions of (a,b) $\delphj$, (c,d) $\mgjn$ and (e,f)
  $|\ctgjn|$. The distributions as functions of $\mgjn$ ($|\ctgjn|$)
  include requirements on $|\ctgjn|$ ($\mgjn$) and
  $|\etagdet+\rapjet|$ (see text). Other details as in the caption to
  Fig.~\ref{fig18}.}
\label{fig19}
\vfill
\end{figure}

\begin{figure}[t]
\vfill
\setlength{\unitlength}{1.0cm}
\begin{picture} (18.0,8.0)
\put (1.0,0.0){\epsfig{figure=\figdir 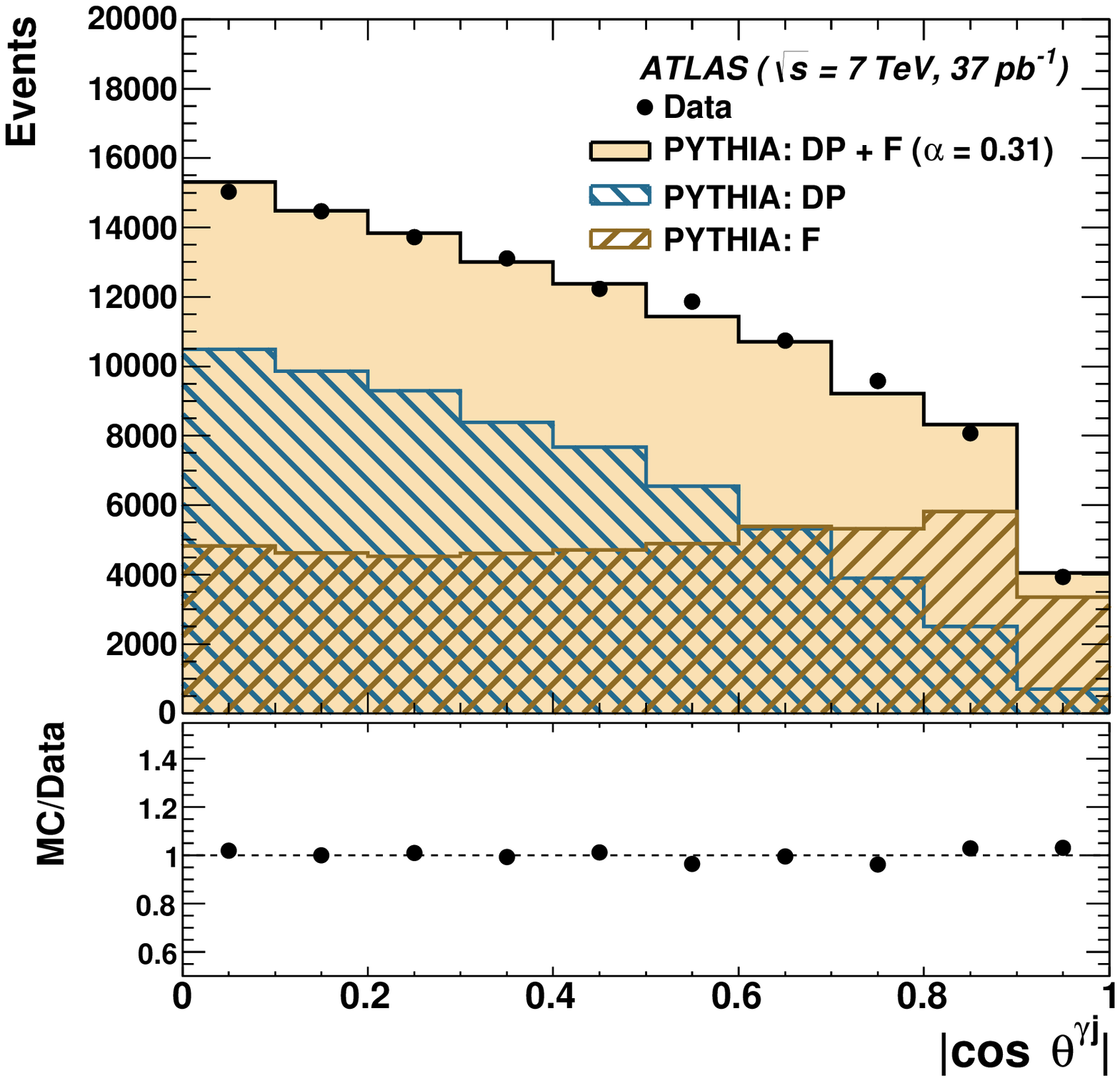,width=7cm}}
\put (9.0,0.0){\epsfig{figure=\figdir 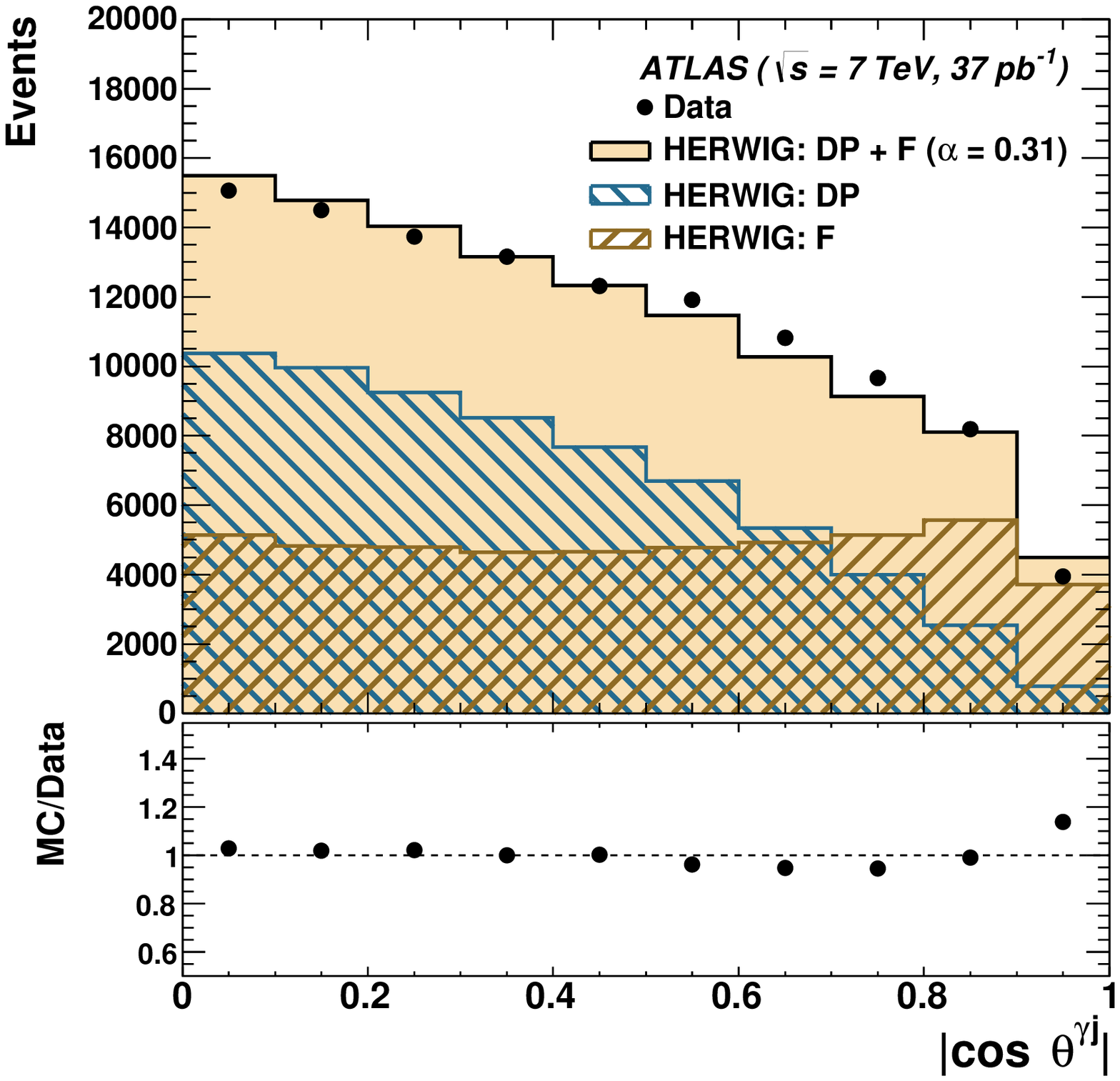,width=7cm}}
\put (4.0,0.55){{\bf\small (a)}}
\put (12.0,0.55){{\bf\small (b)}}
\end{picture}
\caption
{The estimated signal yield in data (dots) using the signal leakage
  fractions from (a) {\sc Pythia} or (b) {\sc Herwig} as functions of
  $|\ctgjn|$. These distributions do not include requirements on
  $\mgjn$ or $|\etagdet+\rapjet|$. Other details as in the caption to
  Fig.~\ref{fig18}.}
\label{fig1202}
\vfill
\end{figure}

After adjusting the fractions of the DP and F components separately
for each distribution, a good description of the data was obtained by
both the {\sc Pythia} and {\sc Herwig} MC simulations for all the
observables (see Figs.~\ref{fig18}--\ref{fig1202}), though the
descriptions of $\delphj$ and $\ptjet$ by {\sc Herwig} are still
somewhat poor. The MC simulations using the optimised admixture for
each observable were used as the baseline for the determination of the
measured cross sections (see Section~\ref{cor}).

To be consistent, the optimisation of the admixture of the two
components should be done simultaneously with the background
subtraction since the signal leakage fractions $\epsilon_K$ also
depend on the admixture. However, such a procedure would result in an
estimated signal yield that would depend on the fitted variable. To
obtain a signal yield independent of the observable, except for
statistical fluctuations, the background subtraction was performed
using the default admixture of the two components and a systematic
uncertainty on the background subtraction due to this admixture was
included (see Section~\ref{syst}).

\subsection{Signal efficiency}
\label{se}
The total selection efficiency, including trigger, reconstruction,
particle identification and event selection, was evaluated from the
simulated signal samples described in Section~\ref{mc}. The integrated
efficiency was computed as 
$\varepsilon=N^{\rm det,part}/N^{\rm part},$ where  $N^{\rm det,part}$
is the number of MC events that pass all the selection requirements at
both the detector and particle levels and $N^{\rm part}$ is the number
of MC events that pass the selection requirements at the particle
level. The integrated efficiency was found to be $68.5\ (67.9)\%$ from
the {\sc Pythia} ({\sc Herwig}) samples. The bin-to-bin efficiency was
computed as $\varepsilon_i=N^{\rm det,part}_i/N^{\rm part}_i,$ where
$N^{\rm det,part}_i$ is the number of MC events that pass all the
selection requirements at both the detector and particle levels and
are generated and reconstructed in bin $i$, and $N^{\rm part}_i$ is
the number of MC events that pass the selection requirements at the
particle level and are located in bin $i$. The bin-to-bin efficiencies
are typically above $60\%$, except for $\ptjet$ and $\delphj$
($\gtrsim 40\%$) due to the limited resolution in these steeply
falling distributions, and are similar for {\sc Pythia} and {\sc Herwig}.

The bin-to-bin reconstruction purity was computed as 
$\kappa_i=N^{\rm det,part}_i/N^{\rm det}_i,$ where $N^{\rm det}_i$ is
the number of MC events that pass the selection requirements at the
detector level and are located in bin $i$. The bin-to-bin
reconstruction purities are typically above $70\%$, except for
$\ptjet$ and $\delphj$ ($\gtrsim 45\%$) due to the limited resolution
in these steeply falling distributions, and are similar for {\sc
  Pythia} and {\sc Herwig}.

The efficiency of the jet-quality criteria (see Section~\ref{jetsel})
applied to the data was estimated using a tag-and-probe method. The
leading photon in each event was considered as the tag to probe the
leading jet. Additional selection criteria, such as $\delphj>2.6$
(probe and tag required to be back-to-back) and 
$|\ptjet-\etg|/p_{\rm T}^{\rm avg}<0.4$, where 
$p_{\rm T}^{\rm avg}=(\ptjet+\etg)/2$ (to have well-balanced probe and
tag), were applied. The jet-quality criteria were then applied to the
leading jet and the fraction of jets accepted was measured as a
function of $\ptjet$ and $|\rapjet|$. The jet-quality selection
efficiency is approximately $99\%$. No correction for this efficiency
was applied, but an uncertainty was included in the measurements (see
Section~\ref{syst}).

\section{Cross-section measurement procedure}
\label{cor}
Isolated-photon plus jet cross sections were measured for photons with
$\etg>45$~GeV, $|\etag|<2.37$ (excluding the region
$1.37<|\etag|<1.52$) and $\etisop<4$~GeV. The jets were reconstructed
using the anti-$\kt$ jet algorithm with $R=0.6$ and selected with
$\ptjet>40$~GeV, $|\rapjet|<2.37$ and 
$\Delta R_{\gamma{\rm j}}>1$. Bin-averaged cross sections were
measured as functions of $\etg$, $\ptjet$, $|\rapjet|$ and
$\delphj$. Bin-averaged cross sections as functions of $\mgjn$ and
$|\ctgjn|$ were measured in the kinematic region
$|\etag+\rapjet|<2.37$, $|\ctgjn|<0.83$ and $\mgjn>161$~GeV. In
addition, the bin-averaged cross section as a function of $|\ctgjn|$
was measured without the requirements on $\mgjn$ or
$|\etag+\rapjet|$.

The data distributions, after background subtraction, were corrected
to the particle level using a bin-by-bin correction procedure. The
bin-by-bin correction factors were determined using the MC samples;
these correction factors took into account the efficiency of the
selection criteria, jet and photon reconstruction as well as migration
effects.

For this approach to be valid, the uncorrected distributions of the
data must be adequately described by the MC simulations at the
detector level. This condition was satisfied by both the {\sc Pythia}
and {\sc Herwig} MC samples after adjusting the relative fractions of
the LO direct-photon and fragmentation components (see
Section~\ref{bgks}). The data distributions were corrected to the
particle level via the formula

$$\frac{d\sigma}{d{\cal O}}(i)=\frac{N_A^{\rm sig}(i)\ C^{\rm MC}(i)}{{\cal L}\
  \Delta {\cal O}(i)},$$
where $d\sigma/d{\cal O}$ is the bin-averaged cross section as a
function of observable ${\cal O}=\etg$, $\ptjet$, $|\rapjet|$,
$\delphj$, $\mgjn$ or $|\ctgjn|$, $N_A^{\rm sig}(i)$ is the number of
background-subtracted data events in bin $i$, $C^{\rm MC}(i)$ is the
correction factor in bin $i$, ${\cal L}$ is the integrated luminosity
and $\Delta {\cal O}(i)$ is the width of bin $i$. The bin-by-bin
correction factors were computed as

$$C^{\rm MC}(i)=\frac{\alpha\
    N^{\rm MC,DP}_{\rm part}(i)+(1\!-\!\alpha)\ N^{\rm MC,F}_{\rm part}(i)}{\alpha\ N^{\rm MC,DP}_{\rm det}(i)+ (1\!-\!\alpha)\  N^{\rm MC,F}_{\rm det}(i)},$$
where $\alpha$ corresponds to the optimised value obtained from the
fit to the data for each observable, as explained in
Section~\ref{bgks}. The final bin-averaged cross sections were
obtained from the average of the cross sections when using $C^{\rm
  MC}$ with MC = {\sc Pythia} or {\sc Herwig}. The uncertainties from
the parton-shower and hadronisation models used for the corrections
were estimated as the deviations from this average when using either
{\sc Pythia} or {\sc Herwig} to correct the data (see
Section~\ref{syst}). The correction factors differ from unity by
typically $20\%$ and are similar for {\sc Pythia} and {\sc Herwig}.

\section{Systematic uncertainties}
\label{syst}
The following sources of systematic uncertainty were considered;
average values, expressed in percent and shown in parentheses,
quantify their effects on the cross section as a function of
$|\ctgjn|$ (with the requirements on $\mgjn$ and $|\etag+\rapjet|$
applied):
\begin{itemize}
\item[$\bullet$] Simulation of the detector geometry. The systematic
  uncertainties originating from the limited knowledge of the material
  in the detector were evaluated by repeating the full analysis using
  a different detector simulation with increased material in front of
  the calorimeter~\cite{epj:c72:1909}. This affects in particular the
  photon-conversion rate and the development of electromagnetic
  showers ($\pm 5\%$).
\item[$\bullet$] Photon simulation and model and fit dependence. The
  MC simulation of the signal was used to estimate (i) the signal
  leakage fractions and (ii) the bin-by-bin correction factors:
  \begin{itemize}
  \item For step (i), both the {\sc Pythia} and {\sc Herwig}
    simulations were used with the admixture of the direct-photon and
    fragmentation components as given by each MC simulation to yield
    two sets of background-subtracted data distributions. The signal
    leakage fractions depend on the relative fraction of the two
    components. The uncertainty related to the simulation of the
    isolated-photon components in the signal leakage fractions was
    estimated (conservatively) by performing the background
    subtraction with only the direct-photon or the fragmentation
    component ($\pm 3\%$).
  \item For step (ii), the effects of the parton-shower and
    hadronisation models in the bin-by-bin correction factors were
    estimated as deviations from the nominal cross sections by using
    either only {\sc Pythia} or only {\sc Herwig} to correct the data
    ($\pm 1\%$).
  \item The bin-by-bin correction factors also depend on the relative
    fractions of the two components; the nominal admixture was taken
    from the fit to the background-subtracted data distributions. A
    systematic uncertainty due to the fit was estimated
    (conservatively) by using the default admixture of the components
    ($\pm 2\%$).
  \end{itemize}
\item[$\bullet$] Jet and photon energy scale and resolution
  uncertainties. These uncertainties were estimated by varying both
  the electromagnetic and the jet energy scales and resolutions within
  their uncertainties~\cite{epj:c72:1909,epj:c73:2304} (photon energy
  resolution: $\pm 0.2\%$; photon energy scale: $\pm 1\%$; jet energy
  resolution: $\pm 1\%$; jet energy scale: $\pm 5\%$).
\item[$\bullet$] Uncertainty on the background correlation in the
  two-dimensional sideband method. In the background subtraction,
  $R^{\rm bg}=1$ was assumed (see Section~\ref{bgks}); i.e. the photon
  isolation and identification variables are uncorrelated for the
  background. This assumption was verified using both the data and
  simulated background samples and was found to hold within a $10\%$
  uncertainty in the kinematic region of the measurements presented
  here. The cross sections were recomputed accounting for possible
  correlations in the background subtraction, and the differences from
  the nominal results were taken as systematic uncertainties ($\pm
  0.6\%$).
\item[$\bullet$] Definition of the background control regions in the
  two-dimensional sideband method. The estimation of the contamination
  in the signal region is affected by the choice of the background
  control regions. The uncertainty due to this choice was estimated by
  repeating the analysis with different identification criteria and by
  changing the isolation boundary from the nominal value of $5$~GeV to
  $4$ or $6$~GeV ($\pm 2\%$).
\item[$\bullet$] Data-driven correction to the photon efficiency. The
  shower shapes of simulated photons in the calorimeter were corrected
  to improve the agreement with the data. The uncertainty on the
  photon-identification efficiency due to the application of these
  corrections was estimated using different simulated photon samples
  and a different detector simulation with increased material in front
  of the calorimeter~\cite{epj:c72:1909} ($\pm 2\%$).
\item[$\bullet$] Uncertainty on the jet reconstruction efficiency. The
  MC simulation reproduces the jet reconstruction efficiencies in the
  data to better than $1\%$~\cite{pr:d86:014022} ($\pm 1\%$).
\item[$\bullet$] Jet-quality selection efficiency. The efficiency of
  the jet-quality criteria was determined to be $99\%$ ($+1\%$).
\item[$\bullet$] Uncertainty on the trigger efficiency ($\pm 0.7\%$).
\item[$\bullet$] Uncertainty arising from the photon-isolation
  requirement. This uncertainty was evaluated by increasing the value
  of $\etisod$ in the MC simulations by the difference ($+500$ MeV)
  between the averages of $\etisod$ for electrons in simulation and
  data control samples~\cite{pl:b706:150} ($+4\%$).
\item[$\bullet$] Uncertainty on the integrated luminosity. The
  measurement of the luminosity has a $\pm 3.4\%$
  uncertainty~\cite{1302.4393} ($\pm 3.4\%$).
\end{itemize}

For $d\sigma/d\etg$, the dominant uncertainties arise from the
detector material in the simulation, the isolation requirement, the
model dependence in the signal leakage fractions and the photon energy
scale, though in some bins the uncertainty from the luminosity
measurement provides the largest contribution. The dominant
uncertainties for the other bin-averaged cross sections come from the
detector simulation, the model dependence in the signal leakage
fractions, the isolation requirement and the jet energy scale. All
these systematic uncertainties were added in quadrature together with
the statistical uncertainty and are shown as error bars in the figures
of the measured cross sections (see Section~\ref{res}).

\section{Next-to-leading-order QCD calculations}
\label{nlo}
The NLO QCD calculations used in this analysis were computed using the
program {\sc Jetphox}~\cite{jhep:0205:028}. This program includes a
full NLO QCD calculation of both the direct-photon and fragmentation
contributions to the cross section.

The number of flavours was set to five. The renormalisation ($\mu_R$),
factorisation ($\mu_F$) and fragmentation ($\mu_f$) scales were chosen
to be $\mu_R=\mu_F=\mu_f=\etg$. The calculations were performed using
the CTEQ6.6~\cite{pr:d78:013004} parameterisations of the proton PDFs
and the NLO photon BFG set II photon fragmentation
function~\cite{epj:c2:529}. The strong coupling constant was
calculated at two-loop order with $\asz=0.118$. Predictions based on
the CT10~\cite{pr:d82:074024} and MSTW2008nlo \cite{epj:c64:653}
proton PDF sets were also computed.

The calculations were performed using a parton-level isolation cut,
which required a total transverse energy below $4$~GeV from the
partons inside a cone of radius $\Delta R=0.4$ around the photon
direction. The anti-$\kt$ algorithm was applied to the partons in the
events generated by this program to define jets of partons. The NLO
QCD predictions were obtained using the photon and these jets of
partons in each event.

\subsection{Hadronisation and underlying-event corrections to the
  NLO QCD calculations}
Since the measurements refer to jets of hadrons with the contribution
from the underlying event included, whereas the NLO QCD calculations
refer to jets of partons, the predictions were corrected to the
particle level using the MC models. The multiplicative correction
factor, $C_{\rm NLO}$, was defined as the ratio of the cross section
for jets of hadrons to that for jets of partons and was estimated by
using the MC programs described in Section~\ref{mc}; a simulation of
the underlying event was only included for the sample of events at
particle level. The correction factors from {\sc Pythia} and {\sc
  Herwig} are similar and close to unity, except at high $\ptjet$; for
$\ptjet>200$~GeV, the value of $C_{\rm NLO}$ is $0.87$ ($0.82$) for
{\sc Pythia} ({\sc Herwig}). The means of the factors obtained from
{\sc Pythia} and {\sc Herwig} were applied to the NLO QCD calculations.

\subsection{Theoretical uncertainties}
The following sources of uncertainty in the theoretical predictions
were considered; average values, expressed in percent and shown in
parentheses, quantify their effects on the cross section as a function
of $|\ctgjn|$ (with the requirements on $\mgjn$ and $|\etag+\rapjet|$
applied):
\begin{itemize}
\item[$\bullet$] The uncertainty on the NLO QCD calculations due to
  terms beyond NLO was estimated by repeating the calculations using
  values of $\mu_R$, $\mu_F$ and $\mu_f$ scaled by the factors $0.5$
  and $2$. The three scales were either varied simultaneously,
  individually or by fixing one and varying the other two. In all
  cases, the condition $0.5\leq\mu_A/\mu_B\leq 2$ was imposed, where
  $A,B=R,F,f$ and $A\neq B$. The final uncertainty was taken as the
  largest deviation from the nominal value among the 14 possible
  variations ($\pm 14\%$) and is dominated by the $\mu_R$ variations.
\item[$\bullet$] The uncertainty on the NLO QCD calculations due to
  those on the proton PDFs was estimated by repeating the calculations
  using the 44 additional sets from the CTEQ6.6 error analysis ($\pm
  3.5\%$).
\item[$\bullet$] The uncertainty on the NLO QCD calculations due to
  that on the value of $\asz$ was estimated by repeating the
  calculations using two additional sets of proton PDFs, for which
  different values of $\asz$ were assumed in the fits, namely
  $\asz=0.116$ and $0.120$, following the prescription of
  Ref.~\cite{pr:d82:054021} ($\pm 2.5\%$).
\item[$\bullet$] The uncertainty on the NLO QCD calculations due to
  the modelling of the parton shower, hadronisation and underlying
  event was estimated by taking the difference of the $C_{\rm NLO}$
  factors based on {\sc Pythia} and {\sc Herwig} from their average
  ($\pm 0.5\%$).
\end{itemize}

For all observables, the dominant theoretical uncertainty is that
arising from the terms beyond NLO. The total theoretical uncertainty
was obtained by adding in quadrature the individual uncertainties
listed above.

\section{Results}
\label{res}
\begin{figure}[p]
\setlength{\unitlength}{1.0cm}
\begin{picture} (10.0,9.0)
\put (0.0,0.0){\centerline{\epsfig{figure=\figdir 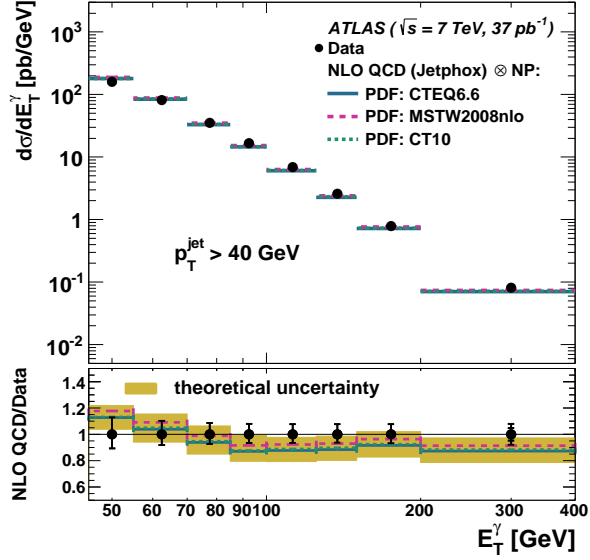,width=9cm}}}
\end{picture}
\vspace{-1cm}
\caption
{The measured bin-averaged cross section for isolated-photon plus jet
  production (dots) as a function of $\etg$. The NLO QCD calculations
  from {\sc Jetphox} corrected for hadronisation and underlying-event
  effects (non-perturbative effects, NP) and using the CTEQ6.6 (solid
  lines), MSTW2008nlo (dashed lines) and CT10 (dotted lines) PDF sets
  are also shown. The bottom part of the figure shows the ratios of
  the NLO QCD calculations to the measured cross section. The inner
  (outer) error bars represent the statistical uncertainties (the
  statistical and systematic uncertainties added in quadrature) and
  the shaded band represents the theoretical uncertainty. For most of
  the points, the inner error bars are smaller than the marker size
  and, thus, not visible.}
\label{fig101}
\end{figure}

\begin{figure}[p]
\setlength{\unitlength}{1.0cm}
\begin{picture} (10.0,9.0)
\put (0.0,0.0){\centerline{\epsfig{figure=\figdir 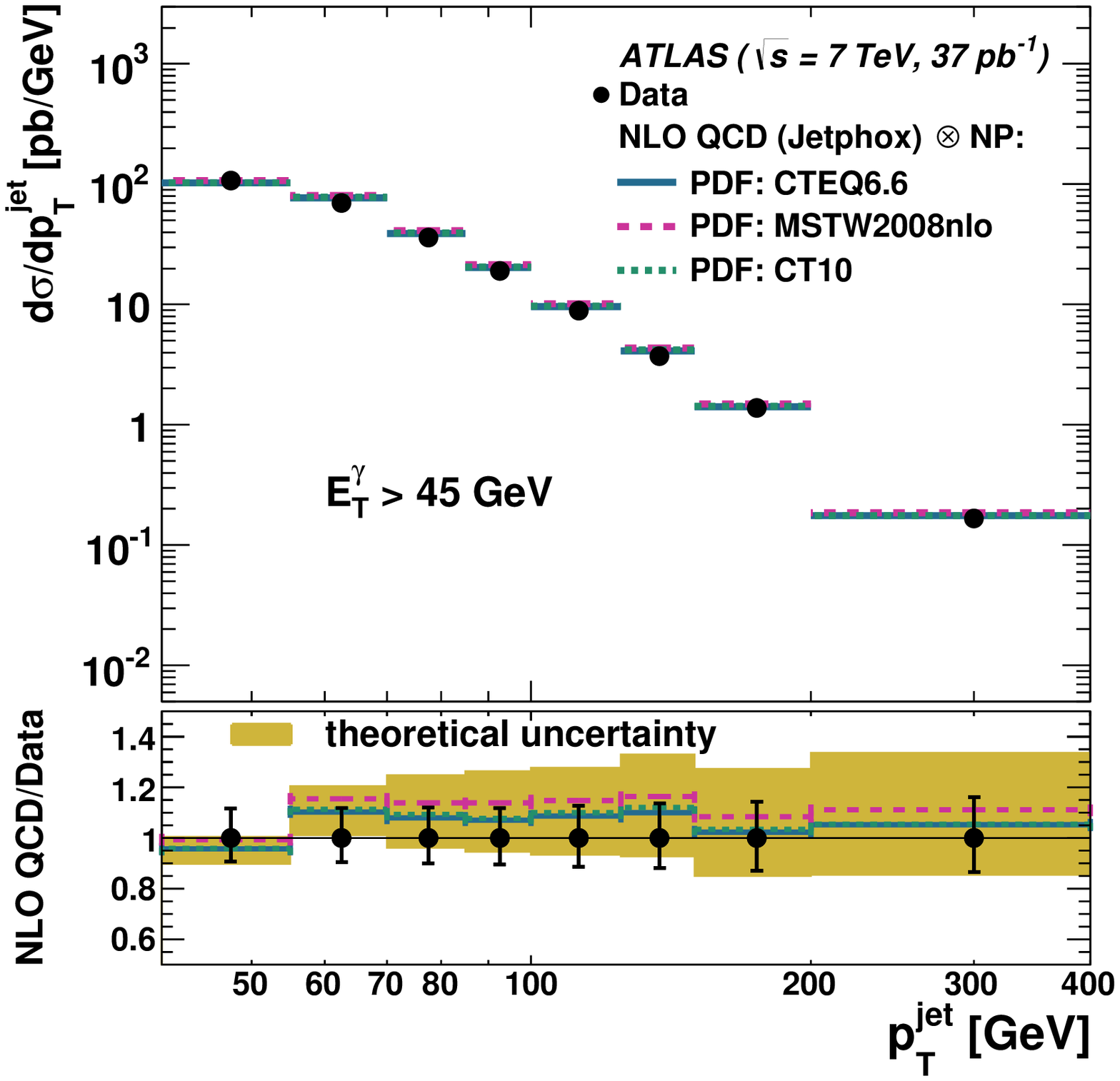,width=9cm}}}
\end{picture}
\vspace{-1cm}
\caption
{The measured bin-averaged cross section for isolated-photon plus jet
  production (dots) as a function of $\ptjet$. Other details as in the
  caption to Fig.~\ref{fig101}.}
\label{fig102}
\end{figure}

\begin{figure}[p]
\setlength{\unitlength}{1.0cm}
\begin{picture} (10.0,9.0)
\put (0.0,0.0){\centerline{\epsfig{figure=\figdir 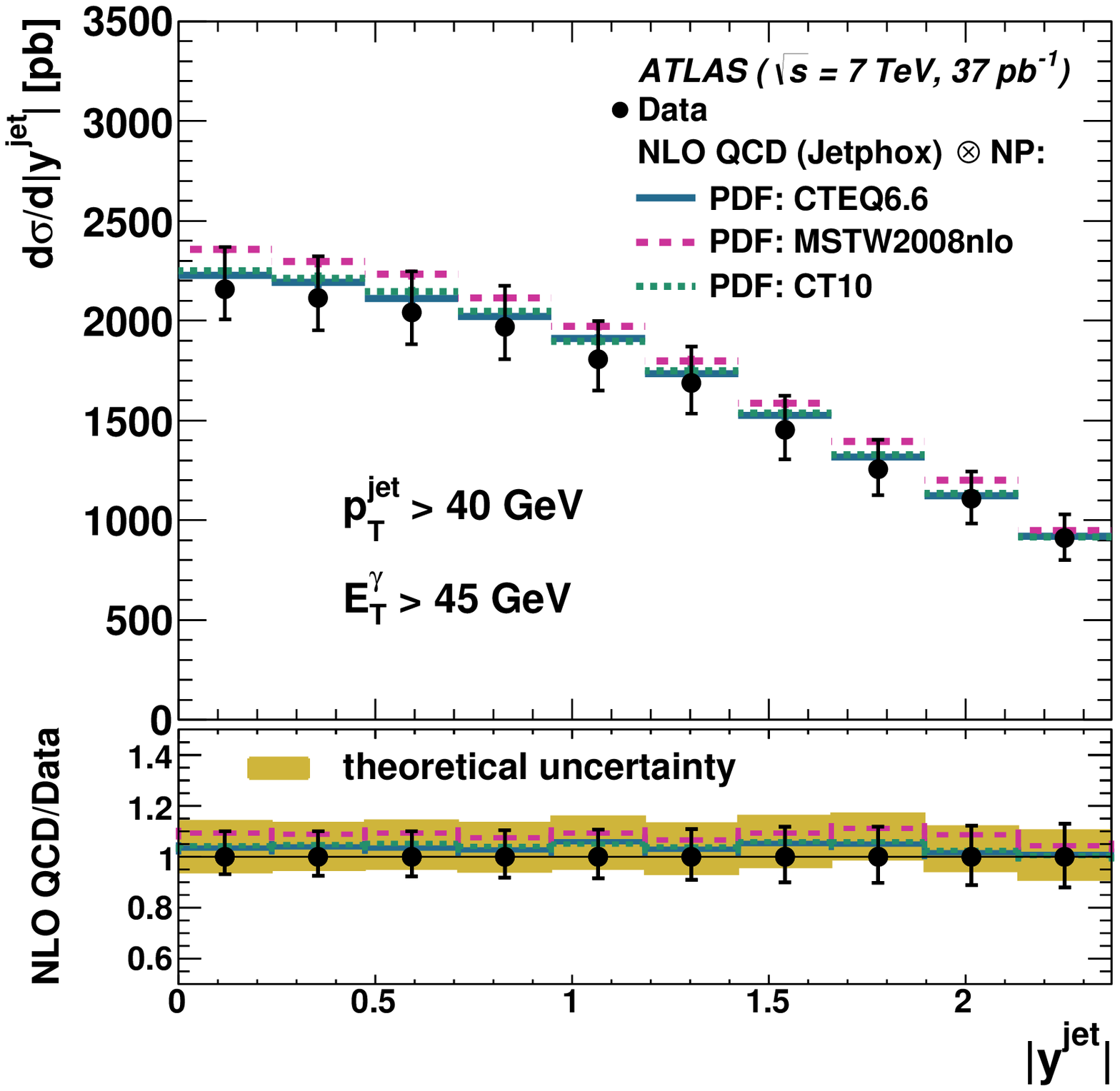,width=9cm}}}
\end{picture}
\vspace{-1cm}
\caption
{The measured bin-averaged cross section for isolated-photon plus jet
  production (dots) as a function of $|\rapjet|$. Other details as in
  the caption to Fig.~\ref{fig101}.}
\label{fig103}
\end{figure}

\begin{figure}[p]
\setlength{\unitlength}{1.0cm}
\begin{picture} (10.0,10.0)
\put (0.0,0.0){\centerline{\epsfig{figure=\figdir 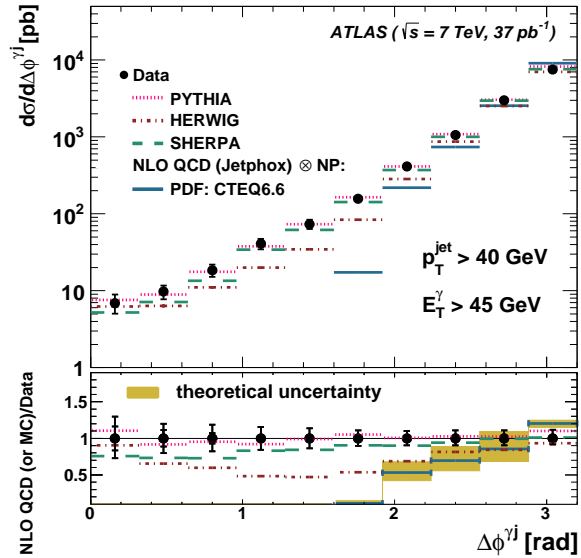,width=9cm}}}
\end{picture}
\vspace{-1cm}
\caption
{The measured bin-averaged cross section for isolated-photon plus jet
  production (dots) as a function of $\delphj$. The predictions from
  the leading-logarithm parton-shower models of {\sc Pythia} (dotted
  lines), {\sc Herwig} (dot-dashed lines) and {\sc Sherpa} (long
  dashed lines) are also shown. Other details as in the caption to
  Fig.~\ref{fig101}.}
\label{fig104}
\end{figure}

\begin{figure}[p]
\setlength{\unitlength}{1.0cm}
\begin{picture} (10.0,10.0)
\put (0.0,0.0){\centerline{\epsfig{figure=\figdir 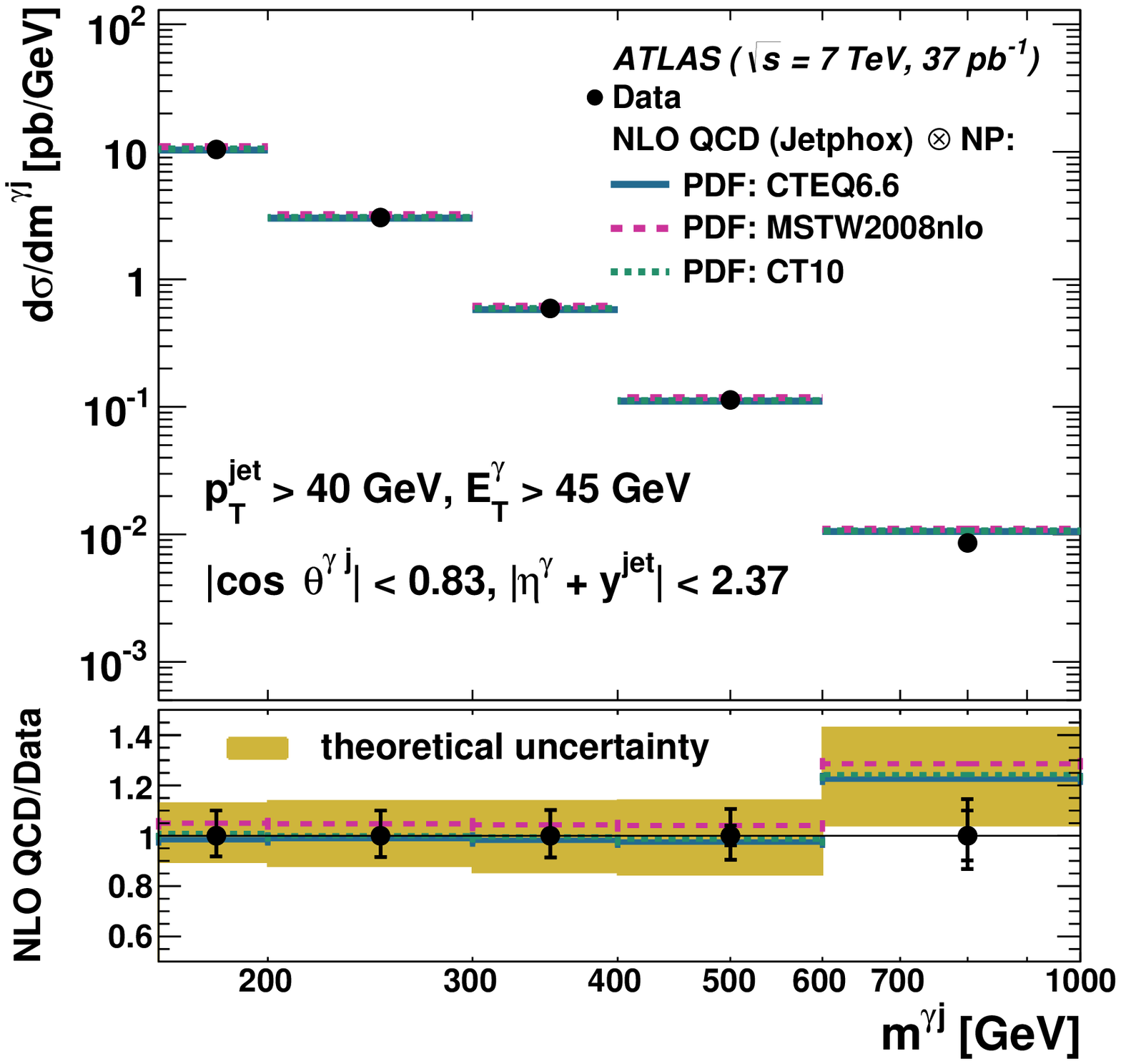,width=9cm}}}
\end{picture}
\vspace{-1cm}
\caption
{The measured bin-averaged cross section for isolated-photon plus jet
  production (dots) as a function of $\mgjn$ including the
  requirements on $|\ctgjn|$ and $|\etag+\rapjet|$. Other details as
  in the caption to Fig.~\ref{fig101}.}
\label{fig105}
\end{figure}

\begin{figure}[p]
\setlength{\unitlength}{1.0cm}
\begin{picture} (10.0,9.0)
\put (0.0,0.0){\centerline{\epsfig{figure=\figdir 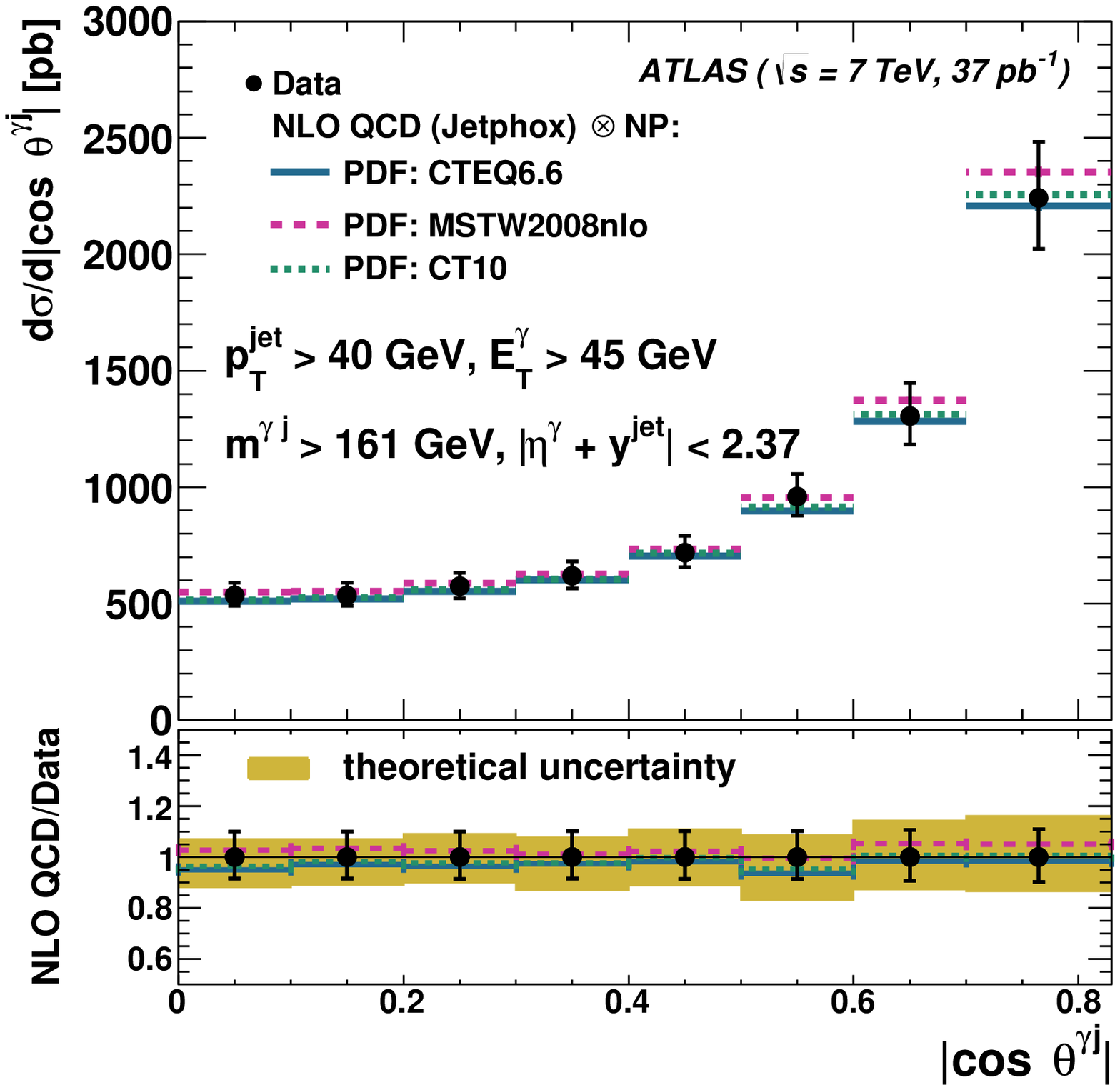,width=9cm}}}
\end{picture}
\vspace{-1cm}
\caption
{The measured bin-averaged cross section for isolated-photon plus jet
  production (dots) as a function of $|\ctgjn|$ including the
  requirements on $\mgjn$ and $|\etag+\rapjet|$. Other details as in
  the caption to Fig.~\ref{fig101}.}
\label{fig106}
\end{figure}

The measured bin-averaged cross sections are presented in
Figs.~\ref{fig101}--\ref{fig106} and
Tables~\ref{tab3}--\ref{tab8}. The measured $d\sigma/d\etg$ and
$d\sigma/d\ptjet$ fall by three orders of magnitude in the measured
range. The measured $d\sigma/d|\rapjet|$ and $d\sigma/d\delphj$
display a maximum at $|\rapjet|\approx 0$ and $\delphj\approx\pi$,
respectively. The measured $d\sigma/d\mgjn$ ($d\sigma/d|\ctgjn|$)
decreases (increases) as $\mgjn$ ($|\ctgjn|$) increases.

The predictions of the NLO QCD calculations from the {\sc Jetphox}
program described in Section~\ref{nlo} and corrected for hadronisation
and underlying-event effects are compared to the data in
Figs.~\ref{fig101}--\ref{fig106}. The predictions give a good
description of the $\etg$ and $\ptjet$ measured cross sections. The
shape and normalisation of the measured cross section as a function of
$|\rapjet|$ is described well by the calculation in the whole range
measured. For the maximum three-body final state of the NLO QCD
calculations, the photon and the leading jet cannot be in the same
hemisphere in the transverse plane, i.e. $\delphj$ is necessarily
larger than $\pi/2$; as a result, it is not unexpected that they fail
to describe the measured $\delphj$ distribution. The leading-logarithm
parton-shower predictions of the {\sc Pythia}, {\sc Herwig} and {\sc
  Sherpa} MC models are also shown in Fig.~\ref{fig104}; {\sc Pythia}
and {\sc Sherpa} give a good description of the data in the whole
range measured whereas {\sc Herwig} fails to do so. The measured cross
sections as functions of $\mgjn$ and $|\ctgjn|$ are described well by
the NLO QCD calculations.

The NLO QCD calculations based on the CT10 and MSTW2008nlo proton PDF
sets are within the uncertainty band of the CTEQ6.6-based
calculations. The shapes of the distributions from the three
calculations are similar. The predictions based on the CTEQ6.6 and
CT10 PDF sets are very similar in normalisation whereas those based on
MSTW2008nlo are approximately $5\%$ higher. All of these comparisons
validate the description of the dynamics of isolated-photon plus jet
production in $pp$ collisions at ${\cal O}(\alpha_{\rm em}\as^2)$.

\begin{figure}[p]
\setlength{\unitlength}{1.0cm}
\begin{picture} (10.0,8.0)
\put (0.0,0.0){\centerline{\epsfig{figure=\figdir 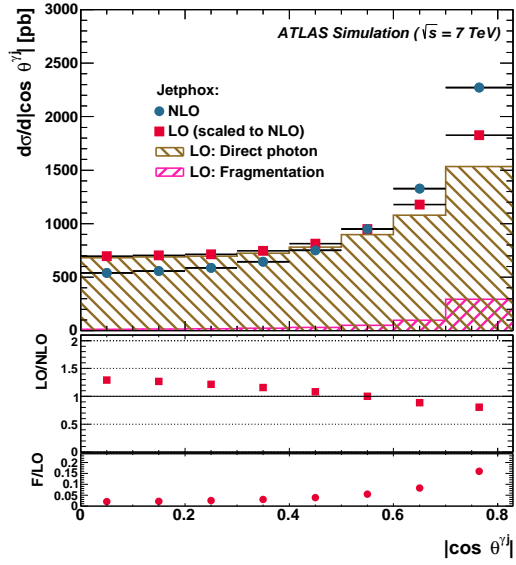,width=8cm}}}
\end{picture}
\caption
{The NLO QCD predicted bin-averaged cross section for isolated-photon
  plus jet production as a function of $|\ctgjn|$ including the
  requirements on $\mgjn$ and $|\etag+\rapjet|$ (dots). The LO QCD
  calculation (squares) scaled to the NLO integrated cross section and
  the contributions of the direct-photon (right-hatched histogram) and
  fragmentation (left-hatched histogram) components are also
  shown. The middle part of the figure shows the ratio of the scaled
  LO to the NLO QCD calculations (squares); the bottom part of the
  figure shows the ratio of the fragmentation component to the full LO
  calculation (dots).}
\label{fig106n}
\end{figure}

To gain further insight into the interpretation of the results, LO QCD
predictions of the direct-photon and fragmentation contributions to
the cross section were calculated. Even though at NLO the two
components are no longer distinguishable, the LO calculations are
useful to identify regions of phase space dominated by the
fragmentation contribution and to illustrate the basic differences in
the dynamics of the two processes. The ratio LO/NLO does (not) show a
strong dependence on $\ptjet$ and $|\ctgjn|$ ($\etg$, $|\rapjet|$ and
$\mgjn$). The LO and NLO QCD calculations as functions of $|\ctgjn|$
are compared in Fig.~\ref{fig106n}. The fragmentation contribution is
observed to decrease as a function of $\etg$, $\ptjet$ and $\mgjn$ and
is approximately constant as a function of $|\rapjet|$. However, it
increases as a function of $|\ctgjn|$ from $2\%$ up to
$16\%$. Therefore, the regions at low $\etg$, $\ptjet$ and $\mgjn$ as
well as large $|\ctgjn|$ are expected to be sensitive to the
fragmentation contribution.

\begin{figure}[p]
\setlength{\unitlength}{1.0cm}
\begin{picture} (9.0,8.0)
\put (0.0,0.0){\centerline{\epsfig{figure=\figdir 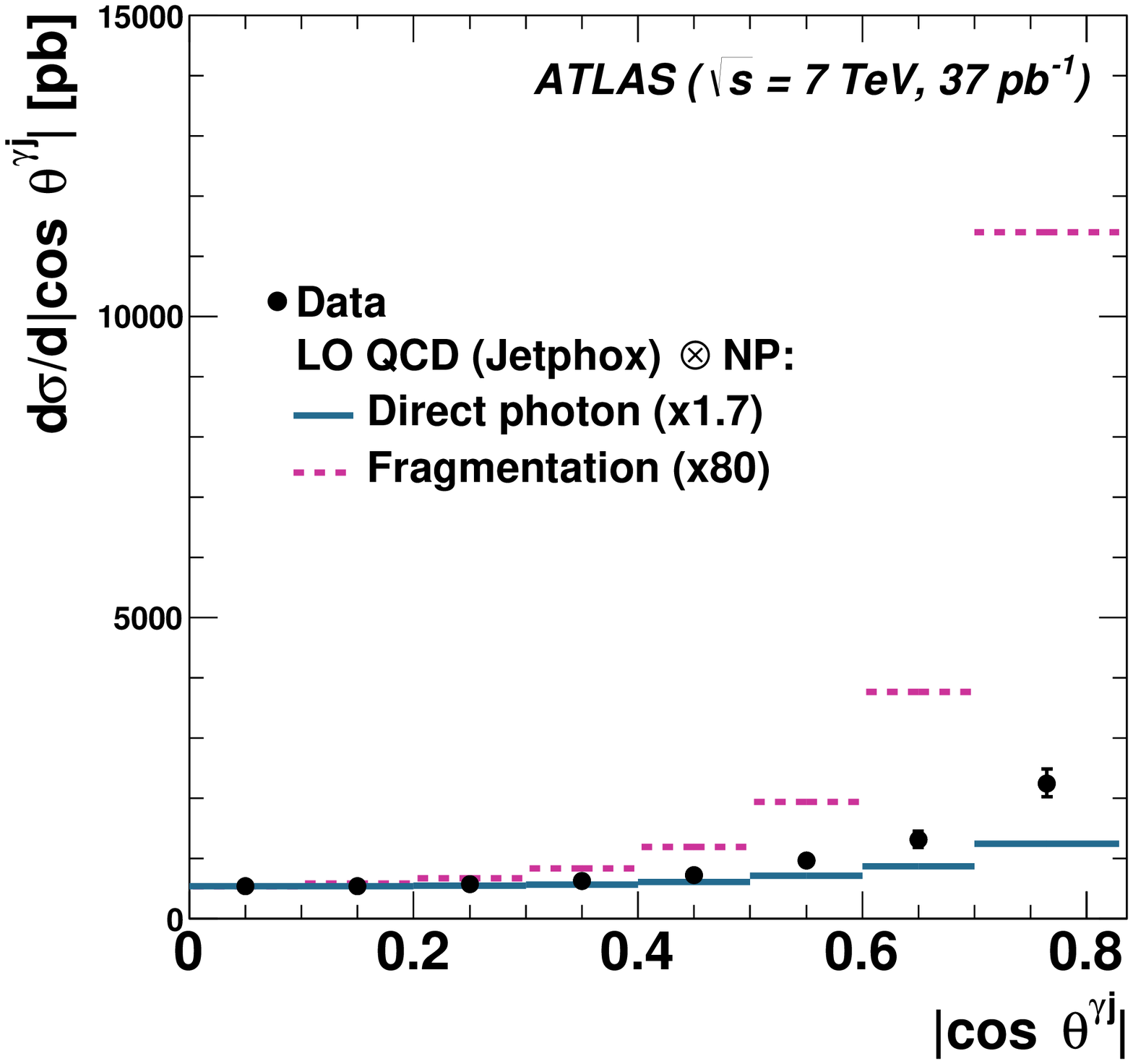,width=7cm}}}
\end{picture}
\caption
{The measured bin-averaged cross section for isolated-photon plus jet
  production (dots) as a function of $|\ctgjn|$ including the
  requirements on $\mgjn$ and $|\etag+\rapjet|$. The direct-photon
  (solid lines) and fragmentation (dashed lines) components of the LO
  QCD prediction are also included. The calculations were normalised
  to the measured cross section for $|\ctgjn|<0.1$; the factors used
  are shown in parentheses.}
\label{fig106nn}
\end{figure}

The shapes of the bin-averaged cross sections for the direct-photon
and fragmentation contributions at LO QCD were compared. The major
difference is seen in the bin-averaged cross section as a function of
$|\ctgjn|$ (see Fig.~\ref{fig106nn}), with the contribution from
fragmentation showing a steeper increase as $|\ctgjn|\rightarrow 1$
than that of direct-photon processes. This different behaviour is due
to the different spin of the exchanged particle dominating each of the
processes: a quark in the case of direct processes and a gluon in the
case of fragmentation processes. Therefore, the distribution in
$|\ctgjn|$ is particularly useful to study the dynamics underlying the
hard process and the relative contributions of direct processes and
fragmentation. The fact that the shape of the measured cross section
$d\sigma/d|\ctgjn|$ is much closer to that of the direct-photon
processes than that of fragmentation is consistent with the dominance
of processes in which the exchanged particle is a quark. Furthermore,
the increase of the cross section as $|\ctgjn|\rightarrow 1$ observed
in the data is milder than that measured in dijet production in $pp$
collisions~\cite{epj:c71:1512}, which is dominated by gluon exchange.

\begin{figure}[p]
\setlength{\unitlength}{1.0cm}
\begin{picture} (10.0,9.0)
\put (0.0,0.0){\centerline{\epsfig{figure=\figdir 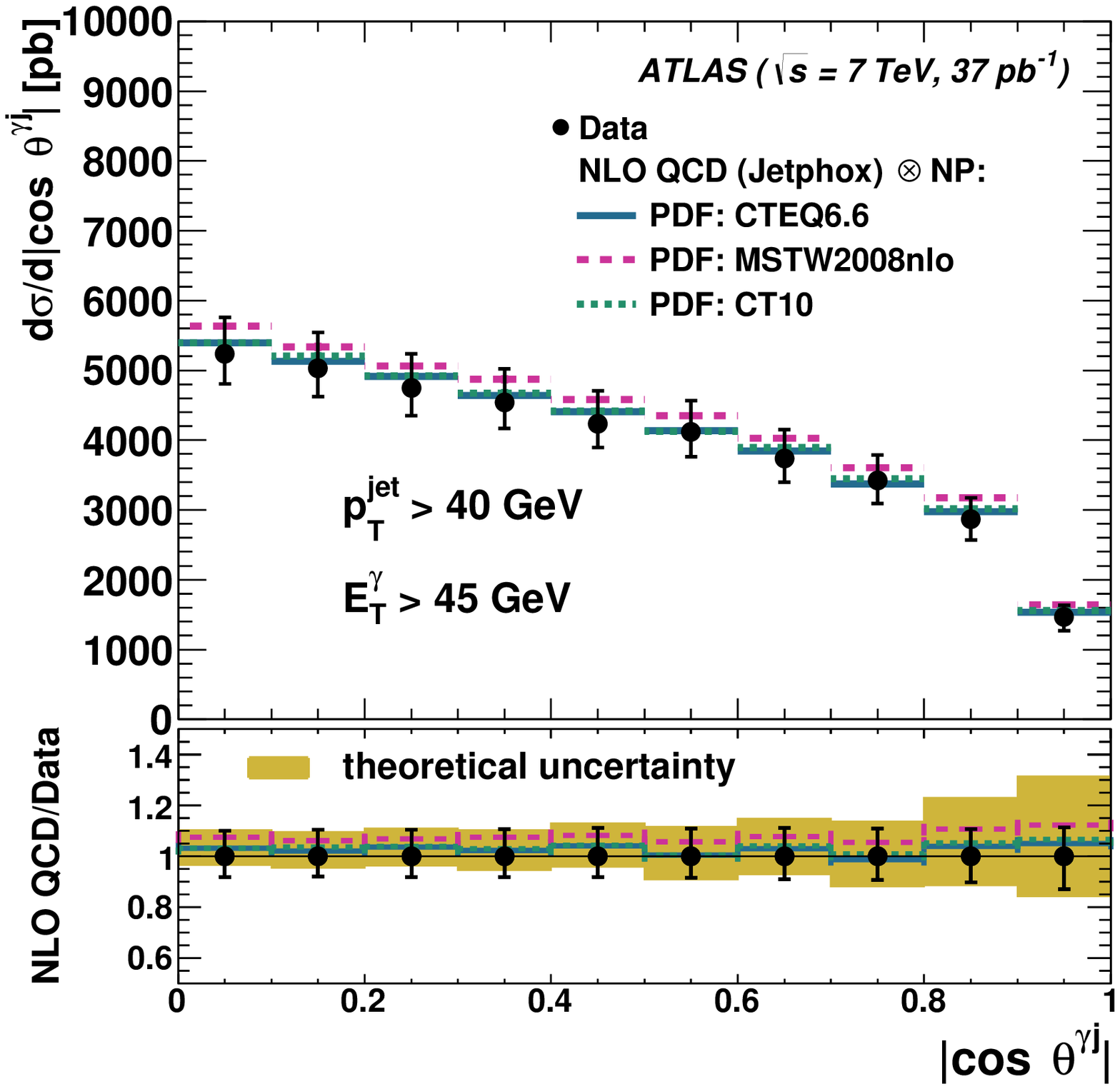,width=9cm}}}
\end{picture}
\vspace{-1cm}
\caption
{The measured bin-averaged cross section for isolated-photon plus jet
  production (dots) as a function of $|\ctgjn|$ without the
  requirements on $\mgjn$ and $|\etag+\rapjet|$. Other details as in
  the caption to Fig.~\ref{fig101}.}
\label{fig138}
\end{figure}

\begin{figure}[p]
\setlength{\unitlength}{1.0cm}
\begin{picture}  (10.0,9.5)
\put (0.0,0.0){\centerline{\epsfig{figure=\figdir 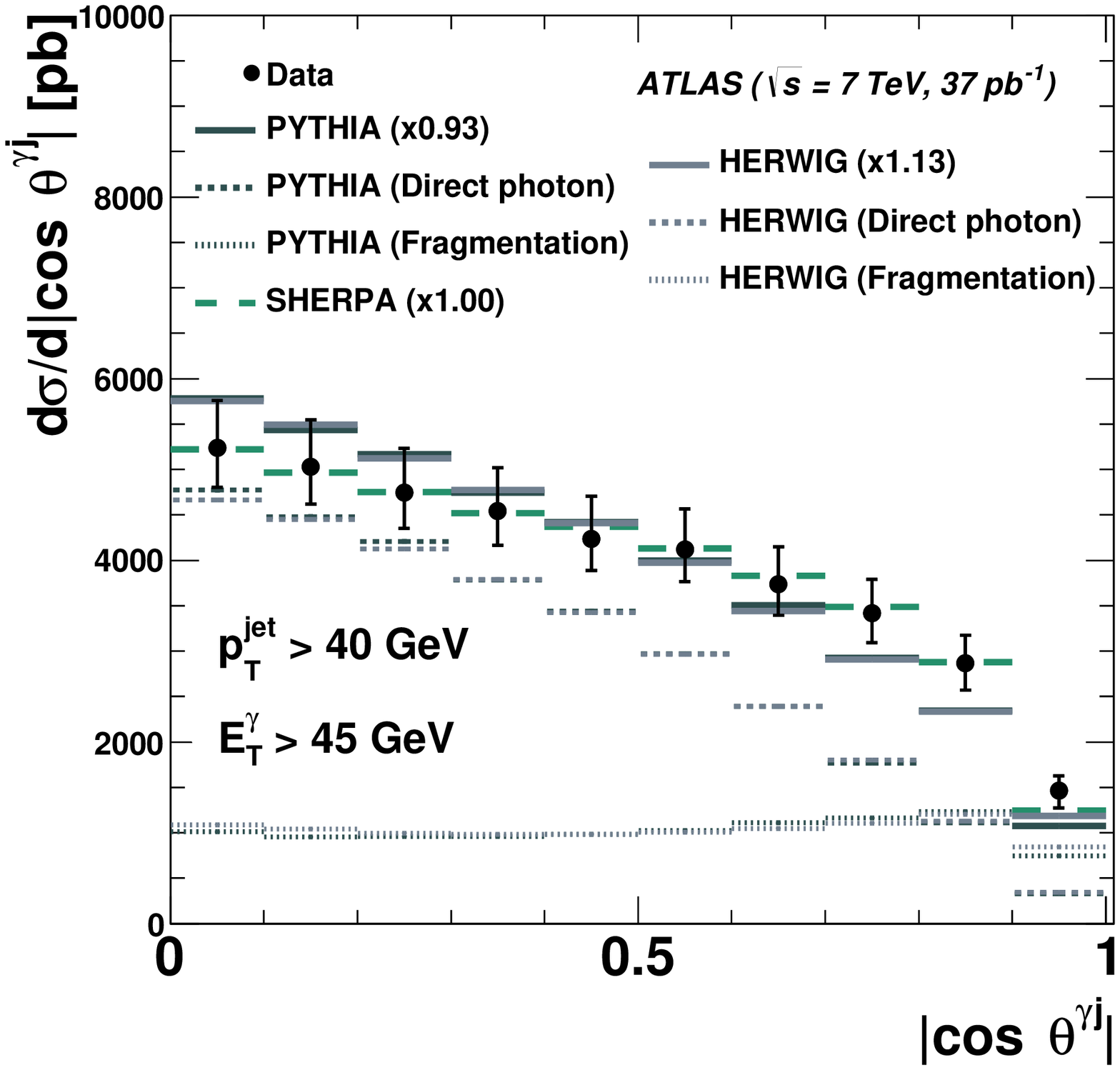,width=7cm}}}
\end{picture}
\caption
{The measured bin-averaged cross section for isolated-photon plus jet
  production (dots) as a function of $|\ctgjn|$ without the
  requirements on $\mgjn$ and $|\etag+\rapjet|$. The {\sc Pythia}
  (dark lines) and {\sc Herwig} (light lines) MC calculations for the
  direct-photon (dashed lines), fragmentation  (dotted lines)
  components and their sum (solid lines) are also shown. The
  prediction from {\sc Sherpa} (long dashed lines) is also
  included. The full MC calculations are normalised to the integrated
  measured cross section. Other details as in the caption to
  Fig.~\ref{fig101}.}
\label{fig138ct1}
\end{figure}

The measurement of the bin-averaged cross section as a function of
$|\ctgjn|$ without the requirements on $\mgjn$ and $|\etag+\rapjet|$
is presented in Fig.~\ref{fig138} and Table~\ref{tab9}. The decrease
of the bin-averaged cross section as $|\ctgjn|$ increases is due to
the non-uniform coverage in $|\ctgjn|$ induced by the requirements on
the photon and jet rapidities and transverse momenta. The NLO QCD
calculations are compared to the data in the same figure; they give a
good description of the measured bin-averaged cross section. The
comparison of the data to the predictions of {\sc Pythia}, {\sc
  Herwig} and {\sc Sherpa} is shown in Fig.~\ref{fig138ct1}; in this
figure, the MC calculations are normalised to the integrated measured
cross section. The shapes of the predictions from {\sc Pythia} and
{\sc Herwig} are very similar and do not describe the measured cross
section. In these predictions, the contributions of direct-photon and
fragmentation processes were added according to the MC default cross
sections. It is possible to improve the description of the measured
cross section by adjusting the relative contribution of the
subprocesses, as demonstrated in Fig.~\ref{fig1202} for the estimated
signal yield. In contrast, the prediction of {\sc Sherpa} gives a good
description of the measured cross section, both in shape and
magnitude; this may be attributable to the inclusion of higher-order
contributions at tree-level in the prediction. The studies summarised
in Figs.~\ref{fig138} and \ref{fig138ct1} give insight into the
characteristics of one of the primary backgrounds in the study of the
new particle discovered by ATLAS~\cite{pl:b716:1} and
CMS~\cite{pl:b716:30} in the search for the Higgs boson.

\section{Summary and conclusions}
\label{sum}
Bin-averaged cross sections for isolated photons in association with a
jet in $7$~TeV proton--proton collisions, 
$pp\rightarrow\gamma+{\rm jet}+{\rm X}$, have been presented using an
integrated luminosity of $37.1$~\pb1. The jets were reconstructed
using the anti-$\kt$ jet algorithm with $R=0.6$. Isolated-photon plus
jet bin-averaged cross sections were measured as functions of $\etg$,
$\ptjet$, $|\rapjet|$, $\delphj$, $\mgjn$ and $\ctgjn$. The
bin-averaged cross sections $d\sigma/d\mgjn$ and $d\sigma/d|\ctgjn|$
were measured with additional selection criteria on $|\etag+\rapjet|$,
$|\ctgjn|$ and $\mgjn$.

Regions of phase space sensitive to the contributions from
fragmentation have been identified. As a result, these measurements
can be used to tune the relative contributions of direct and
fragmentation processes in the description of isolated-photon
production by the Monte Carlo models.

The NLO QCD calculations, based on various proton PDFs and corrected
for hadronisation and underlying-event effects using {\sc Pythia} and
{\sc Herwig}, have been compared to the measurements. The calculations
give a reasonably good description of the measured cross sections both
in shape and normalisation, except for $\delphj$; this distribution is
adequately described by the leading-order plus parton-shower
prediction of {\sc Pythia} or {\sc Sherpa}. The measured dependence on
$|\ctgjn|$ is consistent with the dominance of processes in which a
quark is being exchanged.

A measurement of the bin-averaged cross section as a function of
$|\ctgjn|$ without the requirements on $\mgjn$ and $|\etag+\rapjet|$
was also presented to understand the photon plus jet background
relevant for the studies of the spin of the new particle observed by
ATLAS and CMS in the search for the Higgs boson. The NLO QCD
calculations give a good description of the data.

\section{Acknowledgements}

We thank CERN for the very successful operation of the LHC, as well as
the support staff from our institutions without whom ATLAS could not
be operated efficiently.

We acknowledge the support of ANPCyT, Argentina; YerPhI, Armenia; ARC,
Australia; BMWF and FWF, Austria; ANAS, Azerbaijan; SSTC, Belarus;
CNPq and FAPESP, Brazil; NSERC, NRC and CFI, Canada; CERN; CONICYT,
Chile; CAS, MOST and NSFC, China; COLCIENCIAS, Colombia; MSMT CR, MPO
CR and VSC CR, Czech Republic; DNRF, DNSRC and Lundbeck Foundation,
Denmark; EPLANET, ERC and NSRF, European Union; IN2P3-CNRS,
CEA-DSM/IRFU, France; GNSF, Georgia; BMBF, DFG, HGF, MPG and AvH
Foundation, Germany; GSRT and NSRF, Greece; ISF, MINERVA, GIF, DIP and
Benoziyo Center, Israel; INFN, Italy; MEXT and JSPS, Japan; CNRST,
Morocco; FOM and NWO, Netherlands; BRF and RCN, Norway; MNiSW, Poland;
GRICES and FCT, Portugal; MERYS (MECTS), Romania; MES of Russia and
ROSATOM, Russian Federation; JINR; MSTD, Serbia; MSSR, Slovakia; ARRS
and MIZ\v{S}, Slovenia; DST/NRF, South Africa; MICINN, Spain; SRC and
Wallenberg Foundation, Sweden; SER, SNSF and Cantons of Bern and
Geneva, Switzerland; NSC, Taiwan; TAEK, Turkey; STFC, the Royal
Society and Leverhulme Trust, United Kingdom; DOE and NSF, United
States of America.

The crucial computing support from all WLCG partners is acknowledged
gratefully, in particular from CERN and the ATLAS Tier-1 facilities at
TRIUMF (Canada), NDGF (Denmark, Norway, Sweden), CC-IN2P3 (France),
KIT/GridKA (Germany), INFN-CNAF (Italy), NL-T1 (Netherlands), PIC
(Spain), ASGC (Taiwan), RAL (UK) and BNL (USA) and in the Tier-2
facilities worldwide.

\clearpage
\newpage
\begin{table}
 \caption
{The measured bin-averaged cross-section $d\sigma/d\etg$ for
  isolated-photon plus jet production. The statistical ($\delta_{\rm
    stat}$) and systematic ($\delta_{\rm syst}$) uncertainties are
  shown separately. The corrections for hadronisation and
  underlying-event effects to be applied to the parton-level NLO QCD
  calculations ($C_{\rm NLO}$) are shown in the last column. All
  tables with information on the measured cross sections, their
  uncertainties and correlations are available in HepData.}
\begin{center}
    \begin{tabular}{|c||c|c|c||c|}
\hline
  $\etg$
& $d\sigma/d\etg$
& $\delta_{\rm stat}$
& $\delta_{\rm syst}$
& $C_{\rm NLO}$\\

  [GeV]
& [pb/GeV]
& [pb/GeV]
& [pb/GeV]
& \\
\hline\hline
$45-55$ & $160.2$ & $\pm 0.9$ & $^{+20.6}_{-17.1}$ & $0.97$ \\
$55-70$ & $ 81.1$ & $\pm 0.5$ &   $^{+8.1}_{-6.7}$ & $0.95$ \\ 
$70-85$ & $ 35.39$ & $\pm 0.32$ & $^{+3.00}_{-2.62}$ & $0.94$ \\ 
$85-100$ & $16.75$ & $\pm 0.21$ & $^{+1.30}_{-1.11}$ & $0.92$ \\ 
$100-125$ & $6.89$ & $\pm 0.10$ & $^{+0.52}_{-0.45}$ & $0.92$ \\ 
$125-150$ & $2.58$ & $\pm 0.06$ & $^{+0.19}_{-0.16}$ & $0.92$ \\ 
$150-200$ & $0.789$ & $\pm 0.025$ & $^{+0.054}_{-0.048}$ & $0.90$ \\ 
$200-400$ & $0.081$ & $\pm 0.004$ & $^{+0.005}_{-0.005}$ & $0.91$ \\ 
\hline
    \end{tabular}
 \label{tab3}
\end{center}
\end{table}

\begin{table}
 \caption
{The measured bin-averaged cross-section $d\sigma/d\ptjet$ for
  isolated-photon plus jet production. Other details as in the caption
  to Table~\ref{tab3}.}
\begin{center}
    \begin{tabular}{|c||c|c|c||c|}
\hline
  $\ptjet$
& $d\sigma/d\ptjet$
& $\delta_{\rm stat}$
& $\delta_{\rm syst}$
& $C_{\rm NLO}$\\

  [GeV]
& [pb/GeV]
& [pb/GeV]
& [pb/GeV]
& \\
\hline\hline
$40-55$ & $107.6$ & $\pm 0.6$ & $^{+12.3}_{-10.0}$ & $0.96$ \\ 
$55-70$ & $70.1$ & $\pm 0.5$ & $^{+8.2}_{-6.7}$ & $0.98$ \\ 
$70-85$ & $36.08$ & $\pm 0.31$ & $^{+4.34}_{-3.61}$ & $0.96$ \\ 
$85-100$ & $18.99$ & $\pm 0.22$ & $^{+2.21}_{-1.98}$ & $0.94$ \\ 
$100-125$ & $8.86$ & $\pm 0.11$ & $^{+1.11}_{-1.00}$ & $0.91$ \\ 
$125-150$ & $3.74$ & $\pm 0.07$ & $^{+0.50}_{-0.44}$ & $0.89$ \\ 
$150-200$ & $1.379$ & $\pm 0.031$ & $^{+0.194}_{-0.179}$ & $0.86$ \\ 
$200-400$ & $0.167$ & $\pm 0.005$ & $^{+0.026}_{-0.022}$ & $0.85$ \\ 
\hline
    \end{tabular}
 \label{tab4}
\end{center}
\end{table}

\begin{table}
 \caption
{The measured bin-averaged cross-section $d\sigma/d|\rapjet|$ for
  isolated-photon plus jet production. Other details as in the caption
  to Table~\ref{tab3}.}
\begin{center}
    \begin{tabular}{|c||c|c|c||c|}
\hline
  $|\rapjet|$
& $d\sigma/d|\rapjet|$
& $\delta_{\rm stat}$
& $\delta_{\rm syst}$
& $C_{\rm NLO}$\\
&  [pb]
& [pb]
& [pb]
& \\
\hline\hline
$0.000-0.237$ & $2158$ & $\pm 20$ & $^{+211}_{-148}$ & $0.96$ \\ 
$0.237-0.474$ & $2113$ & $\pm 20$ & $^{+208}_{-161}$ & $0.96$ \\ 
$0.474-0.711$ & $2043$ & $\pm 20$ & $^{+203}_{-159}$ & $0.96$ \\ 
$0.711-0.948$ & $1968$ & $\pm 20$ & $^{+204}_{-160}$ & $0.96$ \\ 
$0.948-1.185$ & $1806$ & $\pm 19$ & $^{+191}_{-153}$ & $0.96$ \\ 
$1.185-1.422$ & $1687$ & $\pm 18$ & $^{+183}_{-153}$ & $0.96$ \\ 
$1.422-1.659$ & $1452$ & $\pm 17$ & $^{+171}_{-147}$ & $0.96$ \\ 
$1.659-1.896$ & $1256$ & $\pm 16$ & $^{+147}_{-130}$ & $0.96$ \\ 
$1.896-2.133$ & $1108$ & $\pm 15$ & $^{+135}_{-123}$ & $0.96$ \\ 
$2.133-2.370$ & $912$ & $\pm 14$ & $^{+117}_{-111}$ & $0.95$ \\ 
\hline
    \end{tabular}
 \label{tab5}
\end{center}
\end{table}

\begin{table}
 \caption
{The measured bin-averaged cross-section $d\sigma/d\delphj$ for
  isolated-photon plus jet production. Other details as in the caption
  to Table~\ref{tab3}.}
\begin{center}
    \begin{tabular}{|c||c|c|c||c|}
\hline
  $\delphj$
& $d\sigma/d\delphj$
& $\delta_{\rm stat}$
& $\delta_{\rm syst}$
& $C_{\rm NLO}$\\

[rad]
&  [pb]
& [pb]
& [pb]
& \\
\hline\hline
$0.00-0.32$ & $6.9$ & $\pm 1.1$ & $^{+1.7}_{-1.5}$ & $-$ \\ 
$0.32-0.64$ & $9.7$ & $\pm 1.1$ & $^{+1.6}_{-1.6}$ & $-$ \\ 
$0.64-0.96$ & $18.5$ & $\pm 1.3$ & $^{+3.2}_{-3.0}$ & $-$ \\ 
$0.96-1.28$ & $41.0$ & $\pm 2.2$ & $^{+5.9}_{-6.1}$ & $-$ \\ 
$1.28-1.60$ & $73.6$ & $\pm 2.9$ & $^{+9.7}_{-9.5}$ & $-$ \\ 
$1.60-1.92$ & $156$ & $\pm 4$ & $^{+16}_{-16}$ & $0.91$ \\ 
$1.92-2.24$ & $412$ & $\pm 8$ & $^{+41}_{-38}$ & $0.96$ \\ 
$2.24-2.56$ & $1063$ & $\pm 12$ & $^{+113}_{-101}$ & $0.95$ \\ 
$2.56-2.88$ & $2985$ & $\pm 21$ & $^{+328}_{-281}$ & $0.96$ \\ 
$2.88-3.20$ & $7518$ & $\pm 34$ & $^{+868}_{-623}$ & $0.95$ \\ 
\hline
    \end{tabular}
 \label{tab6}
\end{center}
\end{table}

\begin{table}
 \caption
{The measured bin-averaged cross-section $d\sigma/d\mgjn$ with the
  requirements on $|\ctgjn|$ and $|\etag+\rapjet|$ for isolated-photon
  plus jet production. Other details as in the caption to
  Table~\ref{tab3}.}
\begin{center}
    \begin{tabular}{|c||c|c|c||c|}
\hline
  $\mgjn$
& $d\sigma/d\mgjn$
& $\delta_{\rm stat}$
& $\delta_{\rm syst}$
& $C_{\rm NLO}$\\

 [GeV]
&  [pb/GeV]
& [pb/GeV]
& [pb/GeV]
& \\
\hline\hline
$161-200$ & $10.46$ & $\pm 0.11$ & $^{+1.03}_{-0.86}$ & $0.97$ \\ 
$200-300$ & $3.069$ & $\pm 0.034$ & $^{+0.303}_{-0.255}$ & $0.95$ \\ 
$300-400$ & $0.594$ & $\pm 0.015$ & $^{+0.058}_{-0.050}$ & $0.92$ \\ 
$400-600$ & $0.114$ & $\pm 0.005$ & $^{+0.011}_{-0.010}$ & $0.91$ \\ 
$600-1000$ & $0.0086$ & $\pm 0.0009$ & $^{+0.0009}_{-0.0008}$ & $0.91$ \\ 
\hline
    \end{tabular}
 \label{tab7}
\end{center}
\end{table}

\begin{table}
 \caption
{The measured bin-averaged cross-section $d\sigma/d|\ctgjn|$ with the
  requirements on $\mgjn$ and $|\etag+\rapjet|$ for isolated-photon
  plus jet production. Other details as in the caption to
  Table~\ref{tab3}.}
\begin{center}
    \begin{tabular}{|c||c|c|c||c|}
\hline
  $|\ctgjn|$
& $d\sigma/d|\ctgjn|$
& $\delta_{\rm stat}$
& $\delta_{\rm syst}$
& $C_{\rm NLO}$\\

&  [pb]
& [pb]
& [pb]
& \\
\hline\hline
$0.00-0.10$ & $536$ & $\pm 14$ & $^{+52}_{-43}$ & $0.94$ \\ 
$0.10-0.20$ & $536$ & $\pm 14$ & $^{+52}_{-44}$ & $0.93$ \\ 
$0.20-0.30$ & $574$ & $\pm 15$ & $^{+55}_{-48}$ & $0.94$ \\ 
$0.30-0.40$ & $619$ & $\pm 15$ & $^{+61}_{-51}$ & $0.93$ \\ 
$0.40-0.50$ & $718$ & $\pm 17$ & $^{+71}_{-60}$ & $0.94$ \\ 
$0.50-0.60$ & $960$ & $\pm 19$ & $^{+94}_{-81}$ & $0.95$ \\ 
$0.60-0.70$ & $1306$ & $\pm 23$ & $^{+137}_{-120}$ & $0.97$ \\ 
$0.70-0.83$ & $2242$ & $\pm 29$ & $^{+239}_{-218}$ & $0.97$ \\ 
\hline
    \end{tabular}
 \label{tab8}
\end{center}
\end{table}

\begin{table}
 \caption
{The measured bin-averaged cross-section $d\sigma/d|\ctgjn|$ without
  the requirements on $\mgjn$ and $|\etag+\rapjet|$ for
  isolated-photon plus jet production. Other details as in the caption
  to Table~\ref{tab3}.}
\begin{center}
    \begin{tabular}{|c||c|c|c||c|}
\hline
  $|\ctgjn|$
& $d\sigma/d|\ctgjn|$
& $\delta_{\rm stat}$
& $\delta_{\rm syst}$
& $C_{\rm NLO}$\\

&  [pb]
& [pb]
& [pb]
& \\
\hline\hline
$0.0-0.1$ & $5240$ & $\pm 50$ & $^{+520}_{-430}$ & $0.95$ \\ 
$0.1-0.2$ & $5030$ & $\pm 50$ & $^{+520}_{-410}$ & $0.95$ \\ 
$0.2-0.3$ & $4750$ & $\pm 50$ & $^{+490}_{-390}$ & $0.95$ \\ 
$0.3-0.4$ & $4540$ & $\pm 50$ & $^{+480}_{-370}$ & $0.96$ \\ 
$0.4-0.5$ & $4240$ & $\pm 40$ & $^{+470}_{-340}$ & $0.95$ \\ 
$0.5-0.6$ & $4120$ & $\pm 40$ & $^{+450}_{-350}$ & $0.95$ \\ 
$0.6-0.7$ & $3740$ & $\pm 40$ & $^{+410}_{-340}$ & $0.96$ \\ 
$0.7-0.8$ & $3420$ & $\pm 40$ & $^{+370}_{-320}$ & $0.95$ \\ 
$0.8-0.9$ & $2870$ & $\pm 40$ & $^{+300}_{-300}$ & $0.96$ \\ 
$0.9-1.0$ & $1460$ & $\pm 30$ & $^{+160}_{-190}$ & $0.95$ \\ 
\hline
    \end{tabular}
 \label{tab9}
\end{center}
\end{table}

\newpage
\clearpage
\providecommand{\etal}{et al.\xspace}
\providecommand{\coll}{Coll.\xspace}
\catcode`\@=11
\def\@bibitem#1{%
\ifmc@bstsupport
  \mc@iftail{#1}%
    {;\newline\ignorespaces}%
    {\ifmc@first\else.\fi\orig@bibitem{#1}}
  \mc@firstfalse
\else
  \mc@iftail{#1}%
    {\ignorespaces}%
    {\orig@bibitem{#1}}%
\fi}%
\catcode`\@=12

\clearpage
\begin{flushleft}
{\Large The ATLAS Collaboration}

\bigskip

G.~Aad$^{\rm 48}$,
T.~Abajyan$^{\rm 21}$,
B.~Abbott$^{\rm 112}$,
J.~Abdallah$^{\rm 12}$,
S.~Abdel~Khalek$^{\rm 116}$,
A.A.~Abdelalim$^{\rm 49}$,
O.~Abdinov$^{\rm 11}$,
R.~Aben$^{\rm 106}$,
B.~Abi$^{\rm 113}$,
M.~Abolins$^{\rm 89}$
% ...
\bigskip
\\
% ...
$^{11}$ Institute of Physics, Azerbaijan Academy of Sciences, Baku, Azerbaijan\\
$^{12}$ Institut de F{\'\i}sica d'Altes Energies and Departament de F{\'\i}sica de la Universitat Aut{\`o}noma de Barcelona, Barcelona, Spain\\
% ...
$^{21}$ Physikalisches Institut, University of Bonn, Bonn, Germany\\
% ...
$^{48}$ Fakult{\"a}t f{\"u}r Mathematik und Physik, Albert-Ludwigs-Universit{\"a}t, Freiburg, Germany\\
$^{49}$ Section de Physique, Universit{\'e} de Gen{\`e}ve, Geneva, Switzerland\\
% ...
$^{89}$ Department of Physics and Astronomy, Michigan State University, East Lansing MI, United States of America\\
% ...
$^{106}$ Nikhef National Institute for Subatomic Physics and University of Amsterdam, Amsterdam, Netherlands\\
% ...
$^{112}$ Homer L. Dodge Department of Physics and Astronomy, University of Oklahoma, Norman OK, United States of America\\
$^{113}$ Department of Physics, Oklahoma State University, Stillwater OK, United States of America\\
% ...
$^{116}$ LAL, Universit{\'e} Paris-Sud and CNRS/IN2P3, Orsay, France\\
% ...
\end{flushleft}


\end{document}